\documentclass[aip, pof, amsmath, amssymb, amsfonts, preprint,unsortedaddress]{revtex4-2}

\usepackage{comment}

\usepackage{tablefootnote} % for table footnotes
\usepackage{graphicx}% Include figure files
\usepackage{dcolumn}% Align table columns on decimal point
\usepackage{bm}% bold math
\usepackage{xfrac}
\usepackage[utf8]{inputenc}
\usepackage[T1]{fontenc}
\usepackage{mathptmx}
\usepackage{etoolbox}

\usepackage[mathscr]{euscript}
\DeclareSymbolFont{rsfs}{U}{rsfs}{m}{n}
\DeclareSymbolFontAlphabet{\mathscrsfs}{rsfs}

\usepackage{caption}
\usepackage{subcaption}
\usepackage{multirow}
\usepackage{enumerate}
\usepackage{bm}
\usepackage[dvipsnames]{xcolor}

\newcommand{\Gp}{G^{\prime}(\omega)}
\newcommand{\Gpp}{G^{\prime\prime}(\omega)}
\newcommand{\De}{\text{De}}
\newcommand{\Wi}{\text{Wi}}
\newcommand{\ex}{\text{ex}}
\newcommand{\nl}{\text{nl}}
\newcommand{\e}{\bm{e}}
\newcommand{\q}{\hat{q}}

\begin{document}

\title{Harmonic Balance for Differential Constitutive Models under Oscillatory Shear}

\author{Shivangi Mittal}%
\email{shmi@iitk.ac.in}
\affiliation{Department of Chemical Engineering, Indian Institute of Technology, Kanpur, INDIA}

\author{Yogesh M. Joshi}%
\email{joshi@iitk.ac.in}
\affiliation{Department of Chemical Engineering, Indian Institute of Technology, Kanpur, INDIA}

\author{Sachin Shanbhag}%
\email{sshanbhag@fsu.edu}
\affiliation{Department of Scientific Computing, Florida State University, Tallahassee, FL 32306. USA}

\begin{abstract}
Harmonic balance (HB) is a popular Fourier-Galerkin method used in the analysis of nonlinear vibration problems where dynamical systems are subjected to periodic forcing. We adapt HB to find the periodic steady-state response of nonlinear differential constitutive models subjected to large amplitude oscillatory shear flow. By incorporating the alternating-frequency-time scheme into HB, we develop a computer program called FLASH (acronym for \textbf{F}ast Large \textbf{A}mplitude \textbf{S}imulation using \textbf{H}armonic balance), which makes it convenient to apply HB to \textit{any} differential constitutive model. We validate FLASH by considering two representative constitutive models, viz., the exponential Phan-Thien Tanner model and a nonlinear temporary network model. In terms of accuracy and speed, FLASH outperforms the conventional approach of solving initial value problems by numerical integration via time-stepping methods often by several orders of magnitude. We discuss how FLASH can be conveniently extended for other nonlinear constitutive models, which opens up potential applications in model calibration and selection, and stability analysis.
\end{abstract}

\keywords{differential constitutive models, large amplitude oscillatory shear, harmonic balance, alternating frequency time}

\maketitle

\section{Introduction}

Oscillatory shear (OS) flow is an important rheological protocol for understanding the viscoelastic behavior of complex fluids. In a typical OS experiment, a sinusoidal strain input $\gamma(t) = \gamma_0 \sin \omega t$ with amplitude $\gamma_0$ and angular frequency $\omega$ is applied to the test material, and the corresponding periodic stress outputs are monitored after all the transients have died out.  The periodic steady state (PSS) response of the shear and normal stresses generated are traditionally analyzed using the concepts of Fourier transform (FT) rheology in which stresses are represented as Fourier series expansions,\cite{Wilhelm2002,Giacomin1993,Hyun2011} 
\begin{equation}
    \sigma_{12} (t) =\gamma_0 \sum\limits_{n \in \text{odd}}   \left[ G_{n}^{\prime}(\omega, \gamma_0) \sin (n \omega t) + G_{n}^{\prime \prime}(\omega, \gamma_0) \cos (n \omega t) \right],
    \label{eqn:shearFourier}
\end{equation}
\begin{equation}
    \sigma_{11}(t)-\sigma_{22}(t)=N_1(t)= \gamma_0^2 \sum\limits_{n \in \text{even}}   \left[ F_{n}^{\prime}(\omega, \gamma_0) \sin (n \omega t) + F_{n}^{\prime \prime}(\omega, \gamma_0) \cos (n \omega t) \right],
    \label{eq:normalFourier}
\end{equation}
where $G_{n}^{\prime} (F_{n}^{\prime}) \text{ and } G_{n}^{\prime\prime}( F_{n}^{\prime \prime})$ are the Fourier moduli representing the in-phase and out-of-phase components of the shear stress (first normal stress difference) with respect to the applied strain deformation. The second normal stress difference $N_2=\sigma_{22}-\sigma_{33}$ is analogous to $N_1$, but is usually difficult to measure experimentally and is ignored in this work. Due to the symmetry of shear flows, even harmonics are absent in expansions of the shear stress, while odd harmonics are absent from normal stresses. 

In the linear limit of small amplitude oscillatory shear (SAOS), the stress response is perfectly sinusoidal,
\begin{equation}
   \sigma_{12}^\text{SAOS}(t) = \gamma_0 \left( \Gp \sin \omega t + \Gpp \cos \omega t\right),
\label{eqn:shearSAOS}
\end{equation}
where $\Gp$ and $\Gpp$ are equivalent to $G_1^\prime$ and $G_1^{\prime \prime}$ in eqn. \eqref{eqn:shearFourier}. The SAOS stress response can be considered as a special case of eqn. \eqref{eqn:shearFourier}. Nevertheless, it rests on sounder analytical foundations: its origin can be traced to the Boltzmann superposition principle, which states that outputs are linear superpositions of independent inputs.\cite{Cho2016} In the SAOS regime, $N_1$ reduces to
\begin{equation}
    N_1^\text{SAOS}(t)= \gamma_0^2 \left( F_0^{\prime \prime} + F_2^{\prime}\sin{2\omega t} + F_2^{\prime\prime}\cos{2\omega t} \right).
    \label{eq:normalSAOS}
\end{equation}

As $\gamma_0$ increases, we transition from SAOS to the large amplitude oscillatory shear (LAOS) limit. The stress response, while still periodic, becomes non-sinusoidal due to the appearance of higher harmonics. Nevertheless, the average energy dissipated in a single cycle of oscillation is controlled by the primary loss modulus $G_1^{\prime \prime}(\gamma_0,\omega)$, and is given by 
\begin{equation}
    \oint \sigma d\gamma = \pi G_1^{\prime\prime}(\gamma_0,\omega)\gamma_0^2.
    \label{eqn:energyDissipation}
\end{equation}
The appeal of OS strain experiments stems from the simultaneous control over two input variables: frequency $\omega$ and amplitude $\gamma_{0}$. This allows us to probe systems at different timescales and deformation scales, which may then be used to construct a Pipkin diagram that acts as a unique material fingerprint.\cite{Hyun2011} OS flows also avoid complications like sudden jumps and sharp ramps associated with other standardized tests such as the stress relaxation or startup flows.

FT rheology is one of several approaches reported in the literature for analyzing LAOS data that includes power series representation of stress,\cite{Pearson1982} weakly nonlinear intrinsic parameters, \cite{Hyun2007,Hyun2009} Chebyshev polynomials, \cite{Ewoldt2008,Ewoldt2013} sequence of physical processes, \cite{rogers2012sequence, Rogers2018, erturk2022comparison} stress decomposition, \cite{Cho2005,Bae2017} characteristic waveforms corresponding to different physical phenomena,\cite{klein2007separation,Hyun2002} etc. Many of these methods require the test material to be probed in the medium amplitude oscillatory shear (MAOS) regime, where the strain amplitude is strong enough to trigger the third harmonic in the shear stress response yet not so large to have contributions from other higher harmonics. A considerable body of work attempts to relate material microstructure with MAOS data by describing it in terms of inter-cycle and intracycle behavior,\cite{Ewoldt2013, renou2010yielding} non-quadratic dependence of third intensity on $\gamma_0$,\cite{Hyun2007, natalia2020questioning} through MAOS solutions of constitutive models (CMs),\cite{Song2020, Martinetti2019, Bharadwaj2015, KateGurnon2012, bharadwaj2015constitutive} etc. In addition, the MAOS moduli associated with the third harmonic obey Kramers-Kronig relations,\cite{kkr1} which allows us to validate experimental data.\cite{kkr2} Despite this theoretical foundation, MAOS analysis suffers from several practical shortcomings. First, it is difficult to identify the range of $\gamma_0$ where the third harmonic is measurable without interference from higher-order harmonics. Second, the process of acquiring MAOS data is laborious, involving the collection and interpolation of large quantities of observations.

The sequence of physical processes approach has emerged as an intuitive alternative to FT rheology.\cite{Rogers2012, Rogers2018, rogers2012sequence, Erwin2010, erturk2022comparison} Here, LAOS data are visualized as a 3D curve in the Frenet-Serret frame defined by the stress, strain, and strain rate axes. The local binormal vector and angle subtended on the osculating plane are used to extract instantaneous and physically meaningful information. While this framework addresses numerous challenges of interpretability and data acquisition, it does not automatically paint a clear microstructural picture, or offer predictions of material behavior under different conditions.

\subsection{Constitutive Models}
\label{sec:intro_cm}

While OS measurements are useful for characterization, we are often interested in predicting how these materials might behave in more complex flow fields that occur during processing. To accomplish this more difficult task, we need to assimilate the results of OS experiments into an appropriate CM that describes a general mathematical relationship between the stress and deformation fields. A variety of integral and differential CMs based on microscopic physics to heuristics have been proposed for modeling different materials.\cite{Bird1987kt, Larson1988, Bird1995, Larson2015} CMs usually have several model parameters that can be estimated using OS data. Once a CM is judiciously chosen and calibrated using OS experiments, it can be plugged into computational fluid dynamics software to model complex flows.\cite{Crochet1984, Owens2002, Favero2010, Alves2021} Differential constitutive relations are used in most computational work involving viscoelastic fluid flows in complex geometries because embedding integral CMs into governing equations for general flow problems incurs higher computational and storage costs. \cite{Keunings2003, Tome2008, Tanner1988, Mitsoulis2013, Hulsen2018}

Estimating the parameters of a CM can be perceived as a Bayesian inference problem or as an optimization or fitting problem. Regardless, this operation involves repeated evaluations of the CM with different guesses for model parameters. Luckily, in homogeneous flow fields associated with OS experiments, partial differential CMs reduce to a system of ordinary differential equations (ODEs). The conventional approach to solving the system of ODEs is to pose it as an initial value problem (IVP). Numerical integration (NI) via a suitable time-stepping method can then be used to solve the IVP by evolving the system until the PSS or alternance solution emerges.

There are several advantages of this conventional approach. It is conceptually simple and mimics the protocol used in experiments. Numerical libraries are already available for solving IVPs, which facilitates the task of modeling arbitrary differential CMs. In general, this approach converges only to stable PSS solutions due to numerical noise injected during time-stepping. However, this method also suffers from several disadvantages. It is difficult to estimate how long it takes to attain PSS for a given set of parameters and initial values. Implicit methods are generally required to avoid numerical instability at large $\gamma_0$ and $\omega$. Such methods are computationally expensive and suffer from numerical damping, a nonphysical decrease in system energy that has purely numerical origins. 

\subsection{Harmonic Balance}

Nonlinear vibration problems are frequently encountered in many engineering applications such as turbomachinery,\cite{Krack2017589, Hartung2019} microwave circuit design,\cite{Suarez20081} structural-acoustic vibrations, \cite{Chaigne2007901} computational fluid dynamics problems,\cite{hall2013harmonic} aerodynamics,\cite{Campobasso2016354, Ekici20121807} cardiovascular flows \cite{Koltukluoglu2019486} etc. Similar to LAOS experiments, these systems produce PSS outputs after the cessation of initial transients.  If only the PSS solution is of interest, an alternative spectrally-accurate approach called \textit{harmonic balance} (HB) can be used, in which the system of nonlinear differential equations is transformed to a system of nonlinear algebraic equations by matching or balancing the harmonics.\cite{Krack2019} This exercise is analogous to previous attempts at finding MAOS solutions to CMs, and can be visualized as a numerical extension of that analytical approach. Unlike MAOS solutions, the calculation and validity of HB solutions are not limited by the magnitude of $\gamma_0$. 

Recently,\cite{Mittal2024} we applied HB to the corotational Maxwell model which is a \textit{linear} CM with a nontrivial LAOS signature for which an exact solution is available as an infinite series.\cite{Saengow2015352}  To our surprise, we found that HB had convergence properties that were superior to the truncated analytical solution. For comparable levels of accuracy, HB could be evaluated about 200x faster than the analytical solution. This is a rare example of a mathematical problem for which numerical solutions are preferable to analytical solutions! We also applied HB to the Giesekus model, which has a quadratic nonlinearity.\cite{Mittal2023} Like all nonlinear CMs, analytical solutions are not available for the LAOS response of the Giesekus model. Nonetheless, comparison of HB with NI showed orders of magnitude outperformance in terms of speed and accuracy.\cite{Mittal2023}

\subsection{Motivation and Layout}

Despite this promise, it is not possible to write HB equations in a self-contained form for most nonlinear CMs. Even when it is feasible, setting up the appropriate equations is tedious and requires substantial manual effort. The goal of this work is to fix both these problems by using a numerical scheme called alternating-frequency-time (AFT).\cite{Krack2019}  HB powered by AFT (i) makes it possible to find the LAOS response of arbitrary differential CMs and (ii) shifts the burden of setting up HB equations from the modeler to the computer. AFT accomplishes this by \textit{numerically} projecting the nonlinear terms in the CM to the frequency space and back during each iteration.

The primary contribution of this work is the conception and development of a computer program called FLASH, which stands for \underline{F}ast \underline{L}arge \underline{A}mplitude  \underline{S}imulations using \underline{H}armonic balance. Incorporating AFT into HB drastically reduces the manual effort required to model arbitrary differential CMs to the point where it is equivalent to setting up the corresponding IVP. 

We start with a brief description of the HB framework, its implementation for differential CMs under OS flow, and the AFT scheme in section \ref{sec:HB}. We then discuss the traditional approach of NI for solving IVPs (section \ref{sec:IVP}), and compare it with the HB method (section \ref{sec:error}). We then validate FLASH for two nonlinear CMs, the exponential Phan-Thein Tanner (PTT) model and a variation of the Ahn-Osaki temporary network model (TNM) in section \ref{sec:validation}. HB is found to be superior to NI in terms of both speed (section \ref{sec:cpu}) and accuracy (section \ref{sec:error2}). Finally, we present a detailed discussion (section \ref{sec:discussion}) on how FLASH may be adapted for other nonlinear differential CMs, its potential applications in theoretical studies and data interpretation, and possibilities for future work.

\section{Constitutive Models in Oscillatory Shear}
\label{sec:cms}
In constitutive modeling, we focus on the dependence of the deviatoric or extra stress tensor $\bm{\sigma}$ on applied deformation.\cite{Bird1987kt, Larson1988} In simple shear flow, the velocity gradient tensor $\bm{\nabla v} = \dot{\gamma} \bm{e}_{1}\bm{e}_{2}$, where the applied oscillatory shear rate $\dot{\gamma} = d\gamma/dt = \gamma_0 \omega \cos \omega t$, and $\bm{e}_{i}$ denotes an orthonormal vector in the $i$th direction. Thus, $\e_{i} \e_{j}$ can be thought of as a $3 \times 3$ matrix whose only nonzero element is a 1 in the $i$th row and $j$th column. In such flows, the symmetric stress tensor has five non-zero components,\cite{morrison2001understanding} of which only four are independent, viz. the shear stress $\sigma_{12}$, and the normal stresses $\sigma_{11}, \sigma_{22},$ and $\sigma_{33}$. 

%Thus, equations \eqref{eqn:shearSAOS} and \eqref{eq:normalSAOS} serve as useful benchmarks. 

A large number of differential CMs are nonlinear extensions of the upper convected Maxwell (UCM) model, and reduce to the UCM model in the SAOS limit. The UCM is a simple but conceptually useful model for dilute polymer solutions. It is derived from the elastic dumbell model, which treats polymers as a pair of beads connected by Hookean springs,\cite{Larson1988, larsoncf} and is given by
\begin{equation}
 \stackrel{\triangledown}{\bm{\sigma}} + \frac{1}{\lambda} \bm{\sigma} = G \dot{\bm{\gamma}}.
\label{eqn:ucm}
\end{equation}
It has two linear viscoelastic model parameters: the relaxation time $\lambda$, and the shear modulus $G$. The symmetric deformation gradient tensor $\dot{\bm{\gamma}} = \bm{\nabla v} + \bm{\nabla v}^T = \dot{\gamma} \bm{e}_{1}\bm{e}_{2} + \dot{\gamma} \bm{e}_{2}\bm{e}_{1}$. Interestingly, the UCM is equivalent to the Lodge integral equation for transient networks with a single relaxation time. \cite{Lodge1956} In homogeneous flows, the stress field is uniform, and the upper convected derivative is given by
\begin{equation}
 \stackrel{\triangledown}{\bm{\sigma}} = \frac{d \bm{\sigma}}{d t} - \bm{\nabla v}^{T} \cdot \bm{\sigma} - \bm{\sigma} \cdot \bm{\nabla v}.
\label{eqn:convectedDvt}
\end{equation}
The UCM does not exhibit a second normal stress difference, i.e., $N_2 = 0$.

\subsection{Phan-Thien Tanner Model}

Phan-Thien and Tanner proposed a nonlinear CM for polymeric fluids based on the generalized theory of transient networks,\cite{Yamamoto1956, Lodge1956} which allowed for the creation and destruction of cross-links.\cite{Thien1977, Thien1978, Cho2016} For affine flow in which there is no slip between the network and the continuous medium, PTT is similar to the UCM but includes a nonlinear term $g^\text{PTT}(\bm{\sigma})$,
\begin{equation}
 \stackrel{\triangledown}{\bm{\sigma}} + \frac{g^\text{PTT}(\bm{\sigma} )}{\lambda} \bm{\sigma} = G \dot{\bm{\gamma}}.
\label{eqn:ptt}
\end{equation}
The nonlinear term is parameterized by an additional dimensionless parameter $\epsilon$,
\begin{equation}
    g^\text{PTT}(\bm{\sigma} ) = \exp{\left(\frac{\epsilon}{G} \text{tr } \bm{\sigma} \right)},
    \label{eq:gptt}
\end{equation}
and depends on the trace of the stress tensor, $\text{tr } \bm{\sigma} = \sigma_{11} + \sigma_{22} + \sigma_{33}$. The exponential form of the nonlinearity was chosen to capture experimental observations in strong flows. For $\epsilon = 0$, PTT becomes equivalent to the UCM model; at higher values of $\epsilon$, the model becomes increasingly nonlinear. Multi-mode versions of the PTT model, can be useful for describing the nonlinear rheology of real materials. \cite{Hatzikiriakos1997,Shiromoto2010,Dietz2015} The relevant set of ODEs under OS are
\begin{align}
\dot{\sigma}_{11} & + \frac{g^\text{PTT}(\bm{\sigma})}{\lambda}\sigma_{11}  - 2\dot{\gamma} \sigma_{12} = 0 \nonumber\\
\dot{\sigma}_{22} & + \frac{g^\text{PTT}(\bm{\sigma})}{\lambda}\sigma_{22} = 0 \nonumber\\
\dot{\sigma}_{33} & + \frac{g^\text{PTT}(\bm{\sigma})}{\lambda}\sigma_{33} = 0. \nonumber\\
\dot{\sigma}_{12} & + \frac{g^\text{PTT}(\bm{\sigma})}{\lambda}\sigma_{12} - \dot{\gamma} \sigma_{22} = G \dot{\gamma}.
\label{eqn:ptt_ode}
\end{align}
In the limit of small $\epsilon$ and $\gamma_0$, the PTT model behaves like the UCM model. At higher values of $\gamma_0$ exact solutions to equation \eqref{eqn:ptt_ode} are not available. However, the MAOS response of the PTT model has been derived as
\begin{align}
\frac{G_{31}^{\prime}(\omega)}{G} & = -\frac{\epsilon \De^{4} (7 + 19 \De^{2})}{2(1+\De^{2})^{3} (1+4 \De^{2})} \notag\\
\frac{G_{31}^{\prime\prime}(\omega)}{G} & =  -\frac{\epsilon \De^{3} (3 + 5 \De^{2} - 10\De^{4})}{2(1+\De^{2})^{3} (1+4 \De^{2})} \notag\\
\frac{G_{33}^{\prime}(\omega)}{G} & = -\frac{\epsilon \De^{4} (7 -17 \De^{2})}{2(1+\De^{2})^{2} (1+4 \De^{2}) (1+9 \De^{2})}\notag\\
\frac{G_{33}^{\prime\prime}(\omega)}{G} & = -\frac{\epsilon \De^{3} (1 -17 \De^{2} + 6 \De^{4})}{2(1+\De^{2})^{2} (1+4 \De^{2}) (1+9 \De^{2})},
\end{align}
where $\De = \omega \lambda$ is the Deborah number.\cite{Bae2017, Song2020}

\subsection{Temporary Network Model}

CMs developed to describe the rheology of associating polymers visualize the material as a network of segments connected at junctions that can be continuously created and destroyed.\cite{Green1946, Yamamoto1956, Tanaka1992, Wang1992, Ahn1995, Vaccaro2000, Tripathi2006, Vernerey2017, Meng2019} An illustrative example of this class of models is the Ahn-Osaki TNM (which we refer to as simply TNM in this work) given by
\begin{equation}
 \stackrel{\triangledown}{\bm{\sigma}} + \frac{d(t)}{\lambda} \bm{\sigma}  = G \dot{\bm{\gamma}} + \frac{G}{\lambda} \left(c(t) - d(t)\right) \bm{I}.
\label{eqn:tnm}
\end{equation}
This TNM was originally developed to understand the mechanism of shear thinning and thickening in complex fluids.\cite{Ahn1995} Here, $c(t)$ and $d(t)$ are dimensionless empirical functions that model the creation and destruction rates of temporary junctions, respectively. They are the primary source of differentiation between different TNMs. In OS flow, the ODEs corresponding to this CM are,
\begin{align}
\dot{\sigma}_{11} & + \frac{d(t)}{\lambda} \sigma_{11} - 2 \dot{\gamma} \sigma_{12} + \frac{G}{\lambda}  \left( d(t) - c(t) \right)= 0 \nonumber\\
\dot{\sigma}_{22} & + \frac{d(t)}{\lambda} \sigma_{22} + \frac{G}{\lambda} \left( d(t) - c(t) \right)= 0 \nonumber\\
\dot{\sigma}_{33} & + \frac{d(t)}{\lambda} \sigma_{33} + \frac{G}{\lambda} \left( d(t) - c(t) \right)= 0 \nonumber\\
\dot{\sigma}_{12} & + \frac{d(t)}{\lambda} \sigma_{12} - \dot{\gamma} \sigma_{22} = G \dot{\gamma}.
\label{eqn:ao_ode}
\end{align}
Originally, Ahn and Osaki assumed $c(t) = e^{a N_{1}/2\sigma_{12}}$ and $d(t) = e^{b N_{1}/2\sigma_{12}}$, where $a$ and $b$ are parameters.\cite{Ahn1995} As $a$ and $b$ approach zero, the TNM effectively reduces to the UCM model. Subsequently, in an attempt to develop a taxonomy of LAOS behavior in complex fluids, Ahn and coworkers used slightly different definitions for creation and loss rates,\cite{Sim2003}  
\begin{gather}
    c(t) = \exp(a|\sigma_{12}(t)|), \label{eq:c_tnm} \\
    d(t) = \exp(b|\sigma_{12}(t)|). \label{eq:d_tnm}
\end{gather}
Here, we consider the Ahn-Osaki TNM with these definitions for creation and loss rates for two reasons: (i) its historical importance in the classification of LAOS behavior,\cite{Sim2003}   and (ii) the functional form for $c(t)$ and $d(t)$ contains the absolute function which is not smooth, and thus presents a difficult benchmark for the HB method. It may be pointed out that in TNMs based on physical mechanisms these two terms are analytic.\cite{Green1946, Yamamoto1956, Tanaka1992, Wang1992, Ahn1995, Vaccaro2000, Tripathi2006, Vernerey2017, Meng2019}

\section{Methods}
\subsection{Harmonic Balance}
\label{sec:HB}
Weighted residual methods are a class of numerical methods used to solve differential equations in which the solution is expressed as a linear combination of judiciously chosen basis functions with unknown coefficients.\cite{Finlayson2013} The residual resulting from inserting this approximation to the true solution into the governing differential equation is then minimized. In the Galerkin method, a widely used weighted residual method, this is accomplished by setting the inner-products of the residual and all the basis functions to zero.\cite{Galerkin1915, heath2018scientific, Finlayson2013} HB belongs to this class of weighted residual methods, wherein the PSS solution is expressed as a Fourier series with trigonometric basis functions $\sin{k\omega t}$ and $\cos{k\omega t}$ (see Appendix \ref{app:Fourier}). 

Any system of first-order differential equations with periodic forcing can be written as
\begin{equation}
\dot{\bm{q}}(t) + \bm{f}_{\nl}(\bm{q},t) -  \bm{f}_{\text{ex}}(t) =  \bm{0} ,
\label{eqn:general_form}
\end{equation}
where $\bm{q}$ represents the desired PSS solution, $\dot{\bm{q}}$ is its time derivative, $\bm{f}_{\nl}$ represents the nonlinear term that subsumes any linear terms, and $\bm{f}_{\text{ex}}(t)$ is the externally applied forcing function which is  sinusoidal in OS flow. $\bm{q}(t)$ can then be expressed as a truncated Fourier series with $H$ harmonics
\begin{equation}
    \bm{q}(t) \approx \bm{q}^{H}(t) = \hat{\bm{q}}^{H} \cdot \bm{B}^{H}(t),
\end{equation}
where $\bm{q}^{H}(t)$ is the truncated Fourier series representation of $\bm{q}(t)$, $\hat{\bm{q}}^{H}$ is a vector of Fourier coefficients (FCs), and $\bm{B}^{H}(t)$ is the vector of corresponding trignometric basis functions (see appendix \ref{app:Fourier}). When $\bm{q}^{H}(t)$ is plugged in eqn. \eqref{eqn:general_form}, we obtain the residual term as
\begin{equation}
    \bm{r}\left(\bm{q}, \dot{\bm{q}}, t \right) \approx \bm{r}^{H}\left(\bm{q}^{H},\dot{\bm{q}}^{H},t \right) \\
    = \dot{\bm{q}}^{H} + \bm{f}_{\nl} \left(\bm{q}^{H},t\right) - \bm{f}_{\ex}(t).
\end{equation}
This can be expressed as a dot product of its FCs ($\hat{\bm{r}}^{H}$) and basis functions as
\begin{equation}
    \bm{r}^{H}\left(\bm{q}^{H},\dot{\bm{q}}^{H},t \right) = \hat{\bm{r}}^{H}\left(\hat{\bm{q}}^{H}\right) \cdot \bm{B}^{H}(t) = \left(\hat{\dot{\bm{q}}}^{H} + \hat{\bm{f}}_{\nl}\left(\hat{\bm{q}}^{H}\right) - \hat{\bm{f}}_{\ex} \right) \cdot \bm{B}^{H}(t).
\end{equation}
In the Galerkin method, the inner-product of the residual $\bm{r}^{H}\left(\bm{q}^{H},\dot{\bm{q}}^{H},t \right)$ and the basis functions is set to zero. From appendix \ref{app:Fourier}, it turns out that these inner products are simply the FCs of the residual. Therefore,
\begin{align}
\frac{1}{T} \int_{0}^{T} 1 \cdot \bm{r}^{H}\left(\bm{q}^{H},\dot{\bm{q}}^{H},t \right) \, dt & = \hat{\bm{r}}^H(0) = \bm{0} \\ 
\frac{2}{T} \int_{0}^{T} \cos(k \omega t) \cdot \bm{r}^{H}\left(\bm{q}^{H},\dot{\bm{q}}^{H},t \right) \, dt & = \hat{\bm{r}}^H_{c}(k) = \bm{0} \\ 
\frac{2}{T} \int_{0}^{T} \sin(k \omega t) \cdot \bm{r}^{H}\left(\bm{q}^{H},\dot{\bm{q}}^{H},t \right) \, dt & =\hat{\bm{r}}^H_s(k) = \bm{0},  
\end{align}
where $T = 2\pi/\omega$ is the period of oscillation. Thus, the orthonormality of the Fourier basis functions leads to HB, which requires that the FCs of the residual  $\hat{\bm{r}}^{H}$ vanish up to the truncation order.

To summarize, in HB, the time domain residual term $\bm{r}^{H}(\bm{q}, \dot{\bm{q}}, t)$ is first projected into frequency space $\hat{\bm{r}}^{H}$ using an \emph{ansatz} of Fourier basis functions, $\bm{q}(t) \approx \bm{q}^H(t) = \hat{\bm{q}} \cdot \bm{B}^H(t)$. Using the Galerkin method, we find that all the components of this frequency-domain residual are equal to zero. Thus, instead of solving an IVP (eqn. \eqref{eqn:general_form}) or the so-called ``strong form'', HB solves the ``weak form'',
\begin{equation}
\hat{\bm{r}}^H(\hat{\bm{q}}^{H}) \equiv \hat{\dot{\bm{q}}}^{H} + \hat{\bm{f}}_{\nl}(\hat{\bm{q}}^{H}) - \hat{\bm{f}}_{\ex}= \bm{0}.
\label{eqn:residualHB}
\end{equation}
Therefore, the HB method can also be called a Fourier-Galerkin method, frequency domain method, or spectral method. 

\subsubsection{HB for Constitutive Modeling} \label{sec:HB_models}

In this section, we illustrate the use of HB for obtaining the PSS stress response of the PTT and TNM models when they are subjected to OS strain. In this and subsequent sections, CMs are presented in dimensionless form to ensure a fair comparison with conventional techniques. Besides the Deborah number $\De =\lambda\omega$, the other dimensionless constant that arises naturally is the Weissenberg number $\Wi = \lambda\omega\gamma_{0}$. Nondimensionalized variables are denoted with a tilde: $\tilde{t}=t/\lambda$, $\tilde{\dot{\gamma}}=\dot{\gamma}/(\gamma_{0}\omega)$, and $\tilde{\bm{\sigma}}=\bm{\sigma}/(G\Wi)$.

In OS flow, the four stress components are packed into a vector
\begin{equation}
    \bm{q}= \begin{bmatrix} q_1 & q_2 & q_3 & q_4 \end{bmatrix} = \begin{bmatrix} \tilde{\sigma}_{11} &  \tilde{\sigma}_{22} & \tilde{\sigma}_{33} & \tilde{\sigma}_{12} \end{bmatrix}, \label{eqn:q}
\end{equation}
and $\dot{\bm{q}} = d\bm{q}/d\tilde{t}$. 
In addition, the external forcing term is given by,
\begin{equation}
    \bm{f}_{\ex}=\left[\begin{array}{cccc}
         0 & 0 & 0 & \tilde{\dot{\gamma}}
    \end{array}\right]. \label{eqn:f_ex}
\end{equation}
It will sometimes be more convenient to write vectors using index notation. Thus, eqns \eqref{eqn:q} and \eqref{eqn:f_ex} can also be written as $\bm{q}= \tilde{\sigma}_{11} \bm{e}_1 + \tilde{\sigma}_{22}\bm{e}_2 + \tilde{\sigma}_{33}\bm{e}_3 + \tilde{\sigma}_{12}\bm{e}_4$, and $\bm{f}_{\ex}= \tilde{\dot{\gamma}}\bm{e}_4$, respectively. These terms appear in all CMs. The remaining nonlinear term $\bm{f}_{\nl}$, which subsumes any linear terms in $\bm{q}$, distinguishes one CM from another. For the UCM model,
\begin{equation}
    \bm{f}_{\nl}^{\text{UCM}} =\left(q_1 - 2\Wi \tilde{\dot{\gamma}} q_{4} \right) \e_1 + q_2 \e_2 + q_3 \e_3 + \left(q_4 - \Wi \tilde{\dot{\gamma}} q_{2} \right) \e_4.
    \label{eqn:fnl_ucm}
\end{equation}
For the PTT and TNM models, the corresponding terms turn out to be,
\begin{align}
    \bm{f}_{\nl} ^{\text{PTT}} & = 
    \left(g^\text{PTT} q_1 - 2\Wi \tilde{\dot{\gamma}} q_{4}\right) \e_1 +
    \left(g^\text{PTT} q_2\right) \e_2 +
    \left(g^\text{PTT} q_3\right) \e_3 + 
    \left(g^\text{PTT} q_4 - \Wi \tilde{\dot{\gamma}} q_{2}\right) \e_4
\label{eqn:fnl_ptt}\\
    \bm{f}_{\nl} ^{\text{TNM}} & =  \left(d(\tilde{t}) q_{1} - 2 \Wi \tilde{\dot{\gamma}}q_{4} + \frac{1}{\Wi}\left( d(\tilde{t}) - c(\tilde{t}) \right)\right) \e_1 +
    \left(d(\tilde{t}) q_{2} + \frac{1}{\Wi}\left( d(\tilde{t}) - c(\tilde{t}) \right)\right) \e_2  
    \notag \\ & +
    \left(d(\tilde{t}) q_{3} + \frac{1}{\Wi} \left( d(\tilde{t}) - c(\tilde{t}) \right)\right) \e_3 + 
    \left(d(\tilde{t}) q_{4} - \Wi \tilde{\dot{\gamma}} q_{2}\right) \e_4 . 
\label{eqn:fnl_tnm}
\end{align}
where $g^{\text{PTT}}$, $c(\tilde{t})$, and $d(\tilde{t})$ are given by eqns. \eqref{eq:gptt}, \eqref{eq:c_tnm} and \eqref{eq:d_tnm}, respectively.

Symmetry constraints dictate that the normal stresses $\{q_i\}_{i=1}^{3}$ contain only even harmonics, while shear stress $q_4$ contains only odd harmonics. Thus, they can be parsimoniously represented using the following sets of basis functions:
\begin{align}
\bm{B}^{H}_{n}(\tilde{t}) & = \{1, \cos{2\De \tilde{t}}, \cos{4\De \tilde{t}}, \cdots, \cos {2H\De \tilde{t}}, \sin{{2\De \tilde{t}}}, \sin{4\De \tilde{t}}, \cdots, \sin{2H\De \tilde{t}}\} \label{eqn:B_n}\\
\bm{B}^{H}_{s}(\tilde{t}) & = \{\cos{\De \tilde{t}}, \cos{3\De \tilde{t}}, \cdots, \cos{(2H+1)\De \tilde{t}}, \sin{\De \tilde{t}}, \sin{3\De \tilde{t}}, \cdots, \sin{(2H+1)\De \tilde{t}}\} \label{eqn:B_s}
\end{align}
\begin{comment}
    \begin{align}
\bm{B}^{H}_{n}(\tilde{t}) & = 1 \e_1 + \sum_{k=1}^H \cos{2k\De \tilde{t}}\text{ } \e_{k+1} + \sin{2k\De \tilde{t}}\text{ } \e_{k+H+1}  \label{eqn:B_n}\\
\bm{B}^{H}_{s}(\tilde{t}) & = \sum_{k=0}^{H} \cos{(2k+1)\De \tilde{t}}\text{ } \e_{k+1} + \sin{(2k+1)\De \tilde{t}}\text{ } \e_{k+H+2} \label{eqn:B_s}
\end{align}
\end{comment}
Here, the subscripts `$n$' and `$s$' stand for normal and shear stress, respectively, and $H$ sets the magnitude of the highest harmonic ($2H+1$) represented in the truncated Fourier series. These basis functions are arranged into a vector $\bm{B}^H(t) = [\bm{B}^{H}_{i}(\tilde{t})]_{i=1}^{4}$, where $\bm{B}^{H}_{1}(\tilde{t}) = \bm{B}^{H}_{2}(\tilde{t}) = \bm{B}^{H}_{3}(\tilde{t}) = \bm{B}^{H}_{n}(\tilde{t})$, and $\bm{B}^{H}_{4}(\tilde{t}) = \bm{B}^{H}_{s}(\tilde{t})$. We  propose an \emph{ansatz} $\bm{q}(t) \approx \bm{q}^H(t)$ in which a linear combination of these basis functions approximate the stresses,
\begin{align}
q_1(\tilde{t}) & \approx q_1^{H}(\tilde{t}) = \hat{q}_1(0) + \sum_{k=1}^{H}\hat{q}_{c,1}(2k)\, \cos{2k\De \tilde{t}} + \hat{q}_{s,1}(2k)\, \sin{2k\De \tilde{t}}, \notag\\
q_2(\tilde{t}) & \approx q_2^{H}(\tilde{t}) = \hat{q}_2(0) + \sum_{k=1}^{H}  \hat{q}_{c,2}(2k)\, \cos{2k\De \tilde{t}} + \hat{q}_{s,2}(2k)\, \sin{2k\De \tilde{t}},  \notag\\
q_3(\tilde{t}) & \approx q_3^{H}(\tilde{t}) = \hat{q}_3(0) + \sum_{k=1}^{H}  \hat{q}_{c,3}(2k)\, \cos{2k\De \tilde{t}} + \hat{q}_{s,3}(2k)\, \sin{2k\De \tilde{t}},  \notag\\
q_4(\tilde{t}) & \approx q_4^{H}(\tilde{t}) =  \sum_{k=0}^{H}  \hat{q}_{c,4}(2k+1)\, \cos{(2k+1)\De \tilde{t}} + \hat{q}_{s,4}(2k+1)\, \sin{(2k+1)\De \tilde{t}}.
\label{eqn:ansatz}
\end{align}
$\hat{q}_{c, i}(n)$ and $\hat{q}_{s, i}(n)$ represent the FCs of $q_i(t)$ corresponding to the cosine and sine terms of the $n$th harmonic, respectively. Thus,
\begin{equation} 
    \hat{\bm{q}}_{i} = [\hat{q}_{i}(0), \hat{q}_{c,i}(2), \cdots, \hat{q}_{c,i}(2H), \hat{q}_{s, i}(2), \cdots, \hat{q}_{s,i}(2H)] \label{eqn:qhat_n}
\end{equation}
for $\{\hat{\bm{q}}_i\}_{i=1}^{3}$ with $2H+1$ terms and
\begin{equation}
    \hat{\bm{q}}_4 = [\hat{q}_{c,4}(1), \cdots, \hat{q}_{c,4}(2H+1), \hat{q}_{s,4}(1), \cdots, \hat{q}_{s,4}(2H+1)]\label{eqn:qhat_s}
\end{equation}
with $2H+2$ terms. This frequency domain representation can be stacked into a vector of FCs $\hat{\bm{q}} = [\hat{\bm{q}}_1, \hat{\bm{q}}_2, \hat{\bm{q}}_3, \hat{\bm{q}}_4]$ with $U = 3(2H+1) + (2H+2) = 8H + 5$ unknowns.    

We are now in a position to define each of the terms in the HB residual (eqn. \ref{eqn:residualHB}). From Appendix \ref{app:Fourier}, $\hat{\dot{\bm{q}}}=\left[\hat{\dot{\bm{q}}}_{1}, \hat{\dot{\bm{q}}}_{2}, \hat{\dot{\bm{q}}}_{3}, \hat{\dot{\bm{q}}}_{4}\right]^{T}$ where,
\begin{equation}
    \hat{\dot{\bm{q}}}_{i} =\De \left[0, 2\hat{q}_{s,i}(2) \cdots 2H\hat{q}_{s,i}(2H), -2\hat{q}_{c,i}(2) \cdots -2H\hat{q}_{c,i}(2H) \right] \label{eqn:qndot_hat}
\end{equation}
for $\{\hat{\bm{q}}_i\}_{i=1}^{3}$ and,
\begin{equation}
    \hat{\dot{\bm{q}}}_{4} =\De \left[ \hat{q}_{s,i}(1) \cdots (2H+1)\hat{q}_{s,i}(2H+1), -\hat{q}_{c,i}(1) \cdots -(2H+1)\hat{q}_{c,i}(2H+1) \right]. \label{eqn:qsdot_hat}
\end{equation}

The external forcing function $\hat{\bm{f}}_{\ex}$ is a vector with $U$ elements, all of which are zero except one: $\hat{\bm{f}}_{\ex}^{(6H+4)}=1$. The nonlinear term for the UCM model $\hat{\bm{f}}_{\nl}=\left[\hat{\bm{f}}_{\nl,1}, \hat{\bm{f}}_{\nl,2}, \hat{\bm{f}}_{\nl,3}, \hat{\bm{f}}_{\nl,4}\right]$ is a vector with $U$ elements. The four components of $\hat{\bm{f}}_{\nl}$ and additional details pertaining to the solution of HB equations for UCM are provided in appendix \ref{app:UCM_HB}.

Unlike the UCM, it is not possible to represent $\hat{\bm{f}}_{\nl}$  explicitly for most nonlinear models because Fourier transforms of exponential terms like $g^{\text{PTT}}=\exp{\left(\epsilon\Wi\left(q_1+q_2+q_3\right)\right)}$ in the PTT model, and $d(\tilde{t})=\exp{\left(a G\Wi\left|q_4\right|\right)}$ and $c(\tilde{t})=\exp{\left(b G\Wi\left|q_4\right|\right)}$ in the TNM cannot be determined analytically. In this respect, the Giesekus model with its polynomial (quadratic) nonlinearity
\begin{align}
    \bm{f}_{\nl}^{\text{G}} = &   \left( q_{1} + \alpha\Wi \left(q_{1}^{2}+q_{4}^{2}\right)-2\Wi q_{1} \tilde{\dot{\gamma}}\right) \e_1 + 
    \left(q_{2} + \alpha\Wi \left(q_{2}^{2}+q_{4}^{2}\right) \right) \e_2 \nonumber
    \\ & + 
    \left(q_{3} + \alpha\Wi q_{3}^{2} \right) \e_3 +
    \left(q_{4} + \alpha\Wi q_{4}\left(q_{1} + q_{2}\right) -  \Wi q_{2}\tilde{\dot{\gamma}} \right) \e_4
    \label{eqn:giesekus}
\end{align}
is an exception.\cite{Mittal2023} Nevertheless, setting up the quadratic nonlinear equation corresponding to HB is tedious. AFT offers a common solution to both these problems: it makes all nonlinear CMs amenable to HB analysis and removes the labor involved in customization.

\subsubsection{Alternating Frequency Time Scheme} \label{sec: AFT}

\begin{figure}
\begin{center}
\includegraphics[scale=0.35]{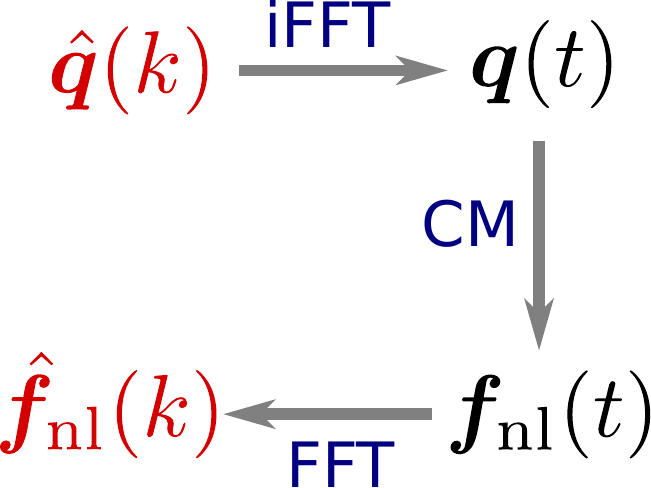}
\end{center}
\caption{Schematic of AFT scheme for evaluating $\hat{\bm{f}}_{\nl}$ from $\hat{\bm{q}}$ using inverse FFT, the constitutive model (CM), and FFT. Frequency (time) domain quantities are shown in red (black). \label{fig:aft}}
\end{figure}

AFT is a versatile and computationally elegant technique for handling arbitrary nonlinear terms in CMs, which shifts the burden from the modeler to the computer.\cite{Urabe1966, Cameron1989, Cardona1998, Krack2019} 
It provides a means to numerically compute $\hat{\bm{f}}_{\nl}$ from $\hat{\bm{q}}$. Given a starting guess for the solution $\hat{\bm{q}}$, AFT involves three steps depicted schematically in Fig. \ref{fig:aft}:
\begin{enumerate}[(i)]
\item $\hat{\bm{q}}(k) \rightarrow \bm{q}(t)$: Consider a dense time-domain grid of equispaced points $\{t_j\}_{j=1}^{N} \in [0, T]$. $N$ should be atleast twice the highest harmonic to avoid aliasing error; thus, $N \geq 2(2H+1)$. We take the vector of FCs $\hat{\bm{q}}$, and use inverse FFT (iFFT) to obtain $\bm{q}_i(\{t_j\})$ at these sample points for all the stress components ($1 \leq i \leq 4$). The result of this frequency $\rightarrow$ time-domain step is a $4N$ dimensional vector $\bm{q}(\{t_j\}) = [\bm{q}_1(\{t_j\}), \bm{q}_2(\{t_j\}), \bm{q}_3(\{t_j\}), \bm{q}_4(\{t_j\})]$ of time-domain stress samples.
\item $\bm{q}(t) \rightarrow \bm{f}_{\nl} (t)$:  Using the time-domain approximation $\bm{q}(\{t_j\})$, we compute the nonlinear term in the time domain by substituting $\bm{q}(t)$ in the expression for $\bm{f}_\nl$ corresponding to the CM. The result of this step is a $4N$ dimensional vector of time-domain samples of the nonlinear terms $\bm{f}_\nl(\{t_j\})$.
\item $\bm{f}_{\nl} (t) \rightarrow \hat{\bm{f}}_{\nl} $: Finally, we transform the time-domain samples of the nonlinear term $\bm{f}_\nl(\{t_j\})$ back to the frequency domain $\hat{\bm{f}}_{\nl} $ using FFT. We discard harmonics greater than $2H$ for the normal stresses and $2H+1$ for the shear stress to be consistent with the ansatz. It can be theoretically shown that AFT preserves the even and odd harmonics required by symmetry.
\end{enumerate}
The $\hat{\bm{f}}_{\nl}$ obtained using AFT is used in each iteration while solving the nonlinear equation system eqn. \eqref{eqn:residualHB}. Luckily, the computational cost of FFT and inverse FFT is $\mathcal{O}(N \log N)$. This is the cost that is shifted from the modeler to the computer.

\subsubsection{Solving HB} \label{sec:solving_hb}

We begin by choosing the number of harmonics $H$ in the \textit{ansatz}. This choice is informed primarily by $\gamma_0$ and $\Wi$. For small values of $\gamma_0$ or $\Wi$, $H \approx 1-3$ is adequate. For larger values of $\gamma_0$ and $\Wi$, larger values of $H$ may be required to resolve the LAOS response fully. In this work, we use $H = 8$ as a standard for all computations unless otherwise specified. Note that this choice resolves shear stress up to the 17th harmonic!

Next, we select a good initial guess $\hat{\bm{q}}^{(0)}$ for the solution. If an analytical solution for the CM in SAOS or MAOS regime is available, then this is usually a good choice, especially for $\gamma_0 \lesssim 1$. Otherwise, the UCM solution may be used as $\hat{\bm{q}}^{(0)}$ instead.
In this work, we use the MAOS solution for the PTT model as the initial guess for $\gamma_0 \leq 1$. For any $\gamma_0>1$ we create a ladder of strain amplitudes, where each subsequent rung of the ladder represents increasing $\gamma_0$. The first rung  starts at $\gamma_0=1$, and uses the MAOS solution as $\hat{\bm{q}}^{(0)}$. The HB solution at each rung is then used as the initial guess for the next step. The spacing of $\gamma_0$ in the interval between 1 and the target strain amplitude
is logarithmic with five rungs per decade. A similar laddered approach is also used for the TNM model. However, since the MAOS solution is not known, the first rung is taken to be $\gamma_0 = 10^{-2}$, and the UCM solution is used as the initial guess. 

With this initial guess for $\hat{\bm{q}}^{(0)}$, we solve the system of nonlinear algebraic equations using the python interface (\texttt{fsolve}) to MINPACK's \texttt{hybrd} method.\cite{More1980, Virtanen2020} The solver uses a modification of the hybrid Powell or dogleg method, in which the Jacobian is approximated by a forward-difference formula.\cite{Powell1970} We use the default value for determining convergence: iteration is terminated when the relative error between consecutive iterates falls below $10^{-8}$.

\subsection{Numerical Integration}
\label{sec:IVP}
Eqn. \eqref{eqn:general_form} specifies a periodic boundary value problem since the initial conditions $\bm{q}(0)$ are not explicitly specified. Instead, we seek a PSS solution for which,
\begin{equation}
\lim_{t \rightarrow \infty} \, q(t + T) = q(t).
\label{eqn:alternance}
\end{equation}

One approach is to treat it as a boundary value problem and use the shooting method where we guess the initial condition $\bm{q}(0)$ and iteratively update it until the condition $\bm{q}(T) = \bm{q}(0)$ is met.\cite{heath2018scientific} Each iteration of the shooting method involves solving the IVP for $t \in [0, T]$. A more conventional method of determining the alternance solution is to mimic the experimental setup and pose the ODE as an IVP with arbitrary initial conditions (e.g., $\bm{q}(0) = \bm{0}$). Fortunately, these initial conditions do not matter, and a PSS profile emerges after initial transients decay.

We use a 5th order implicit Runge-Kutta scheme of the Radau IIA family, implemented in the python package \texttt{scipy} with absolute and relative tolerances of $10^{-10}$ and $10^{-8}$, respectively.\cite{Hairer1996, Virtanen2020} We use an implicit scheme because explicit schemes tend to struggle with high frequencies $\omega$ and strong nonlinearities. By default, we solve the problem for $t \in [0, n_{p} T]$ where $n_p = 10$ is the number of periods.

For each period $p$, we consider the response over that period, $\bm{q}_p(s)$ where $s = t \text{ mod } T$, and `mod' is the modulo or remainder operator. The response over the first period $\bm{q}_1(s)$ is typically nonperiodic because of the arbitrary choice of initial conditions. As $p$ increases, the $\bm{q}_p(s)$ approaches the alternance state. To quantify the approach to PSS, we evaluate $\bm{q}_p(s)$ over $N$ equispaced points $\{s_i\}_{i=1}^{N} \in [0, T]$. We compare the difference in $\bm{q}_p(s)$ over the last two of these periods ($n_p$ and $n_{p} - 1$), by computing the metric, 
\begin{equation}
E_p = \frac{1}{4N} \sqrt{\sum_{i = 1}^{N} \left(\bm{q}_{n_p}(s_i) - \bm{q}_{n_p - 1}(s_i)\right)^2}
\end{equation}
If $E_p$ falls below the threshold $10^{-10}$, we take the solution obtained over the last period as $\bm{q}_\text{NI} \equiv \bm{q}_{n_p}$. Otherwise, we update initial conditions using the last cycle $\bm{q}_{n_p + 1}(t=0) = \bm{q}_{n_p}(T)$ and repeat the calculation over $n_p$ more cycles until the threshold is met.

\subsection{Quantifying Error}
\label{sec:error}
How can we determine the accuracy of numerical solutions when benchmark analytical solutions are not available? How can we compare the accuracy of the IVP and HB methods? One idea is to develop a time-domain error metric $\epsilon_t$ based on the residual corresponding to the IVP, $\bm{r}(\bm{q}, \dot{\bm{q}}, t) = \dot{\bm{q}}(t) + \bm{f}_{\nl} (\bm{q}, t) - \bm{f}_\ex(t)$, and a frequency-domain error metric $\epsilon_\omega$ based on the residual corresponding to HB, $\hat{\bm{r}}^H(\hat{\bm{q}}) \equiv \hat{\dot{\bm{q}}} + \hat{\bm{f}_{nl}}(\hat{\bm{q}}) - \hat{\bm{f}}_{\ex}$. If $\bm{q}(t)$ is an exact solution with FCs $\hat{\bm{q}}$, both the residuals $\bm{r}(\bm{q}, \dot{\bm{q}}, t)$ and $\hat{\bm{r}}^H(\hat{\bm{q}})$ would be identically zero. However, when $\bm{q}(t)$ is numerically approximated using the IVP or HB methods, the residuals are nonzero but hopefully small.

Let $\bm{q}_\text{NI}(t)$ and $\hat{\bm{q}}_\text{HB}$ be the solutions obtained using the IVP and HB methods for a given problem. We can interpolate $\bm{q}_\text{NI}(t)$ at a large number (say, $N=2^{6}$) of equispaced points $t_i \in [0, T]$, and take the FFT to obtain $\hat{\bm{q}}_\text{NI}$. To facilitate comparison, we only select even (odd) harmonics up to order $2H$ ($2H+1$) for normal (shear) stress. Typically, the harmonics that are not included are negligible. We can compute a frequency-domain error metric by directly substituting $\hat{\bm{q}} = \hat{\bm{q}}_\text{NI}$ or $\hat{\bm{q}} = \hat{\bm{q}}_\text{HB}$ into the expression for $\hat{\bm{r}}^H(\hat{\bm{q}})$. Since $\hat{\bm{r}}^H(\hat{\bm{q}})$ is a vector with $U$ elements, we set $\epsilon_\omega$ based on the root mean squared error (RMSE) of the residuals,
\begin{equation}
\epsilon_{\omega}(\hat{\bm{q}}) = \dfrac{1}{\sqrt{U}} ||\hat{\bm{r}}^H(\hat{\bm{q}})||_{2}.
\end{equation}
To compute $\epsilon_t$, we use $\hat{\bm{q}} = \hat{\bm{q}}_\text{NI}$ or $\hat{\bm{q}} = \hat{\bm{q}}_\text{HB}$ and first estimate $\dot{\bm{q}}(t_i)$ and $\bm{q}(t_i)$ at $t_i \in [0, T]$, $1 \leq i \leq N$, using eqn. \eqref{eqn:ansatz}. Again, we use the RMSE of the time-domain residuals as,
\begin{equation}
\epsilon_t(\bm{q}) = \dfrac{1}{\sqrt{4N}} ||\bm{r}(\bm{q})||_{2}.
\end{equation} 
Using this procedure, we can obtain $\epsilon_t$ and $\epsilon_\omega$ for both the NI and HB methods.

\section{Results}

\subsection{Validation of FLASH}
\label{sec:validation}
We test the validity of FLASH by finding the LAOS response of the PTT model and the TNM at different operating conditions $(\gamma_0, \omega)$. All calculations using FLASH or NI are performed with nondimensionalized equations. However, the solutions obtained are converted back to dimensional units when results are reported in this section and in supplementary material. We assume that the modulus $G = 1$ Pa so that the moduli reported in figures \ref{fig:PTT_fig} and \ref{fig:TNM_fig} are expressed in units of $G$.

\subsubsection{Phan-Thien Tanner model}

\begin{figure}
    \begin{tabular}{cc} 
    {\large \bm{$\gamma_0=0.1$}} & {\large \bm{$\gamma_0=10$}} \\ \includegraphics[scale=0.25]{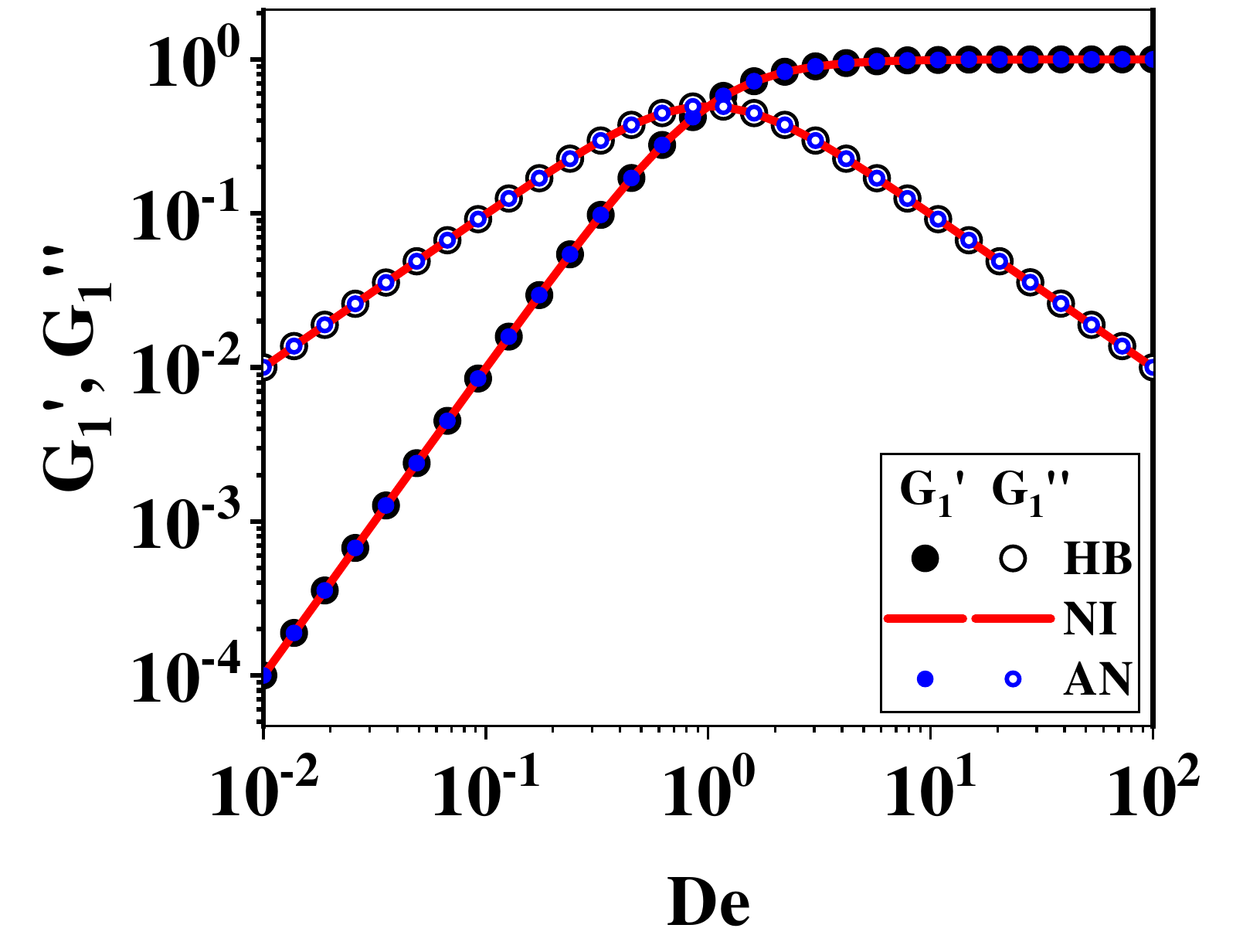} & \includegraphics[scale=0.25]{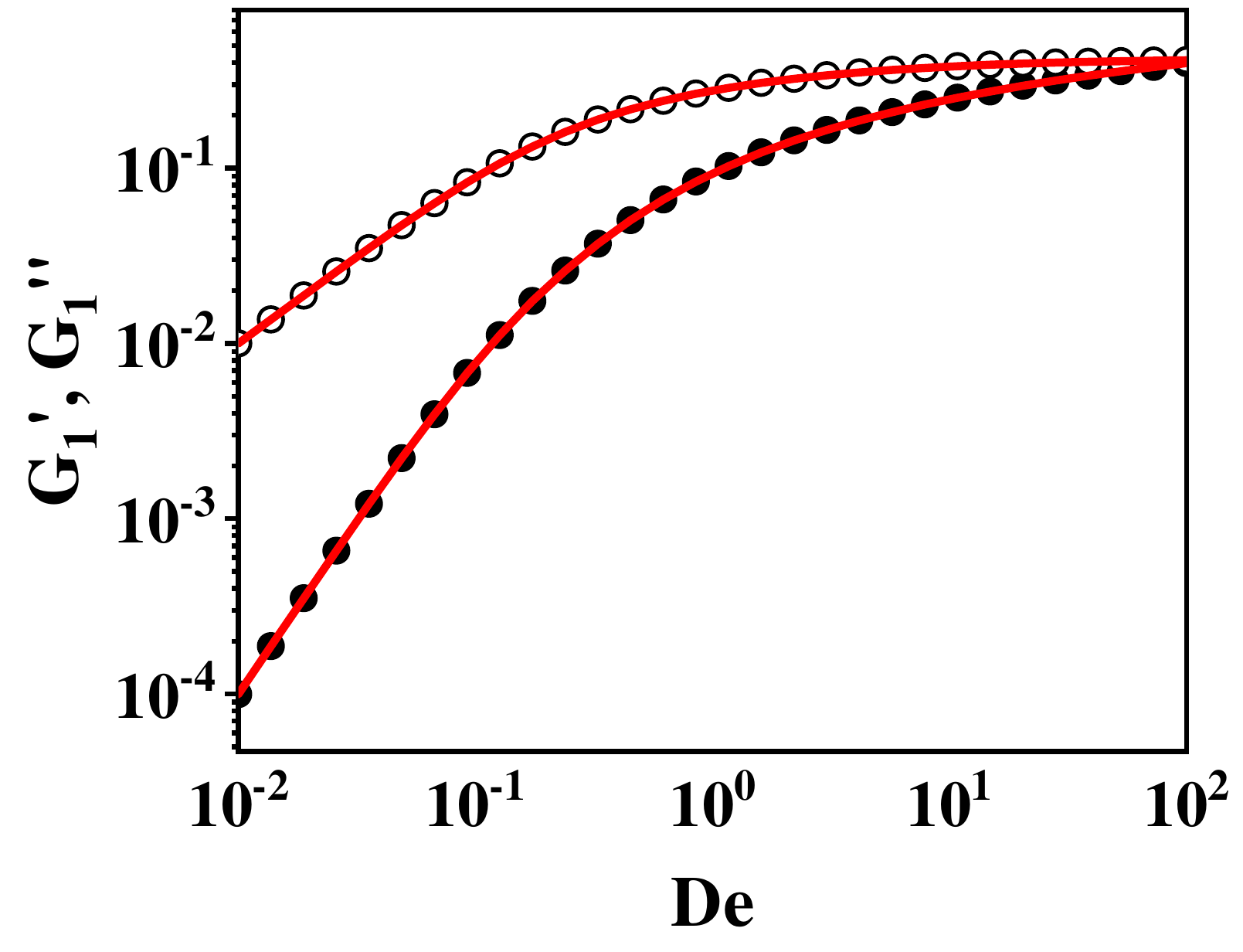} \\
    (a) & (b) \\
    \includegraphics[scale=0.25]{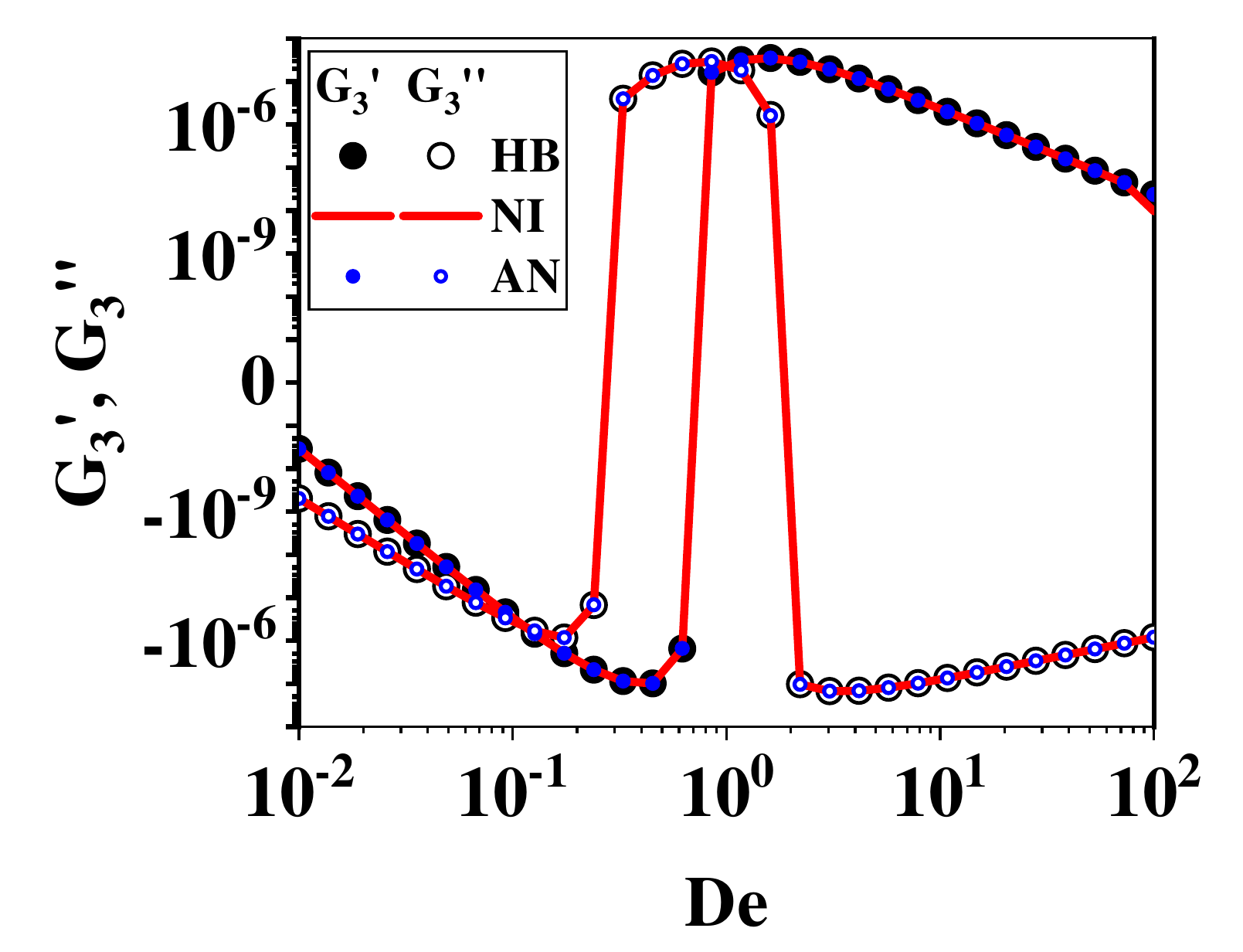} & \includegraphics[scale=0.25]{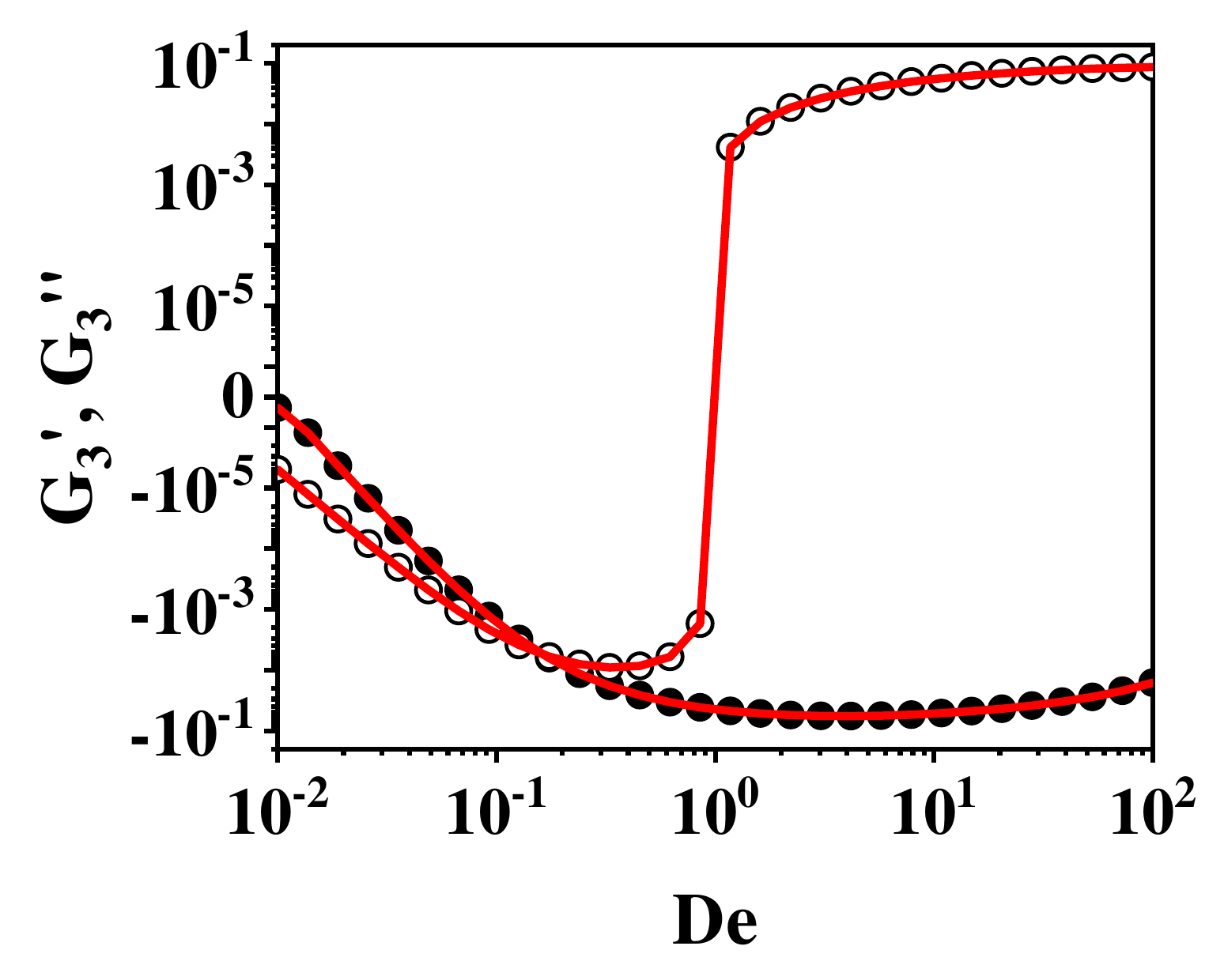} \\
    (c) & (d)
    \end{tabular} 
    \caption{\textbf{PTT model}: The moduli corresponding to the first harmonic $G_{1}'$ and $G_{1}''$ for (a) $\gamma_0=0.1$ and (b) $\gamma_0=10$, and the third harmonic $G_{3}'$ and $G_{3}''$ for (c) $\gamma_0=0.1$ and (d) $\gamma_0=10$ obtained by NI and FLASH agree with each other. For $\gamma_0=0.1$, these solutions also agree with the analytical MAOS solution (AN).}
    \label{fig:PTT_fig}
\end{figure}

We set $\epsilon = 0.1$ for this part. Fig. \ref{fig:PTT_fig} depicts the shear response of the PTT model computed using FLASH at two different strain amplitudes in the MAOS $\left(\gamma_0=0.1\right)$ and LAOS $\left(\gamma_0=10\right)$ regimes. The first $(G_1',G_1'',)$ and third $(G_3',G_3'',)$ harmonics are plotted as a function of applied frequency. Fig. \ref{fig:PTT_fig}a and \ref{fig:PTT_fig}c show that the MAOS response obtained using FLASH agrees with the numerical solution of the IVP and the MAOS analytical solution (AN).\cite{Song2020} Similar agreement between FLASH and NI is observed in fig. \ref{fig:PTT_fig}b and \ref{fig:PTT_fig}d for the LAOS response. Additional plots of stress waveforms and the leading normal stress moduli are provided in the supplementary material. %\highlight{Note, that the dimensions of these moduli are consistent with eqn. \eqref{eqn:shearFourier} for the shear moduli and eqn. \eqref{eq:normalFourier} for the norma stress difference coefficients. }

\subsubsection{Temporary Network Model}

\begin{figure}
    \begin{tabular}{cc} 
    {\large  \textbf{Type I: Strain Softening}} &  {\large \textbf{Type III: Weak Strain Overshoot}} \\
    {\large $a$ = -1.0, $b$ = 1.0} & {\large $a$ = 0.5, $b$ = 1.0} \\ 
    \includegraphics[scale=0.25]{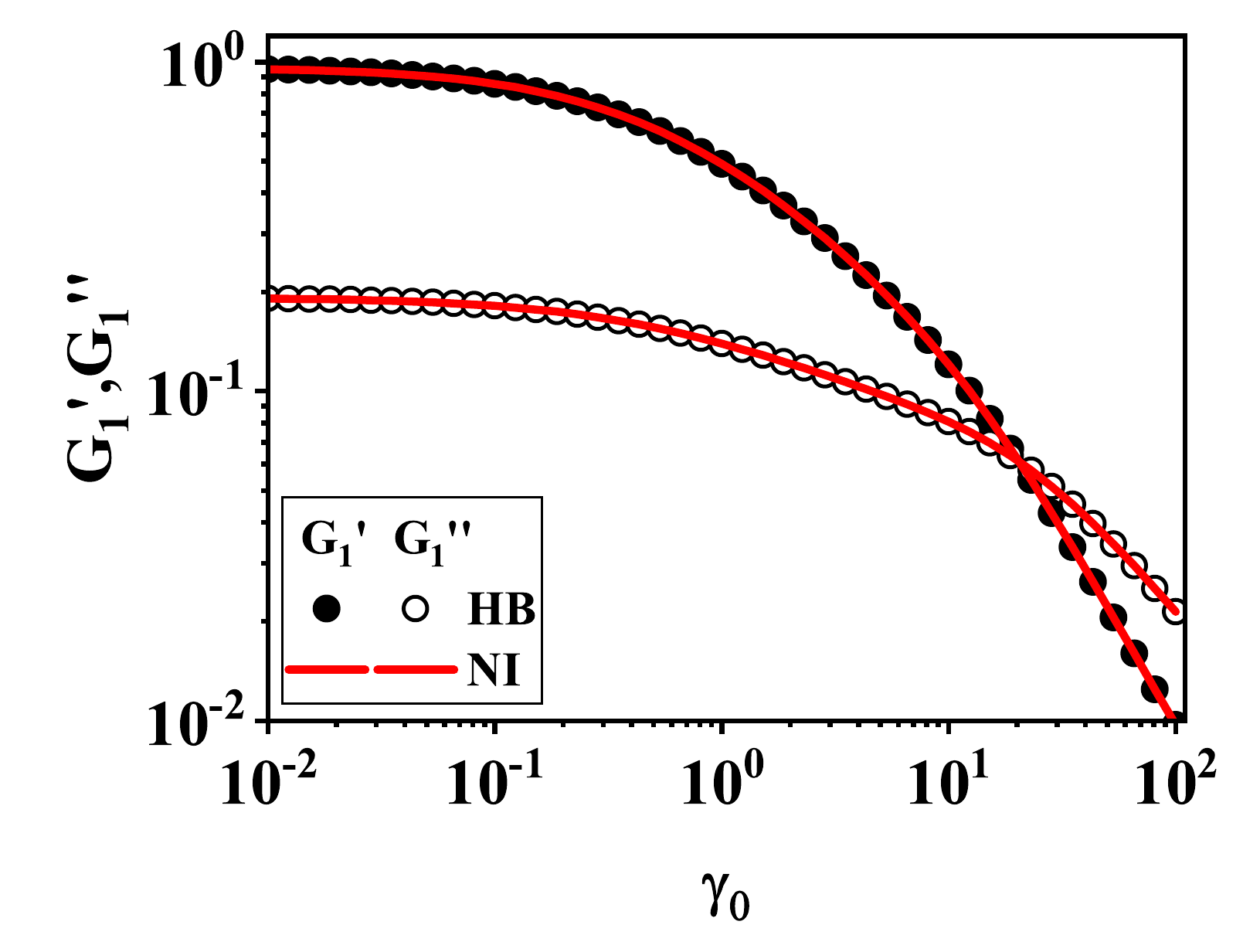} & \includegraphics[scale=0.25]{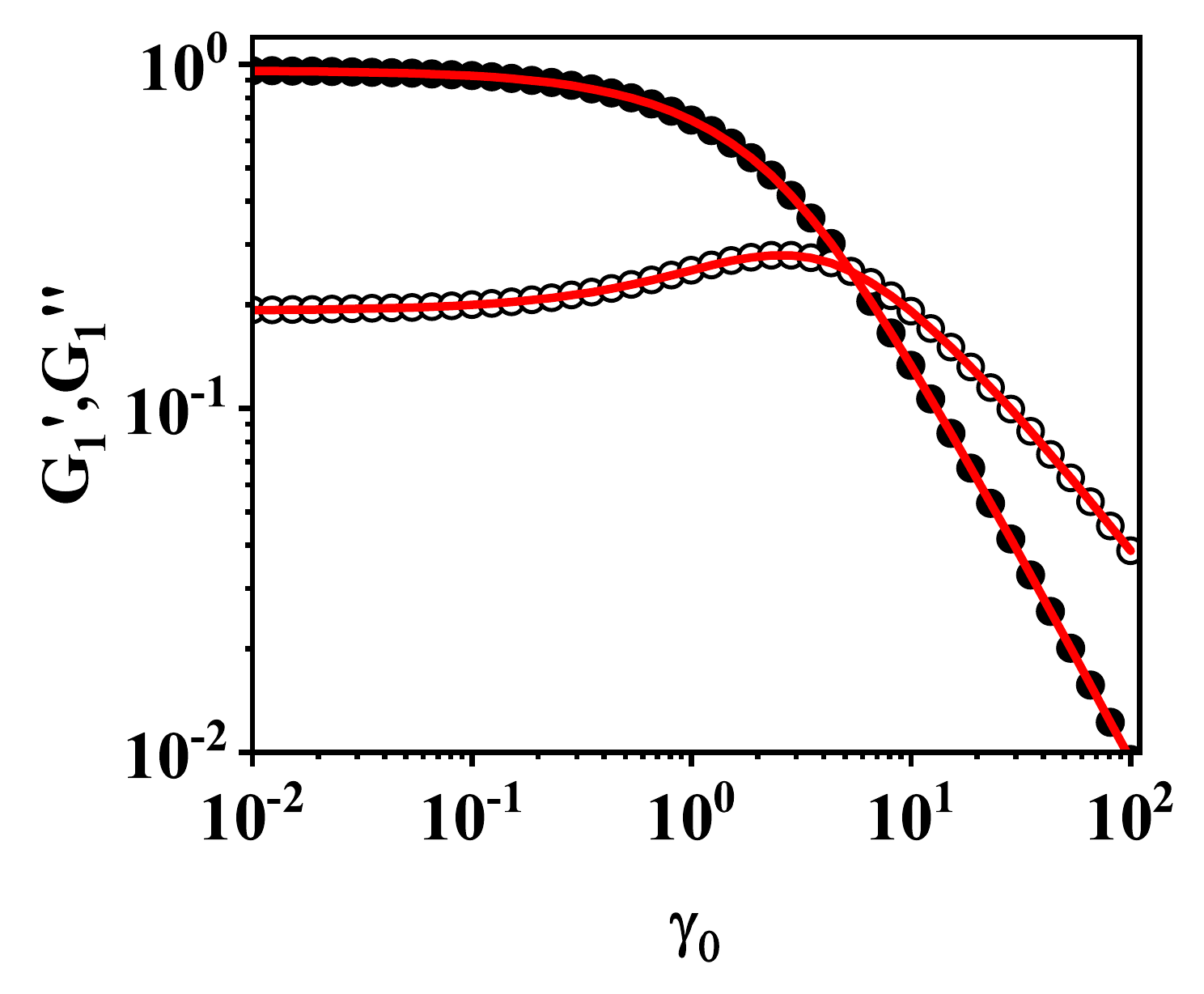} \\
    (a) & (b) \\
    \includegraphics[scale=0.25]{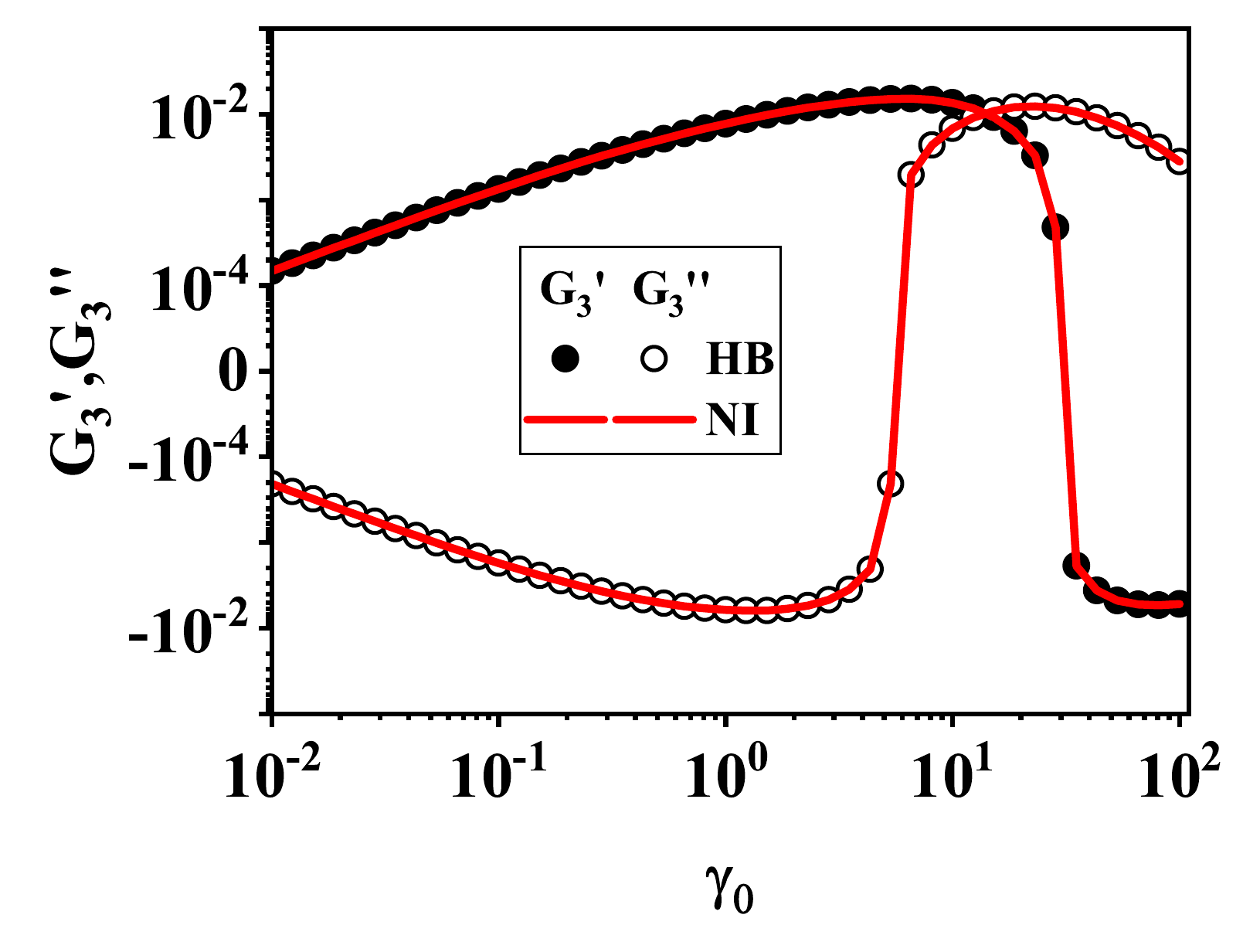} & \includegraphics[scale=0.25]{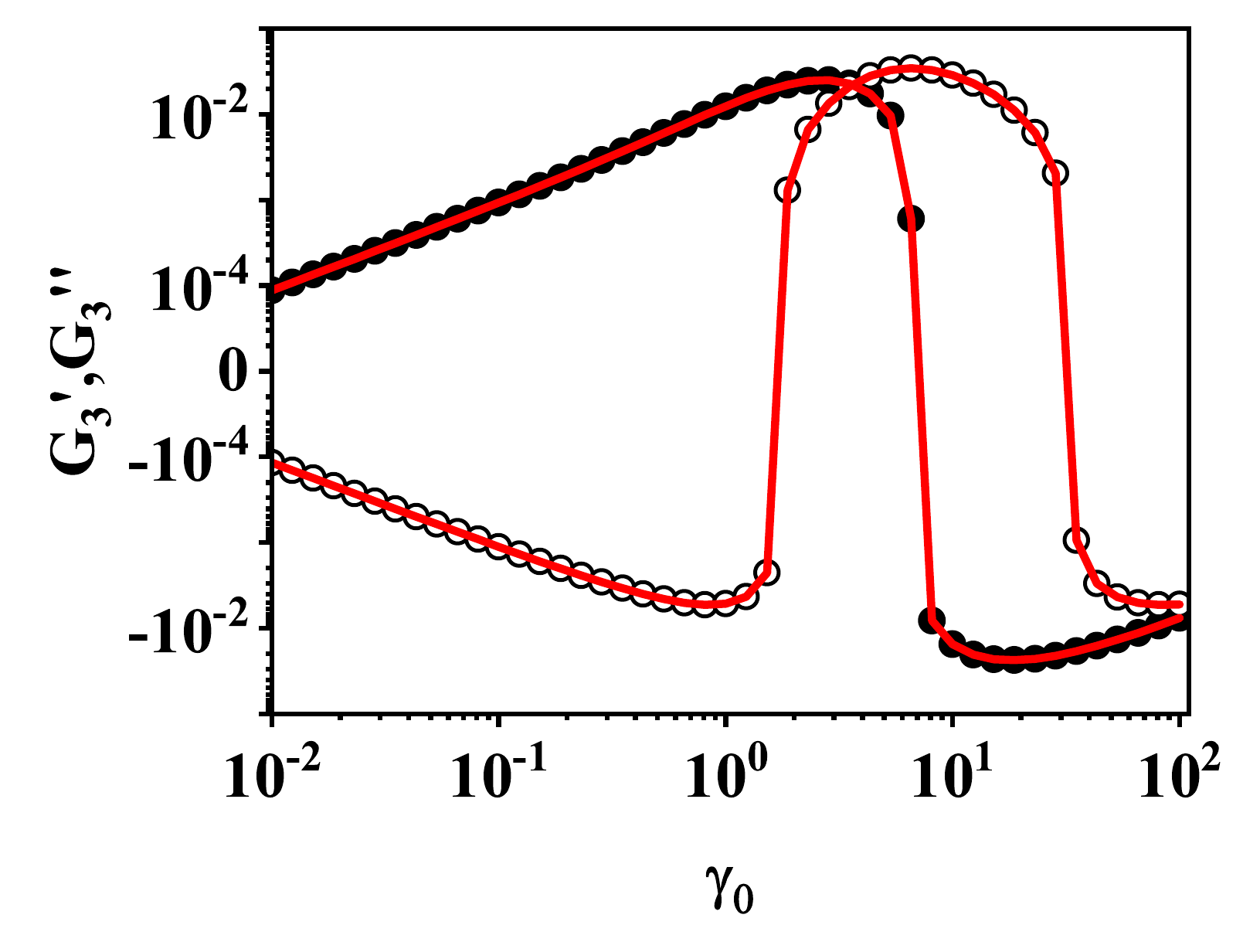} \\
    (c) & (d)
    \end{tabular} 
    \caption{First and third harmonic moduli for the TNM with parameters chosen to illustrate type I (a and c), and type 3 (c and d) behavior at $\De=5$ and different $\gamma_0$. The agreement between FLASH and NI is excellent.\label{fig:TNM_fig}}
\end{figure}

The Ahn-Osaki TNM captures four canonical types of LAOS behavior depending on the values of the parameters $a$ and $b$.\cite{Sim2003, Hyun2002,Hyun2011} To validate FLASH, we illustrate two of these four categories, namely, type I or strain softening ($a$ = -1.0, $b$ = 1.0), and type III or weak strain overshoot ($a$ = 0.5, $b$ = 1.0). Figures \ref{fig:TNM_fig}a and \ref{fig:TNM_fig}b show the moduli associated with the first harmonic, while figures \ref{fig:TNM_fig}c and \ref{fig:TNM_fig}d show the moduli associated with the third harmonic. In type I behavior, $G_1^{\prime}$ and $G_1^{\prime\prime}$  decrease monotonically as $\gamma_0$ is increased, while in type III behavior, $G_1^{\prime \prime}$ exhibits a local maximum at intermediate $\gamma_0$. Calculations are performed at a fixed frequency corresponding to $\De=5$ over four decades of $\gamma_0$. The agreement between FLASH and NI solutions is satisfactory. The non-differentiable absolute value function in the creation ($c(t)$) and destruction ($d(t)$) terms of the TNM introduces a first-order discontinuity and that presents some issues for FLASH, which we shall examine shortly. Nevertheless, these examples illustrate that FLASH is not only a flexible framework but also quite robust.

\subsection{Computational Cost}
\label{sec:cpu}

\begin{figure}
    \begin{subfigure}[b]{0.45\textwidth}
        \includegraphics[width=\textwidth]{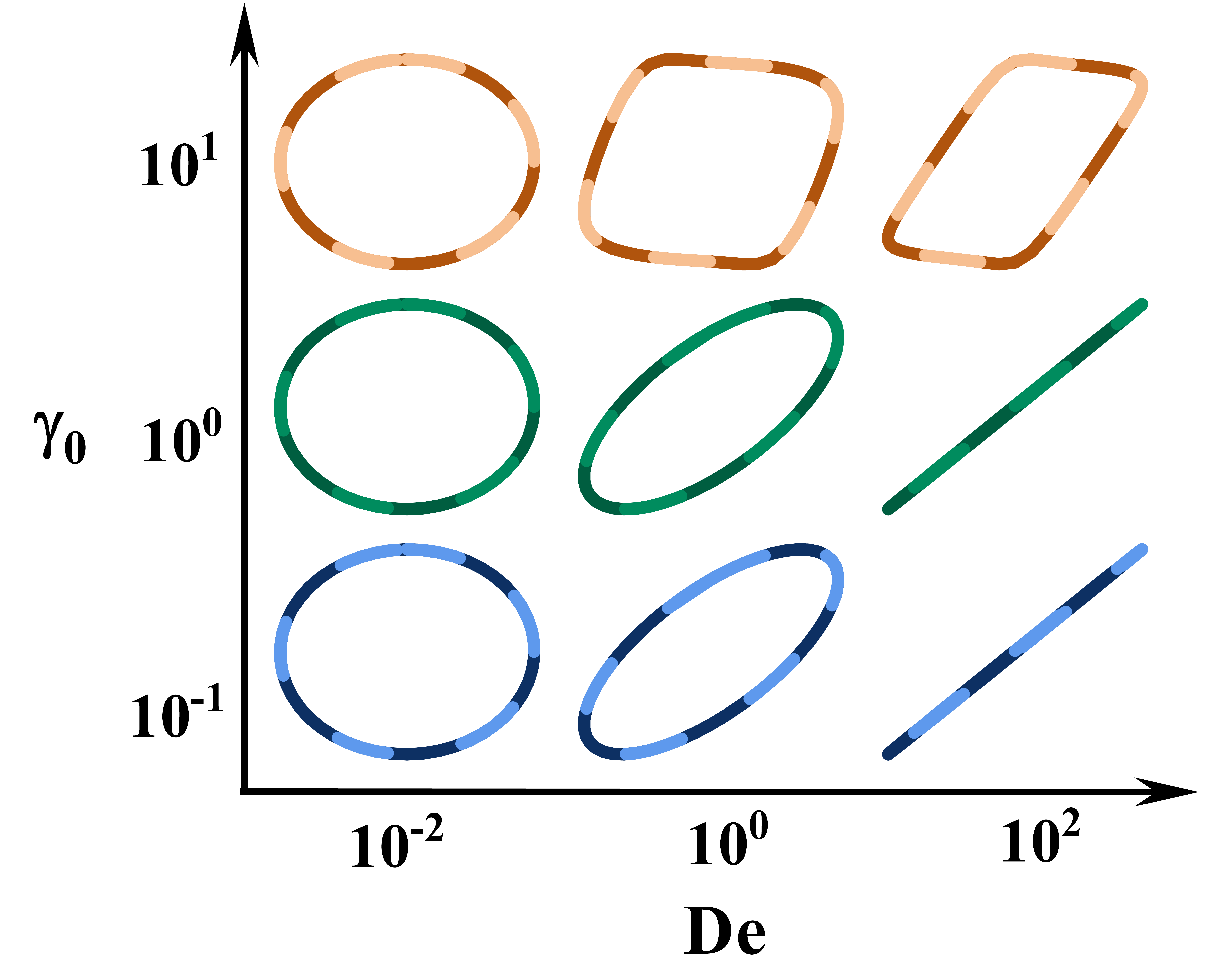}
        \caption{Lissajous plots}
        \label{fig:PTT_Lissajous}
    \end{subfigure}
    \begin{subfigure}[b]{0.45\textwidth}
        \includegraphics[width=\textwidth]{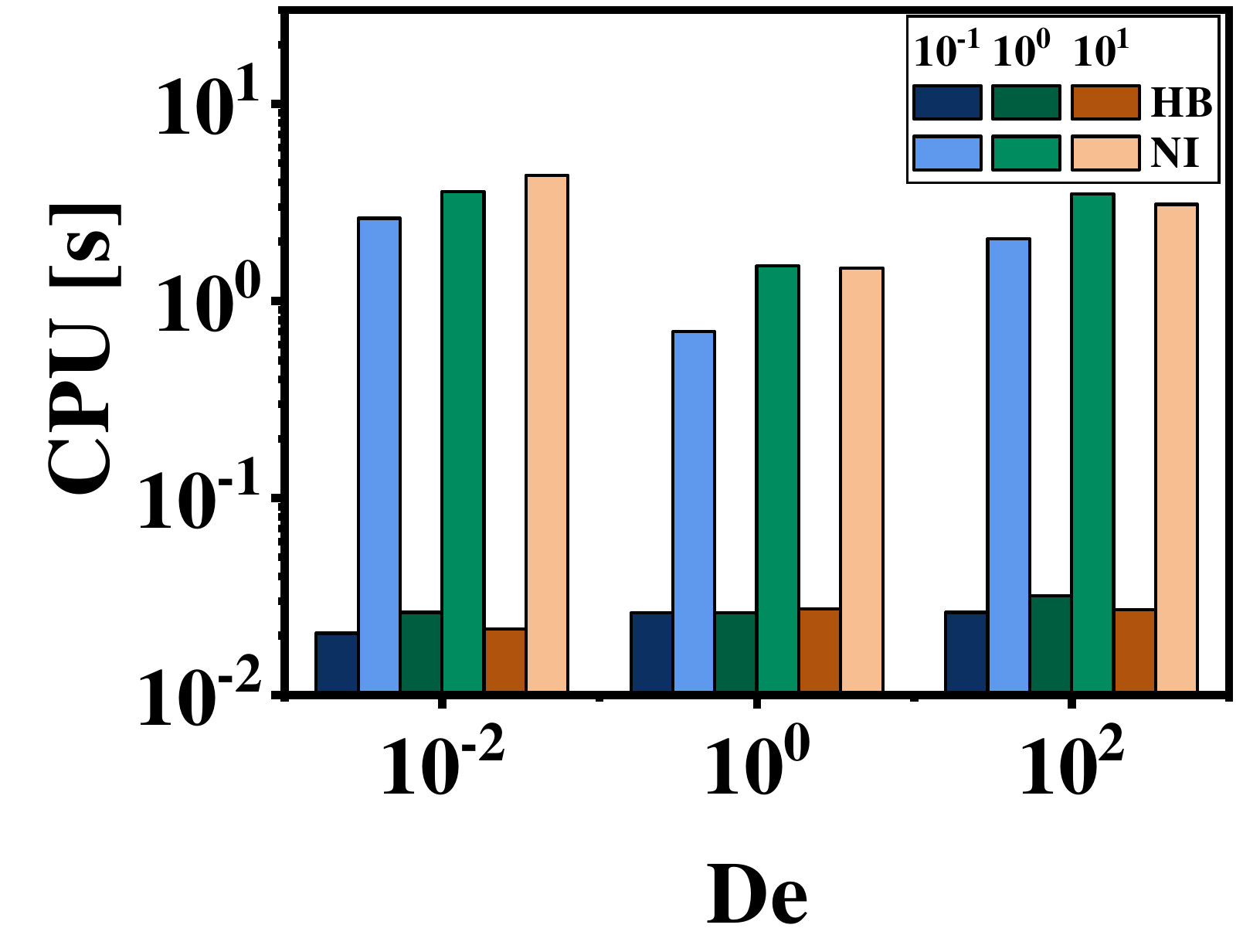}
        \caption{CPU}
        \label{fig:PTT_cpu}
    \end{subfigure}
    \caption{\textbf{PTT model}: (a) Normalized elastic Lissajous curves of the shear stress $\sigma_{12}$ vs $\gamma$, and (b) comparison of the computational time (in seconds) at three different values of $\gamma_0$ and $\text{De}$.}
\end{figure}

\begin{figure}
    \begin{subfigure}[b]{0.45\textwidth}
        \includegraphics[width=\textwidth]{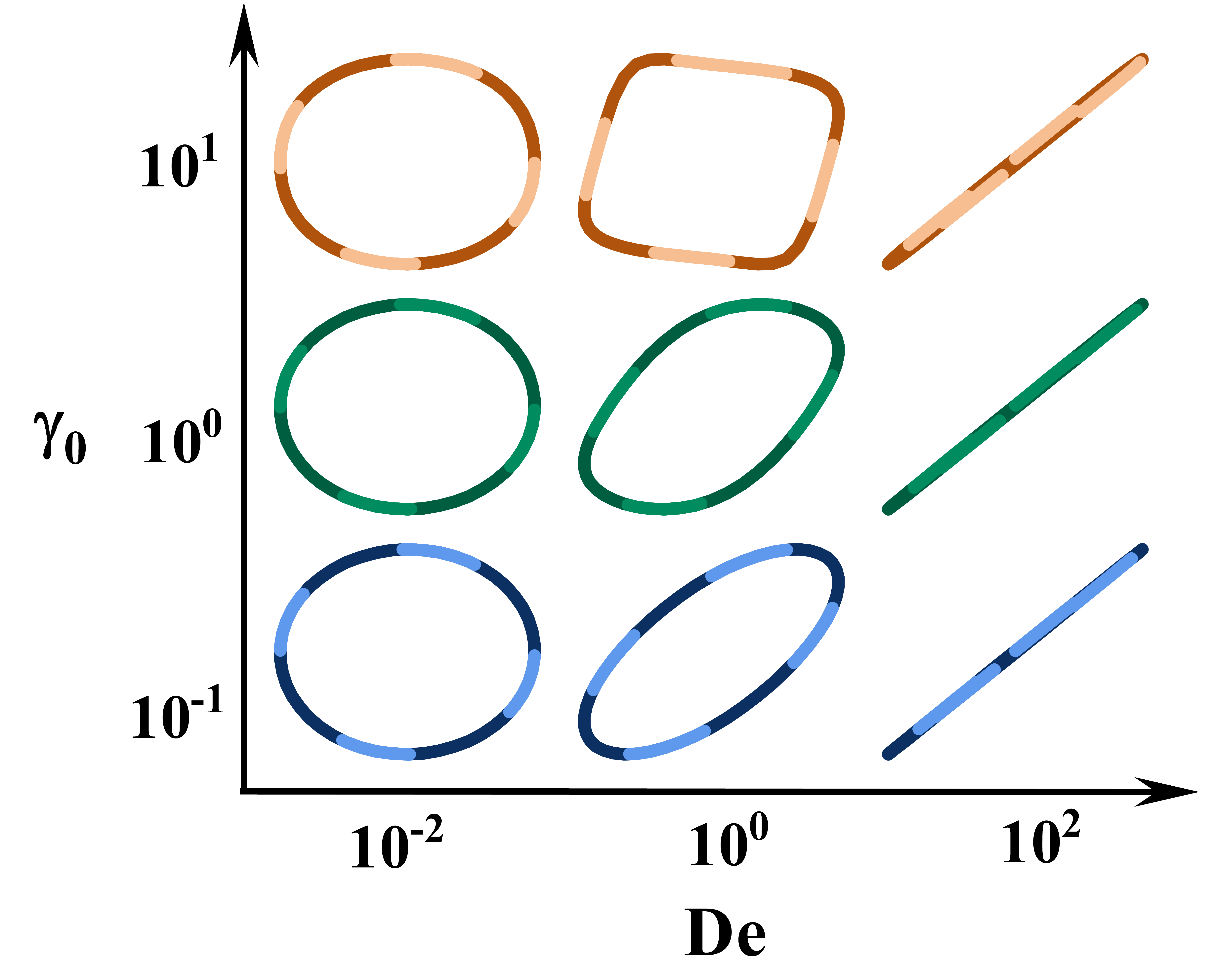}
        \caption{Lissajous plots}
        \label{fig:TNM_Lissajous}
    \end{subfigure}
    \begin{subfigure}[b]{0.45\textwidth}
        \includegraphics[width=\textwidth]{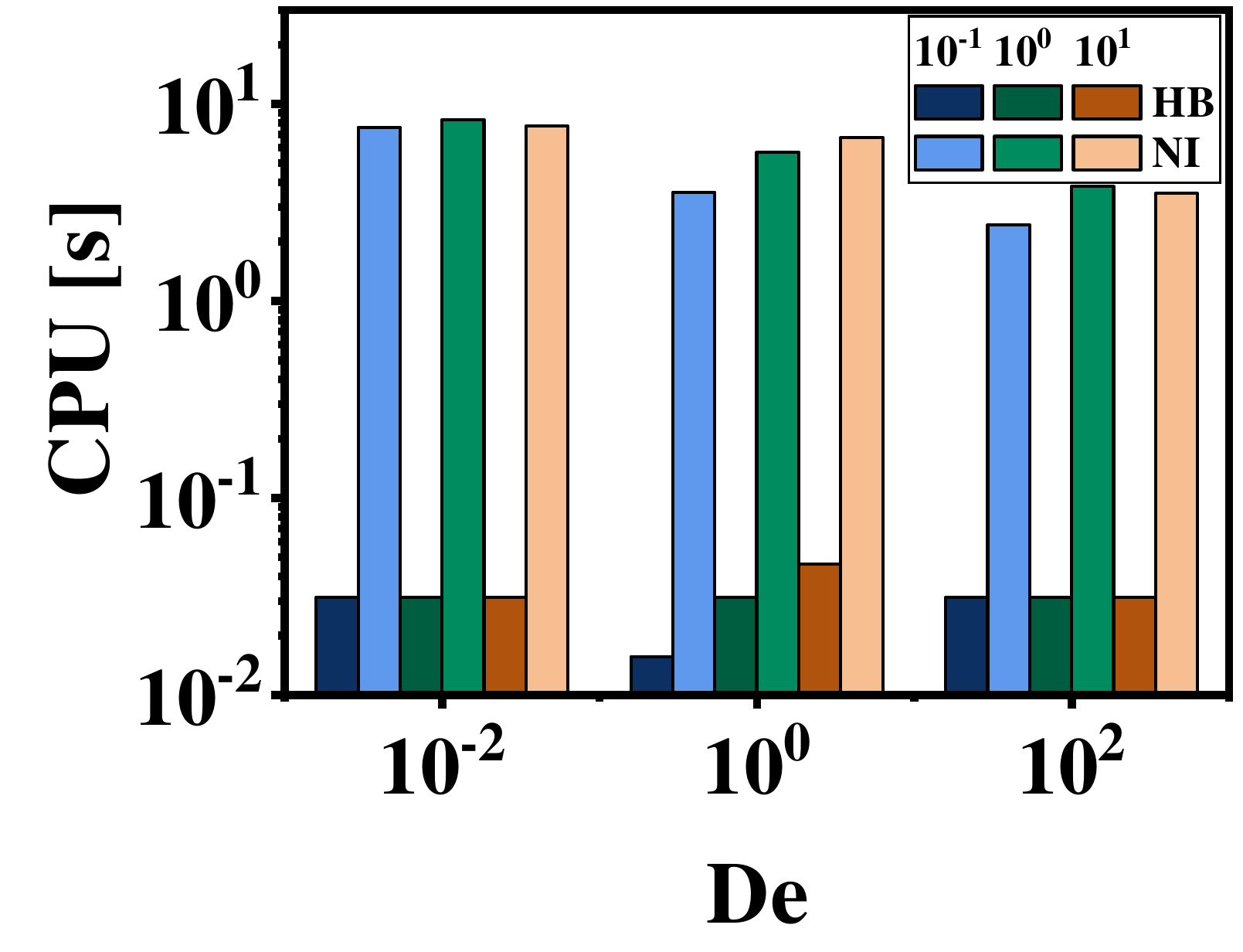}
        \caption{CPU}
        \label{fig:TNM_cpu}
    \end{subfigure}
    \caption{\textbf{TNM - Type I}: Normalized elastic Lissajous curves of the shear stress $\sigma_{12}$ vs $\gamma$, and (b) comparison of the computational time (in seconds) at three different values of $\gamma_0$ and $\text{De}$. }
\end{figure}

Figures \ref{fig:PTT_Lissajous} and \ref{fig:TNM_Lissajous} show a representative Pipkin diagram for the PTT model and type I case of the TNM model, respectively. These figures depict the elastic Lissajous curves for the shear stress, i.e. $\sigma_{12}(t)$ vs $\gamma(t)$, at three strain amplitudes $\left(10^{-1},10^{0},10^{1}\right)$ and three frequencies $\left(10^{-2},10^{0},10^{2}\right)$. Since the agreement between the two methods appears reasonable, we compare the CPU time required to compute these solutions in figures \ref{fig:PTT_cpu} and fig. \ref{fig:TNM_cpu} for the PTT model and the TNM, respectively. Calculations were performed on Intel i7-6700 3.4 GHz Windows workstation running Python 3.8.10, \texttt{numpy} 1.24.4, and \texttt{scipy} 1.10.1. On average, FLASH takes roughly 30 ms per evaluation for both CMs. NI takes longer: about 0.7 -- 4.55 s for the PTT model and  2 -- 8 s per evaluation for the TNM. Thus, it is evident that FLASH is between 1-3 orders of magnitude faster than NI, regardless of the operating condition. It is useful to note that even NI struggles with the first-order discontinuity in the TNM.

\subsection{Accuracy}
\label{sec:error2}

\begin{figure}
    \begin{subfigure}[b]{0.45\textwidth}
        \includegraphics[width=\textwidth]{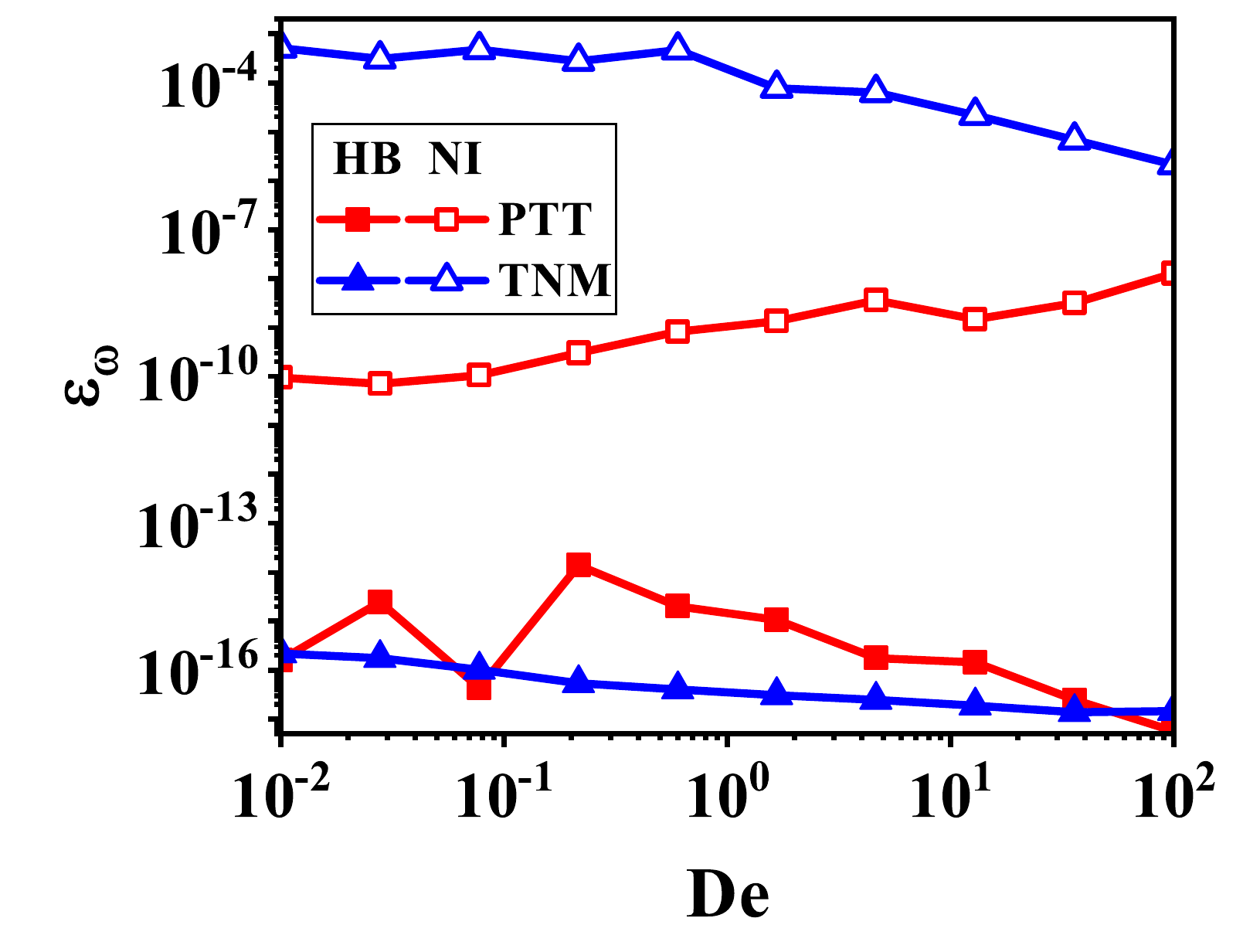}
        \caption{frequency domain error}
        \label{fig:eFC_De}
    \end{subfigure}
    \begin{subfigure}[b]{0.45\textwidth}
        \includegraphics[width=\textwidth]{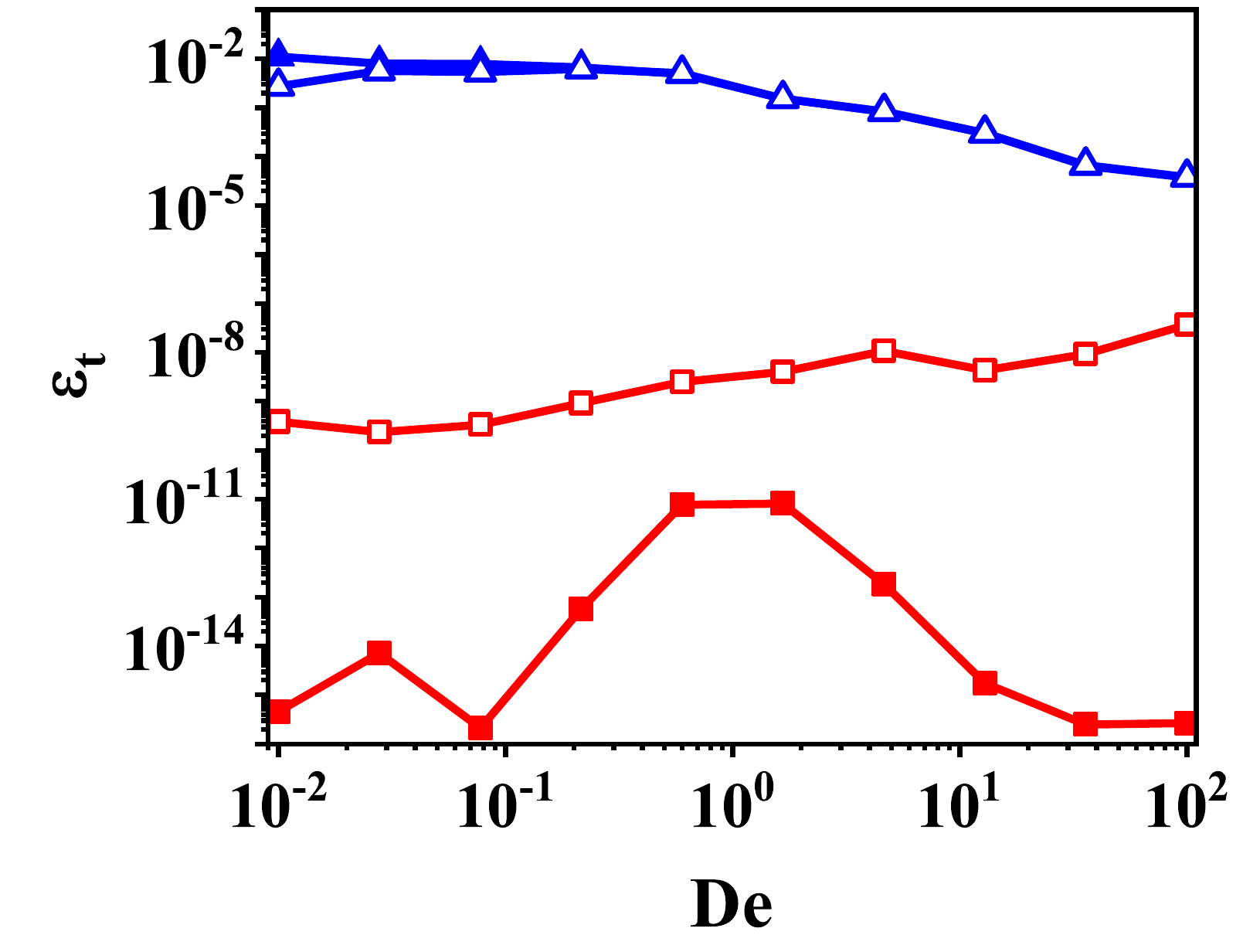}
        \caption{time domain error}
        \label{fig:eODE_De}
    \end{subfigure}
    \caption{The frequency and time domain error metrics for the HB (filled symbols) and NI (open symbols) methods. Calculations are performed at $\gamma_0=10$ for the PTT model (squares) and the type 1 TNM (triangles) with $H=8$.}  
    \label{fig:error}
\end{figure}

We compare the relative accuracy of the HB and IVP methods. Due to the absence of analytical solutions for nonlinear CMs, we examine accuracy using error metrics based on residuals in the time ($\epsilon_t$) and frequency domains ($\epsilon_\omega$) as described in section \ref{sec:error}. Figures \ref{fig:eFC_De} and \ref{fig:eODE_De} represent the frequency domain $\epsilon_{\omega}$ and time domain $\epsilon_{t}$ error metrics, respectively. Results for both the PTT model and the TNM are shown at a large strain amplitude of $\gamma_0=10$ over a range of frequencies. The parameters used for the TNM were $a=-1.0$ and $b=1.0$ corresponding to type I behavior, while $\epsilon=0.1$ was used for the PTT model.

From fig. \ref{fig:eFC_De}, it is evident that FLASH is able to reach its desired target of minimizing the frequency space residual nearly to the level of machine precision. In terms of $\epsilon_\omega$, FLASH outperforms NI for both models, as expected. However, the accuracy of NI, as quantified by this metric, is also quite satisfactory.  It is, therefore mildly surprising that in terms of $\epsilon_{t}$, FLASH leads to a lower error than NI for the PTT model, although the error is quite small for both methods. For the type I TNM in Fig. \ref{fig:eODE_De}, the performance of both methods degrades to approximately the same level. %As observed previously, this loss of accuracy originates from the non-smooth form of the creation and destruction terms, $c(\tilde{t})=\exp{\left(a\text{Wi}\left|\tilde{\sigma}_{12}\right|\right)}$ and $d(\tilde{t})=\exp{\left(b\text{Wi}\left|\tilde{\sigma}_{12}\right|\right)}$.

\begin{figure}
    \centering
    \begin{subfigure}[b]{0.49\textwidth}
        \centering
        \includegraphics[width=\textwidth]{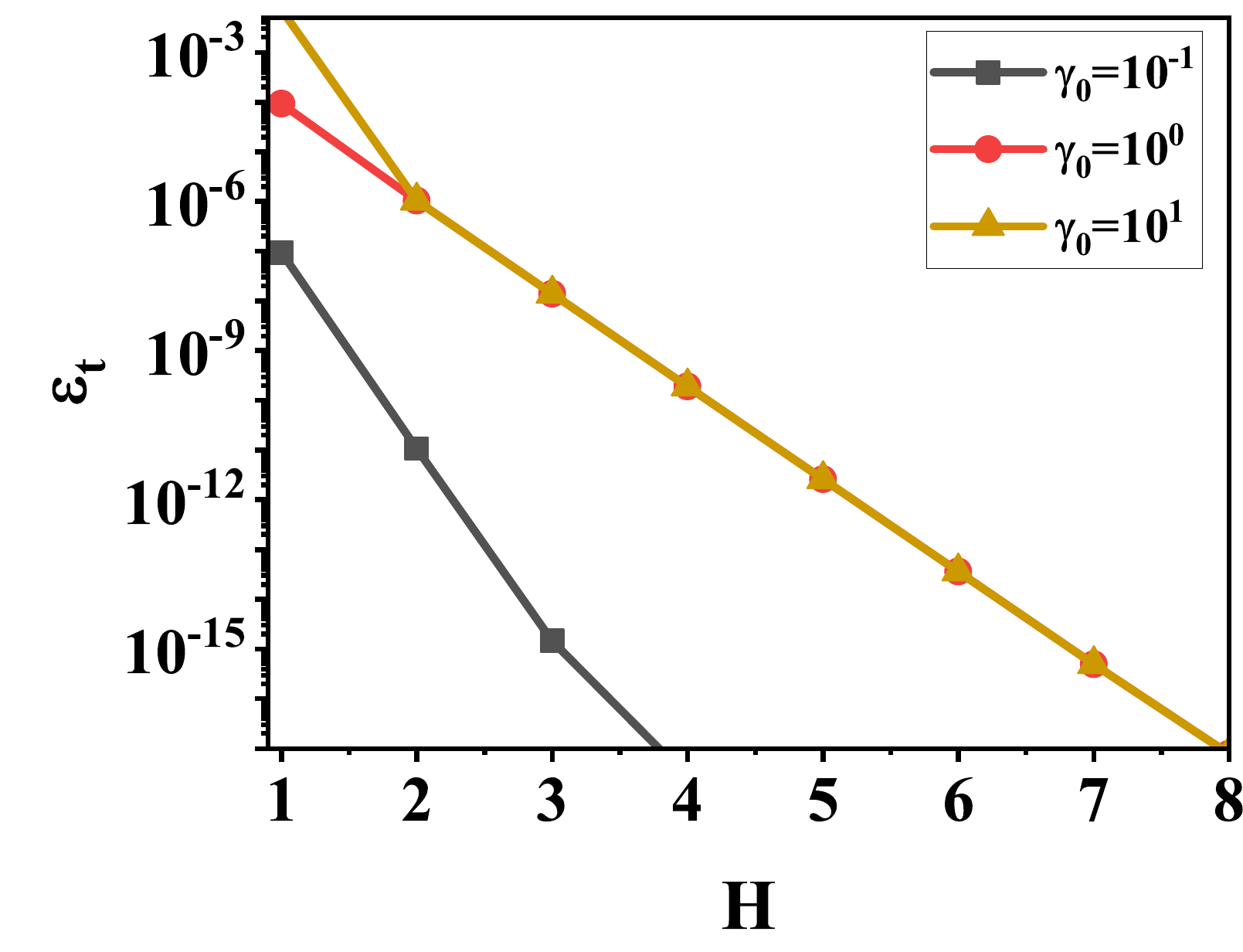}
        \caption{PTT}
        \label{fig:eODE_ptt}
    \end{subfigure}
    \begin{subfigure}[b]{0.49\textwidth}
        \centering
        \includegraphics[width=\textwidth]{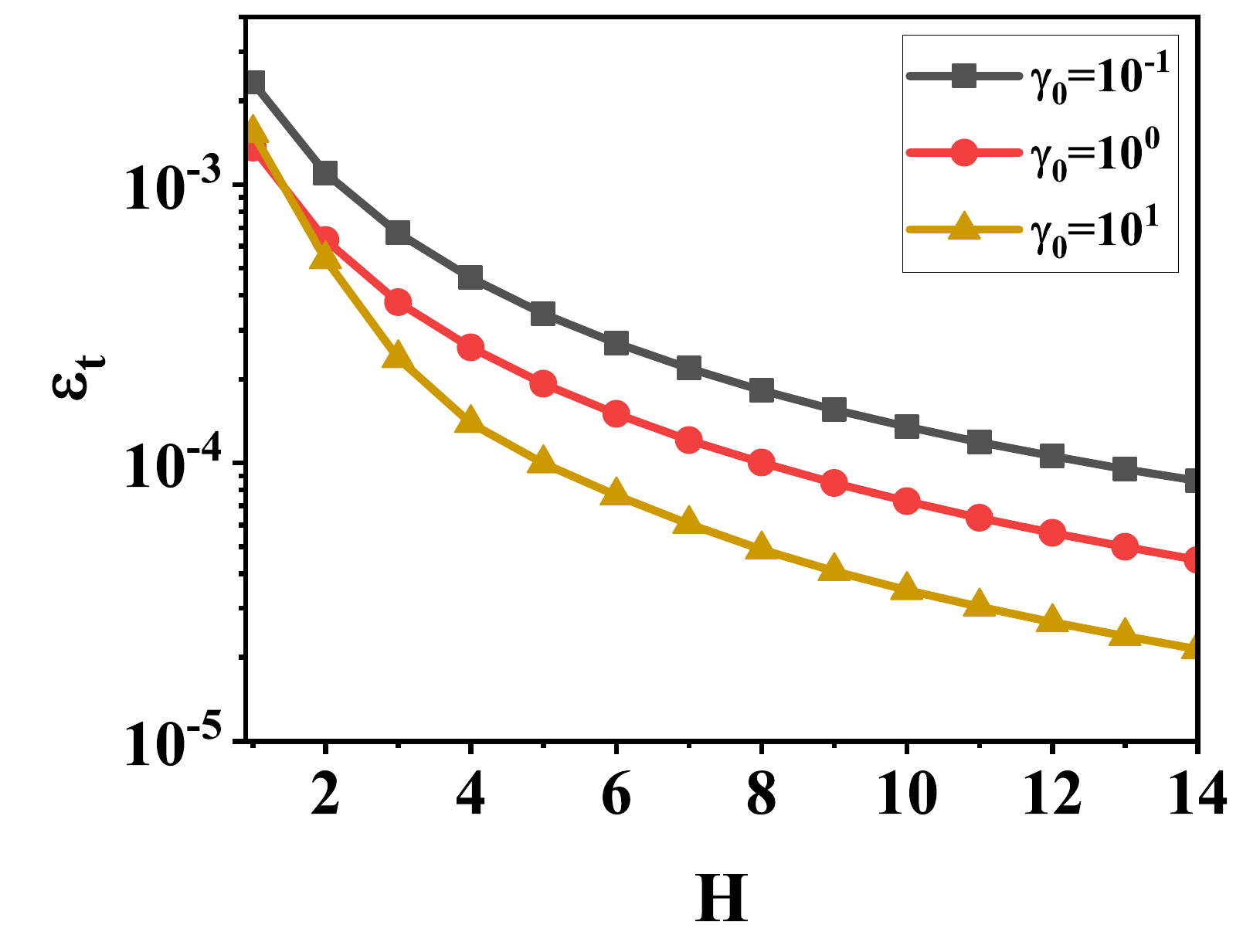}
        \caption{TNM}
        \label{fig:eODE_tnm}
    \end{subfigure}
    \caption{$\epsilon_t$ at different $H$ for the PTT and the TNM model at $\De = 1$ and different $\gamma_0=[10^{-1},10^{0},10^{1}]$. }
    \label{fig:epsilon_t}
\end{figure}

Two user-specified parameters in the FLASH algorithm affect accuracy: the number of data points $N$ in the AFT step and the number of harmonics $H$ in the \emph{ansatz}. The choice of $N$ does not affect accuracy once it is large enough to avoid aliasing error, i.e. $N \ge 2(2H+1)$. Since the AFT scheme is based on the FFT, it is recommended to select $N$ as a power of two. A larger value of $N$ increases the computational cost asymptotically as $\mathcal{O}(N \log N)$. 

The influence of $H$, on the other hand, is more interesting. For the PTT model, fig. \ref{fig:eODE_ptt} shows that $\epsilon_{t}$ decreases to machine precision as $H$ is increased to 8, even at large strain amplitudes.  The linear and super-linear decay observed on the semilog plot indicates fast exponential convergence towards the true solution since the nonlinearity in the PTT model is analytic. In contrast, convergence is sluggish for the TNM. $\epsilon_{t}$ does not approach machine precision even for large $H$, as shown in fig. \ref{fig:eODE_tnm}. Unsurprisingly, this loss of performance originates from the first-order discontinuity in the CM. Gibbs phenomenon is a well-known observation in which Fourier series representations struggle with sharp discontinuities. Such a Fourier series produces oscillations or ``ringing artifacts'' near the point of discontinuity, and the magnitude of the maximum overshoot cannot be reduced by simply increasing $H$. Thus, spectral convergence falters as $H$ is increased and the accuracy of the solution improves gradually. However, it is useful to emphasize that the creation and destruction terms used in this TNM are artificial; these terms are analytic in physically motivated TNMs and present no special issues for FLASH. 

\section{Discussion} 
\label{sec:discussion}

These results establish that FLASH is a fast and accurate method. We discuss how it can be conveniently generalized to other nonlinear CMs, and the potential applications that are unlocked by these attributes.

\subsection{General nonlinear differential constitutive equations}
\label{sec:nonlinear_cms}

%Thus far, the use of HB for LAOS flow problems was substantiated for the single-parameter exponential PTT model and the two-parameter Temporary Network model. However, the AFT algorithm integrated with the HB framework provides great flexibility to handle any form of nonlinearity in a differential CM. 

To adapt FLASH to any arbitrary nonlinear CM, we only need to adapt the nonlinear term $\bm{f}_{\text{nl}}$ corresponding to the CM in eqn. \eqref{eqn:general_form}. Thus far, we already presented $\bm{f}_{\text{nl}}$ for the UCM, PTT, and TNM models by eqns. \eqref{eqn:fnl_ucm}, \eqref{eqn:fnl_ptt} and \eqref{eqn:fnl_tnm}, respectively. Here, we offer $\bm{f}_{\text{nl}}$ for some commonly encountered differential CMs. Additional CMs are entertained in supplementary material.

For clarity, we revert to the dimensional form of CMs in this section. Although non-dimensionalization is recommended for both FLASH and NI as it partially mitigates numerical difficulties encountered at high $\text{Wi}$, it is not a pre-condition for numerical computation. Let us define the unknown variables as $\bm{q} = [\sigma_{11}, \sigma_{22}, \sigma_{33}, \sigma_{12}]$,
%\begin{equation}
%    \bm{q} = \left[ \begin{matrix}
%        \sigma_{11} & \sigma_{22} & \sigma_{33} & \sigma_{12}
%    \end{matrix} \right],
%    \label{eqn:q_dim}
%\end{equation}
and the external forcing term $\bm{f}_{\text{ex}} = G \dot{\gamma}\e_4$ in dimensional form.
%\begin{equation}
%    \bm{f}_{\text{ex}} = \left[ \begin{matrix}
%        0 & 0 & 0 & G \dot{\gamma}
%    \end{matrix} \right].
%    \label{eqn:fex_dim}
%\end{equation}
%\begin{comment}
%    \begin{equation}
%   \bm{f}_{\text{ex}} = G \dot{\gamma}\e_4.
%   \label{eqn:fex_dim}
%\end{equation}
%\end{comment}
For OS flow, we can usually drop $\sigma_{33}$ and all terms associated with it in $ \bm{f}_{\text{nl}}$ and $ \bm{f}_{\text{ex}}$.

Consider the Oldroyd 8-constant linear CM,\cite{Bird1987kt} 
\begin{align}
    \bm{\sigma} + \lambda_{1}\stackrel{\triangledown}{\bm{\sigma}} 
    + \frac{1}{2}\lambda_{3} \left(\bm{\gamma}_{(1)} \cdot \bm{\sigma} 
    + \bm{\sigma} \cdot \bm{\gamma}_{(1)} \right) 
    + \frac{1}{2} \lambda_{5}\left( \text{tr}(\bm{\sigma}) \right)\bm{\gamma}_{(1)}
    + \frac{1}{2} \lambda_{6} \left( \bm{\sigma}: \bm{\gamma}_{(1)} \right) \bm{\delta}
     \nonumber \\ 
     - G \left( \bm{\gamma}_{(1)}
    + \lambda_{2}\bm{\gamma}_{(2)}
    + \lambda_{4} \left(\bm{\gamma}_{(1)} \cdot \bm{\gamma}_{(1)}\right) 
    + \frac{1}{2} \lambda_{7}\left(\bm{\gamma}_{(1)}:\bm{\gamma}_{(1)}  \right)\bm{\delta} \right)
     = \bm{0}
     \label{eqn:oldroyd8constant}
\end{align}
where the derivatives of strain, indicated by subscripts in parenthesis, are given by,
\begin{align*}
    \bm{\gamma}_{(1)} & = \dot{\gamma} \e_{1} \e_{2} + \dot{\gamma} \e_{2} \e_{1}\\
   \bm{\gamma}_{(2)} & =  - 2\dot{\gamma}^2 \e_{1} \e_{1} + \ddot{\gamma} \e_{1} \e_{2} + \ddot{\gamma} \e_{2} \e_{1}
\end{align*}
for OS flow. Here, $\ddot{\gamma} = d\dot{\gamma}/dt$. The $\{\lambda_{i}\}_{i=1}^{7}$ and $G$ constitute the 8 model parameters, lending the model its name. Many popular CMs are special cases of this model. A short list is shown in Table \ref{tab:oldroyd_cases}; more exhaustive lists are available elsewhere.\cite{Saengow2018, Bird1987kt}

For OS flow, the nonlinear term needed by FLASH is found to be,
\begin{align}
    \bm{f}_{\text{nl}}^{\textbf{OD8}} & = \left( \frac{\sigma_{11}}{\lambda_1} - \dot{\gamma}\sigma_{12} \left( 2-\frac{\lambda_3}{\lambda_{1}} - \frac{\lambda_6}{\lambda_{1}} \right) - \frac{1}{2}\frac{\lambda_6}{\lambda_{1}}\dot{\gamma} \left(\sigma_{11} + \sigma_{22} \right) + G \dot{\gamma}^{2} \left( 2\lambda_{2} -\lambda_4 - \frac{1}{2}\lambda_7 \right) \right) \text{ }\e_1   
    \nonumber \\ & + 
    \left( \frac{\sigma_{22}}{\lambda_1}  - \dot{\gamma}\sigma_{12} \left( \frac{\lambda_3}{\lambda_{1}} + \frac{\lambda_6}{\lambda_{1}} \right) - \frac{1}{2} \frac{\lambda_6}{\lambda_{1}} \dot{\gamma} \left(\sigma_{11}+\sigma_{22}\right) - G \dot{\gamma}^{2} \left( \lambda_4 + \frac{1}{2}\lambda_7 \right)  \right) \text{ }\e_2 
    \nonumber \\ & + 
    \left(\frac{\sigma_{33}}{\lambda_1} + \frac{1}{2}\frac{\lambda_6}{\lambda_{1}} \dot{\gamma} \left(\sigma_{11}+\sigma_{22}+2\sigma_{12} \right)  - \frac{1}{2} G \lambda_7 \dot{\gamma}^{2} \right) \text{ }\e_3
    \nonumber \\ & +
    \left(\frac{\sigma_{12}}{\lambda_1} + \dot{\gamma} \sigma_{11} \left(\frac{\lambda_3+\lambda_5}{2\lambda_{1}}\right)  + \dot{\gamma} \sigma_{22} \left(\frac{\lambda_3+\lambda_5-2\lambda_{1}}{2\lambda_{1}}\right) 
    - G \lambda_{2} \frac{d\dot{\gamma}}{dt} \right) \text{ }\e_4.
    \label{eqn:oldroyd_fnl}
\end{align}
%\begin{comment}
%    \begin{equation}
%    \bm{f}_{\text{nl}}^{\textbf{O8}} = \left[ \begin{matrix}
%        \frac{\sigma_{11}}{\lambda_1} - \dot{\gamma}\sigma_{12} \left( 2-\frac{\lambda_3}{\lambda_{1}} - \frac{\lambda_6}{\lambda_{1}} \right) - \frac{1}{2}\frac{\lambda_6}{\lambda_{1}}\dot{\gamma} \left(\sigma_{11} + \sigma_{22} \right) + G \dot{\gamma}^{2} \left( 2\lambda_{2} -\lambda_4 - \frac{1}{2}\lambda_7 \right)   
%        \\
%        \frac{\sigma_{22}}{\lambda_1} -  \dot{\gamma}\sigma_{12} \left( \frac{\lambda_3}{\lambda_{1}} + \frac{\lambda_6}{\lambda_{1}} \right) - \frac{1}{2} \frac{\lambda_6}{\lambda_{1}} \dot{\gamma} \left(\sigma_{11}+\sigma_{22}\right) - G \left( \lambda_4 + \frac{1}{2}\lambda_7 \right) \dot{\gamma}^{2}
%        \\
%        \frac{\sigma_{33}}{\lambda_1} + \frac{1}{2}\frac{\lambda_6}{\lambda_{1}} \dot{\gamma} \left(\sigma_{11}+\sigma_{22}+2\sigma_{12} \right)  - \frac{1}{2} G \lambda_7 \dot{\gamma}^{2}
%        \\
%        \frac{\sigma_{12}}{\lambda_1} + \dot{\gamma} \sigma_{11} \left(\frac{\lambda_3+\lambda_5}{2\lambda_{1}}\right)  + \dot{\gamma} \sigma_{22} \left(\frac{\lambda_3+\lambda_5-2\lambda_{1}}{2\lambda_{1}}\right) 
%         - G \lambda_{2} \ddot{\gamma}
%    \end{matrix} \right].
%    \label{eqn:oldroyd_fnl}
%\end{equation}
%\end{comment}
With this information, ideally coupled with nondimensionalization, it is straightforward to determine the LAOS response of the Oldroyd 8-constant model and its special cases using FLASH. Interestingly, for CMs that are \textit{linear} in $\bm{\sigma}$ such as the Oldroyd 8-constant model, it is possible to develop a specialized HB method that does not require AFT to calculate $\hat{\bm{f}}_{\text{nl}}$. Although this approach requires additional manual effort for setup, it handily outperforms even analytical solutions, as recently reported for the corotational Maxwell model.\cite{Mittal2024}

\begin{table}
    \begin{tabular}{ll}
    \hline 
    \textbf{Constitutive Model}  &  \textbf{Special Case}
    \\
    \hline
    Upper Convected Maxwell &
          $\lambda_2 = \lambda_3 = \lambda_4 = \lambda_5 = \lambda_6= \lambda_7 = 0$ \\
     
   Oldroyd A fluid &
        $\lambda_3 = 2\lambda_1, \lambda_4 =2\lambda_2,\lambda_5=\lambda_6=\lambda_7=0$\\
   Oldroyd B fluid (convected Jeffreys) &
        $\lambda_3=\lambda_4=\lambda_5=\lambda_6=\lambda_7=0$ \\
    
   Corotational Jeffreys Model & $\lambda_3 = \lambda_1, \lambda_4 =\lambda_2,\lambda_5=\lambda_6=\lambda_7=0$\\

        Johnson-Segalman model${}^\text{a}$ &
         $\lambda_1=\frac{\eta_{s}\lambda}{\eta_0}, \lambda_2=\zeta_\text{JS} \lambda , \lambda_3=\frac{\zeta_\text{JS}\eta_s\lambda}{\eta_0}$ \\
&         $\lambda_4=\lambda_5=\lambda_6=\lambda_7=0$\\
    \hline
    \end{tabular}
    \caption{Selected special cases of the Oldroyd 8 constant model. ${}^\text{a}$ For the Johnson-Segalman model, $G = \eta_0^{2}/(\lambda \eta_{s})$ and $\eta_0=\eta_s+\eta_p$, where $\eta_s$ and $\eta_p$ represent the solvent and polymer viscosity contributions. $\lambda $ is the relaxation time and $\zeta_\text{JS}$ is an additional model parameter.}
    \label{tab:oldroyd_cases}
\end{table}

Analogous to the Oldroyd 8-constant model, the generalized Maxwell-like CM presented by Song et al.\cite{Song2020} has several nonlinear CMs for polymer melts and solutions as special cases. It reduces to the UCM model in the linear limit. For uniformity, we use the stress tensor $\bm{\sigma}$ in our exposition instead of the conformation tensor used by the authors. This CM is then given by
\begin{equation}
    \stackrel{\triangledown}{\bm{\sigma}} 
    + \zeta \left( \bm{\gamma}_{(1)} \cdot \bm{\sigma} + \bm{\sigma} \cdot \bm{\gamma}_{(1)}  \right)
    + \frac{1}{\lambda}\bm{H}(\bm{\sigma}) + \bm{J}(\bm{\sigma}) - (1-\zeta)G\bm{\gamma}_{(1)} = \bm{0}
    \label{eqn:general_nonlinear}
\end{equation}
where $G$ and $\lambda$ are linear viscoelastic parameters, $0<\zeta<1$, and different expressions for $\bm{H}(\bm{\sigma})$ and $\bm{J}(\bm{\sigma})$ correspond to different popular nonlinear CMs (see Table  \ref{tab:gn_fnl_table} for some examples). In OS flow, the nonlinear term required by FLASH is
\begin{align}
    \bm{f}_{\text{nl}}^{\textbf{GM}} & =
    \left (2\zeta\sigma_{12}\dot{\gamma} + \frac{1}{\lambda}H_{11}(\bm{\sigma}) + J_{11}(\bm{\sigma}) \right) \text{ }\e_1
    +
    \left( 2\zeta\sigma_{12}\dot{\gamma} + \frac{1}{\lambda}H_{22}(\bm{\sigma}) + J_{22}(\bm{\sigma}) \right) \text{ }\e_2
   \nonumber \\ & +
    \left( \frac{1}{\lambda}H_{33}(\bm{\sigma}) + J_{33}(\bm{\sigma}) \right) \text{ }\e_3
     +
    \left( \zeta\dot{\gamma} \left(\sigma_{11}+\sigma_{22}\right) + \frac{1}{\lambda} H_{12}(\bm{\sigma}) + J_{12}(\bm{\sigma}) + G\zeta\dot{\gamma} \right) \text{ }\e_4.
    \label{gn_fnl}
\end{align}

%\begin{comment}
%    \begin{equation}
%    \bm{f}_{\text{nl}}^{\textbf{GM}} = \left[
%    \begin{matrix}
%        2\zeta\sigma_{12}\dot{\gamma} + \frac{1}{\lambda}H_{11}(\bm{\sigma}) + J_{11}(\bm{\sigma}) 
%        \\
%        2\zeta\sigma_{12}\dot{\gamma} + \frac{1}{\lambda}H_{22}(\bm{\sigma}) + J_{22}(\bm{\sigma}) 
%        \\
%        \frac{1}{\lambda}H_{33}(\bm{\sigma}) + J_{33}(\bm{\sigma}) 
%        \\
%        \zeta\dot{\gamma} \left(\sigma_{11}+\sigma_{22}\right) + \frac{1}{\lambda} H_{12}(\bm{\sigma}) + J_{12}(\bm{\sigma}) + G\zeta\dot{\gamma} 
%    \end{matrix}
%    \right].
%    \label{gn_fnl}
%\end{equation}
%\end{comment}

\begin{table}
	\begin{center}
    \begin{tabular}{ll}
    \hline
      \textbf{Constitutive Equation}  &  \textbf{Special Case}
      \\
      \hline
       Giesekus Model${}^{a}$ &
            $\zeta=0, \bm{J}(\bm{\sigma})=0$, 
            $\bm{H}(\bm{\sigma})=\bm{\sigma}+\frac{\alpha}{G}\bm{\sigma \cdot \sigma}$\\
       
            Phan Thien Tanner Model${}^{b}$ &
            $\zeta \neq 0, \bm{J}(\bm{\sigma})=0,\bm{H}(\bm{\sigma})=Y(\text{tr }\bm{\sigma})\bm{\sigma}$ \\
        Rolie-Poly model${}^{c}$ &
        $\zeta = 0, \bm{H}(\bm{\sigma})=\bm{\sigma}, \bm{J}(\bm{\sigma}) \neq 0$\\
    \hline
    \end{tabular}
    \end{center}
    \caption{Some commonly used nonlinear differential CMs. ${}^{a}$ For the Giesekus model this assumes solvent viscosity $\eta_s=0$. ${}^{b}$ For the PTT model $Y(\text{tr }\bm{\sigma}) = 1+\epsilon \left(\text{tr } \bm{\sigma} \right)$ or $Y(\text{tr }\bm{\sigma}) = \exp{\left(\text{tr }\bm{\sigma}\right)}$ are commonly used forms.\cite{Song2020} ${}^{c}$ For the Rolie-Poly model $\bm{J}(\bm{\sigma})$ takes different forms based on the character of chains.\cite{Song2020}}
    \label{tab:gn_fnl_table}
\end{table}

Several nonlinear CMs, such as the pom-pom model, do not fit the mold of equation \eqref{eqn:general_nonlinear}. In such cases, we track additional dynamical variables by supplementing the $\bm{q}$ vector, as demonstrated in supplementary material. 

\subsection{Model calibration and model selection}

As mentioned in section \ref{sec:intro_cm}, one way to digest and interpret LAOS data is to first assimilate it into a suitable CM by fitting the model parameters. This step is called model calibration. Examples of model calibration based on OS data include studies of colloidal gels,\cite{suman2023large, merger2015large} wormlike micellar solutions,\cite{KateGurnon2012, JACOB201440} cross-linked hydrogel systems,\cite{Bharadwaj2015, Bharadwaj2017} polymer melts and solutions,\cite{Giacomin1993, Jeyaseelan1993267, Nam2008, Hoyle2014} etc. Typically, SAOS data are fitted first to obtain linear viscoelastic parameters. For real systems, this may entail extraction of discrete or continuous relaxation spectra using software like DISCRETE, IRIS, or pyReSpect.\cite{provencher76, baumgaertel89a, Takeh2013, Shanbhag2019respect, Shanbhag2020}

Subsequently, LAOS data is used to determine the nonlinear model parameters.\cite{calin2010determination} Model calibration requires multiple evaluations of the CM with different parameter choices to determine the best fit. The number of model evaluations required can range between $\mathcal{O}(10^3) - \mathcal{O}(10^5)$ depending on the number of $\gamma_0$ and $\omega$ explored, required accuracy, and desire for uncertainty quantification. In such cases, the speed and accuracy advantage of FLASH over NI can practically mean the difference between an hour or several days of computation.

For many materials, such as polymers, multiple CMs may serve as equally attractive candidates at the outset. The goal of model selection is to select the best of these potential candidates, perhaps for inclusion in computational fluid dynamics software. Model selection seeks to balance goodness-of-fit and model complexity to avoid over-fitting. Simple techniques for model selection include the Akaike or Bayesian information criteria, which penalize models with many parameters.  Furthermore, a well-chosen CM aids both scientific discovery (statistical inference) and prediction (statistical prediction). For example, Suman et al. used the functional dependence of damping factor on shear rate obtained from steady shear experiments to predict the LAOS behavior of a critical gel system and make inferences about the change in material microstructure.\cite{suman2023large}  This study used a spectral method for integral CMs to expedite the process.\cite{Shanbhag2021} FLASH brings the same power to differential CMs, which are far more plentiful than integral CMs.

\subsection{Thermodynamic Studies}

The speed and accuracy of FLASH not only aid in the interpretation of experimental LAOS data but also enable theoretical studies that have only been performed rarely due to the associated computational cost. One such application is marking the region of thermodynamic stability of CMs in LAOS flow. For demonstration, we employ the Ziegler criterion, which is rooted in nonequilibrium thermodynamics, as formulated by Saengow and Giacomin to study the corotational Maxwell model.\cite{Saengow2018} The objective here is only to provide a glimpse of the potential that FLASH unleashes, without necessarily commenting on whether it is the best way to study stability.

Saengow and Giacomin \cite{Saengow2018} observed that polymeric liquids subjected to large oscillatory deformations tend to become aperiodic due to the formation of new thermodynamic phases. According to the Ziegler criterion (simplified version of eqn. 83  in their paper), thermodynamic instability is encountered when
%\begin{equation}
%    -\Wi\frac{\partial \eta_1'}{\partial \Wi} - \eta_1' \geq 0,
%    \label{eqn: ziegler1}
%\end{equation}
%where $\eta_1^{\prime}$ is the real part of the complex viscosity $\eta_1^*=\eta_1^\prime - i\eta_1 ^{ \prime \prime}$ which is related to the complex version of the primary modulus as,
%\begin{gather*}
%    \eta_1^*=G_1^*/ (i\omega)  \\
%    \eta_1 ^\prime -i\eta_1^{\prime \prime} = \frac{G_1^{\prime \prime}}{\omega} - \frac{G_1^\prime}{\omega}
%\end{gather*}
%
%Thus, eqn. \eqref{eqn: ziegler1} can further be simplified to give,
\begin{equation}
    Z = -\gamma_0 \left .\frac{\partial G_1''}{\partial \gamma_0}\right|_{\De} - G_1''  \geq 0,
    \label{eqn:Z}
\end{equation}
where $G_1''$ is primary loss modulus in equations \eqref{eqn:shearFourier} and \eqref{eqn:energyDissipation}. Note that eqn. \eqref{eqn:Z} can be further simplified as, 
\begin{equation}
    \left. \frac{\partial \log{G_1''}}{\partial \log{\gamma_0}}\right|_{\De}  \leq -1,
\end{equation}
which links thermodynamic instability to extreme strain-softening. 

In order to make phase diagrams of thermodynamic stability, the CM has to be evaluated on a dense grid in Pipkin space ($\gamma_0$ versus $\omega$). The ladder-based approach for generating initial guesses in FLASH is well-suited for systematically exploring Pipkin space. The PSS solution at a particular value of $\gamma_0$ automatically serves as an excellent initial guess for the solution at a slightly higher value of $\gamma_0$.
Besides improving the precision of the boundary between stable and unstable phases, a dense $\gamma_0$ grid is also required to evaluate the gradient $\partial G_1''/\partial \gamma_0$. However, for smooth curves such $\log{G_1''}$ versus $\log{\gamma_0}$, it is reasonable to obtain $G_1''$ for some specific values of $\gamma_0$ using FLASH and interpolate the rest of the data to obtain a dense grid for calculating the gradient. 
Nevertheless, creating contour maps of $Z$ using eqn \eqref{eqn:Z} can become computationally expensive and has thus far been restricted to CMs with analytical OS solutions such as the corotational Maxwell model. FLASH frees us from this constraint. 

\begin{figure}
    \centering
    \begin{subfigure}[b]{0.45\textwidth}
        \centering
        \textbf{PTT Model}
        \includegraphics[width=\textwidth]{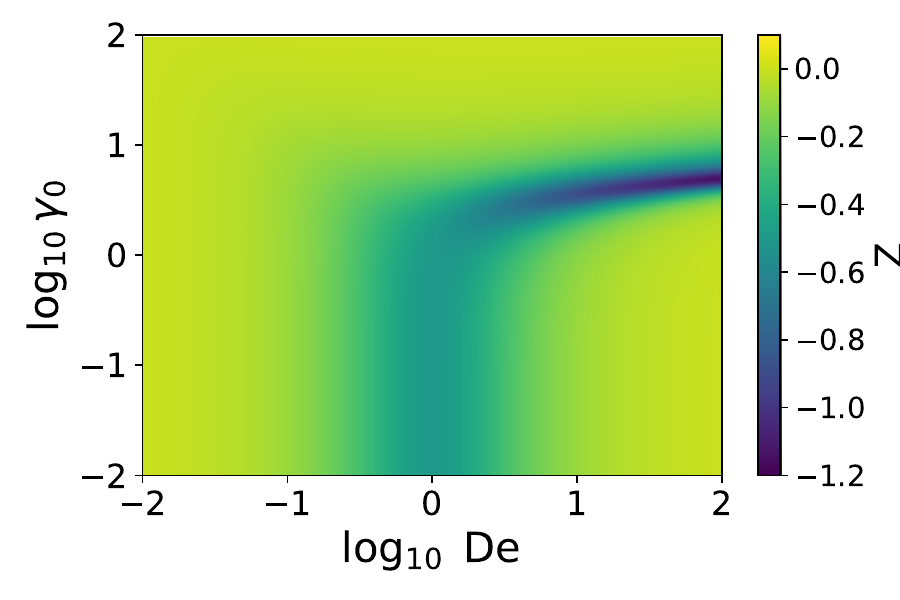}
        \caption{$\epsilon=0.1$, CPU $\approx$ 93s}
        \label{fig:pttThermoPipkin}
    \end{subfigure}
    \begin{subfigure}[b]{0.45\textwidth}
        \centering
        \textbf{Giesekus Model}
        \includegraphics[width=\textwidth]{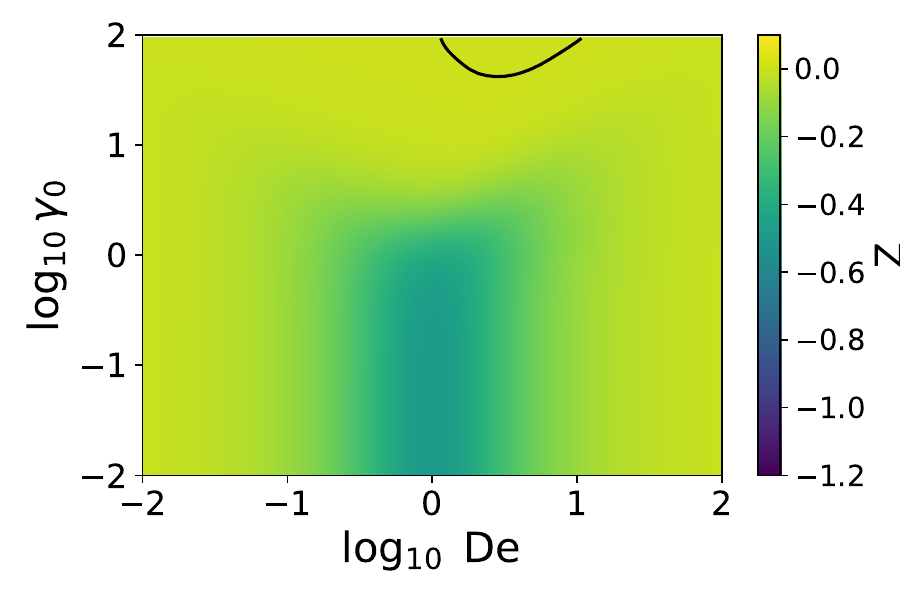}
        \caption{$\alpha=0.5$, CPU $\approx$ 96s}
        \label{fig:giesekusThermoPipkin}
    \end{subfigure}
    \begin{subfigure}[b]{0.45\textwidth}
        \centering
        \includegraphics[width=\textwidth]{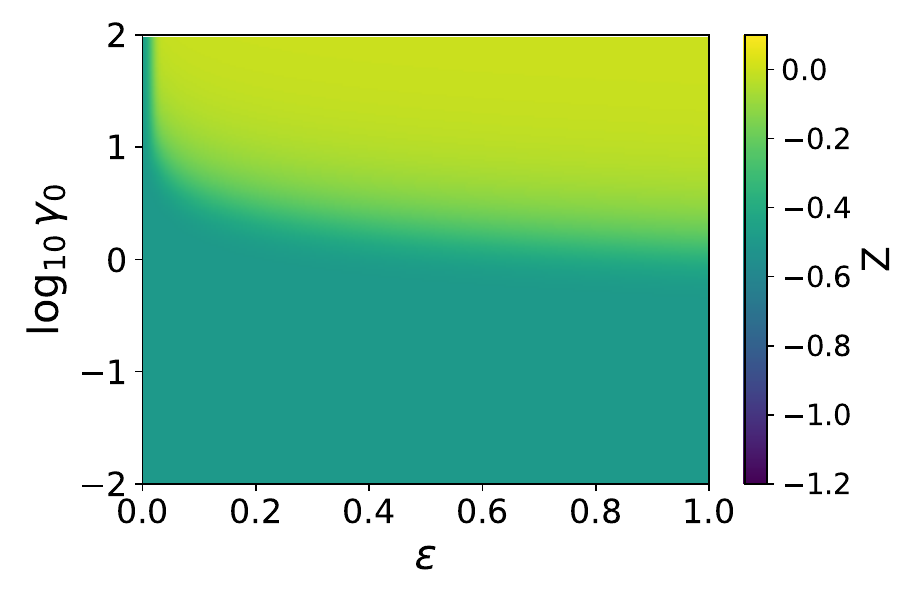}
        \caption{$\De=1$, CPU $\approx$ 62 s}
        \label{fig:pttThermoEPS}
    \end{subfigure}
    \begin{subfigure}[b]{0.45\textwidth}
        \centering
        \includegraphics[width=\textwidth]{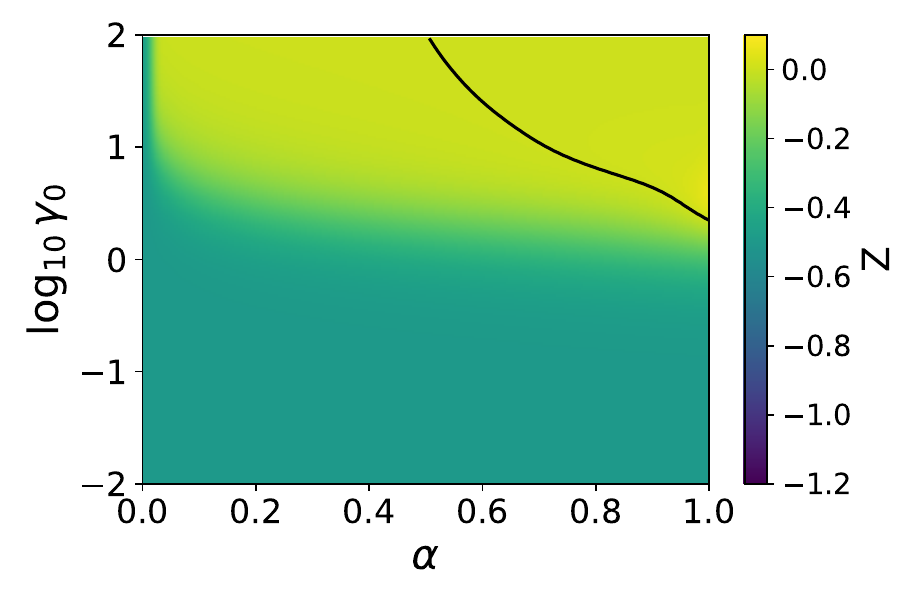}
        \caption{$\De=1$ CPU $\approx$ 66 s}
        \label{fig:giesekusThermoAlpha}
    \end{subfigure}
    \caption{Contour plots of $Z$ for the PTT (left column) and Giesekus (right column) models. The top row spans a range of frequencies, holding the nonlinear model parameter ($\epsilon$ or $\alpha$) fixed. The bottom row spans a range of values for the model parameters at a particular frequency. The thick black lines in (b) and (d) correspond to $Z=0$ and mark the boundary of stability using the Ziegler criterion.}
    \label{fig:thermo_plots}
\end{figure}

Figures \ref{fig:pttThermoPipkin} and \ref{fig:giesekusThermoPipkin} show contour maps based on the Ziegler criteria for the PTT model $(\epsilon=0.1)$ and the Giesekus model $(\alpha=0.5)$ respectively. We use the Giesekus model in lieu of the TNM because it is a single parameter model, which allows us to visualize the phase diagram more easily. In order to generate these plots, $G_1''$ is obtained on a logarithmically equispaced grid of $\De$ and $\gamma_0$. We use 65 different value of $\De \in [10^{-2},10^2]$, and 80 different values of $\gamma_0 \in [10^{-2},10^2]$. 

For $\Wi<1$, a small number of harmonics $(H=3)$ is sufficient to obtain an accurate solution, while $H = 6$ was used when $\Wi \geq 1$. In accordance with $H$, $N=2^5$ was used for all the computations. Subsequently, the data are interpolated using cubic splines, from which the gradient $d\log{G_1''}/d\log{\gamma_0}$ is analytically computed. A similar approach was used to obtain Figures \ref{fig:pttThermoEPS} and \ref{fig:giesekusThermoAlpha}, where the model parameters for the PTT and the Giesekus models were varied, keeping the frequency fixed ($\De=1$). Plots analogous to figure \ref{fig:thermo_plots} for the CMM and the TNM are provided in the supplementary information. 

It is evident from \ref{fig:pttThermoPipkin} and \ref{fig:pttThermoEPS} that the PTT model is stable across the entire range of strains, frequencies, and model parameters probed. On the other hand, figure \ref{fig:giesekusThermoPipkin}  indicates that the Giesekus model tends to become thermodynamically unstable in the large $\gamma_0$ regime $(\gamma_0 > 10)$. This is an expected outcome since the model shows a strong strain softening in this regime. Furthermore, figure \ref{fig:giesekusThermoAlpha} shows that this transition from stable to unstable region happens at a lower $\gamma_0$ as the model parameter is increased. This is likely a consequence of the increasing importance of the quadratic nonlinear term. The CPU time required for generating each of the top two figures was about 1.5 min, where FLASH was evaluated nearly  $5000$ (65 $\times$ 80) times. The bottom two plots required about 1.25 min, for a similar number of model evaluations. It is worthwhile to point out that similar thermodynamic stability plots for the corotational Maxwell model on a coarser $(\gamma_0, \omega)$ mesh
took about 12 - 24h using NI.\cite{Saengow2018}  %Thus, the study of thermodynamic stability of nonlinear CMs subjected to LAOS flow fields is an example of the vast opportunities FLASH provides in theoretical studies on LAOS.

\subsection{Stable and Unstable Solutions}

\begin{figure}
    \centering
    \includegraphics[width=0.9\textwidth]{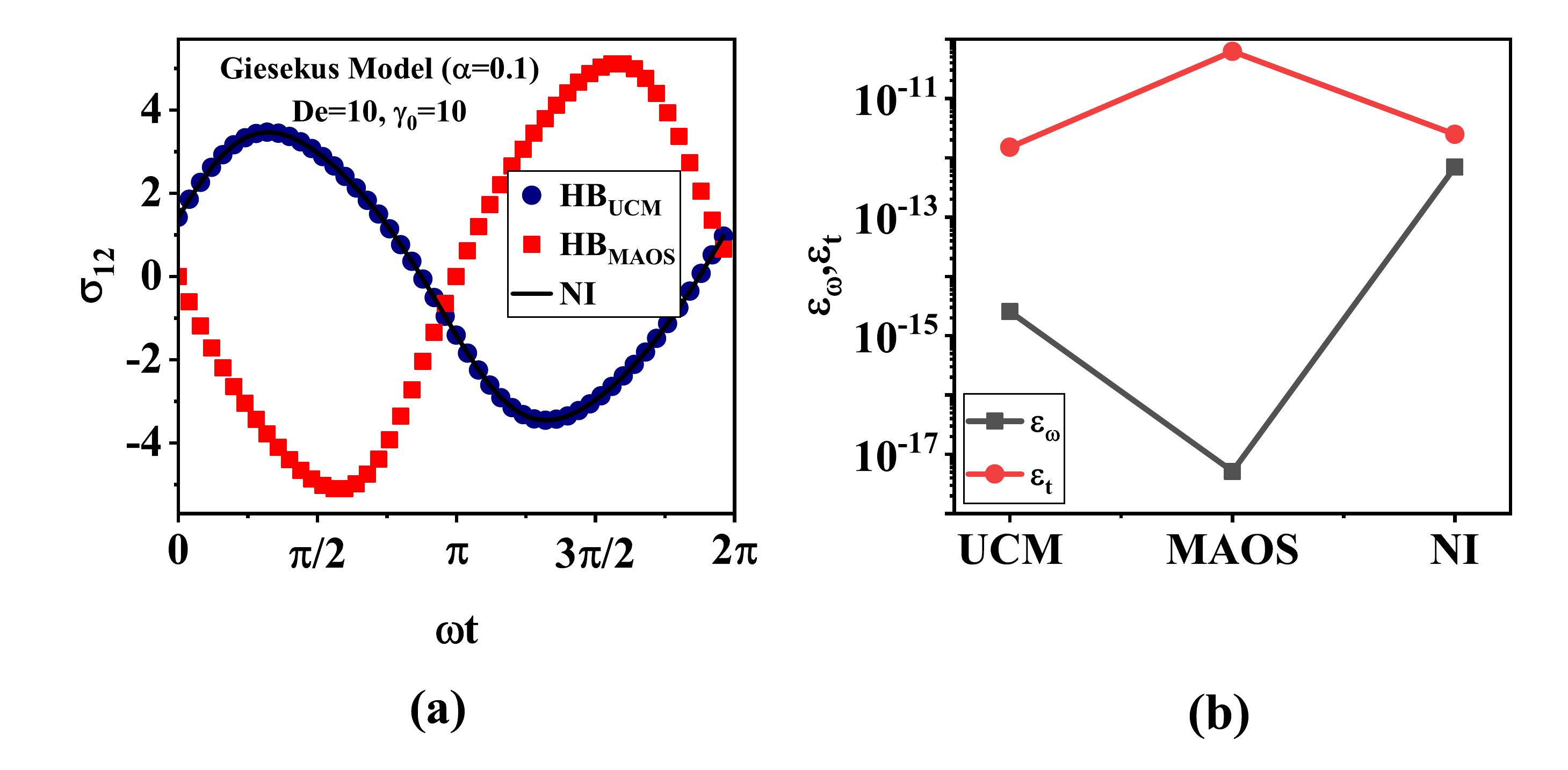}
    \caption{(a) LAOS shear stress response of the Giesekus model $(\alpha=0.1)$ at $\gamma_0=10$ and $\De=10$ obtained by using (i) the UCM solution, and (ii) the MAOS solution as initial guess values to FLASH, and (iii) NI. (b) the time and frequency domain error metrics for the three solutions.}
    \label{fig:unstable_curves}
\end{figure}

Nonlinear CMs, like other nonlinear dynamical systems, can exhibit multiple solutions when subjected to large perturbations. Some of these solutions may be stable, while others may be unstable. Depending on the initial condition, a dynamical system follows a unique trajectory to reach a time-invariant state called the attractor. Each attractor has its basin of attraction, i.e., the set of initial conditions that lead to that solution. Depending on the nature of the attractor, the asymptotic trajectory can be classified as fixed point, periodic oscillation, quasi-periodic, or chaotic.\cite{Krack2019} If an asymptotic solution is mathematically unstable, we do not expect to observe it in experiments or even in NI solutions due to environmental or numerical perturbations. Nevertheless, it is essential to study the stability characteristics of dynamical systems because, in addition to theoretical insights, they provide robust limits that are directly relevant to computational fluid dynamics calculations.

Unlike thermodynamic stability analysis, which can also be carried out using NI, albeit at a significant computational cost, only HB allows us to track stable and unstable solution branches. Since HB solves nonlinear algebraic equations that have multiple solutions,  we can potentially describe all possible solutions simply by tweaking the initial guess.

We illustrate this capability in figure \ref{fig:unstable_curves}a. Here, the LAOS solution of the Giesekus model $(\alpha=0.1)$ at $\gamma_0=10$ and $\De=10$ is obtained using FLASH for two different initial guesses: (i) the UCM model solution, and (ii) the MAOS solution. We also present the `stable' solution obtained using NI, which visually overlaps with the HB solution that uses the UCM response as the initial guess. This suggests that the HB solution using the MAOS response as the initial guess is unstable. To confirm that both the solutions obtained using FLASH are valid, we show the error metrics based on the frequency and time domain residuals in Fig. \ref{fig:unstable_curves}b. The low values of these error metrics ($\lesssim 10^{-10}$) indicates that all three solutions plotted in Fig. \ref{fig:unstable_curves}a are reasonable.

A PSS solution is said to be asymptotically stable if all the trajectories that start in its vicinity remain there. A standard procedure for asymptotic stability analysis involves computing the eigenvalues of a monodromy matrix and using the Floquet theorem to assess the stability.\cite{floquet1883equations,Moore2005}  An alternative frequency domain method, that is more compatible with HB, is Hill's method.\cite{Hill1886,LAZARUS2010510} Bifurcations, where the stability of a periodic solution changes character in response to changes in model parameters or operating conditions, can also be studied with HB.\cite{Krack2019} Hill's method can be combined with numerical path continuation to generate phase diagrams for mathematical stability.

\section{Summary and Conclusions}

HB is a Fourier-Galerkin method that transforms a system of nonlinear ODEs defined by the CM to a system of nonlinear algebraic equations. However, this transformation cannot be performed analytically for all nonlinear CMs. AFT is a numerical scheme that overcomes this obstacle by numerically shuttling between time and frequency domains. This work presents FLASH, a flexible method and computer program that combines HB with the AFT scheme to compute the PSS LAOS response of arbitrary nonlinear differential CMs. This offers a convenient interface for adapting HB to arbitrary CMs by shifting the burden of setting up equations in the right format from the modeler to the computer.

FLASH is validated using two CMs: the exponential PTT model, and a variant of the TNM. Its speed and accuracy is compared with the standard method of solving IVPs using NI. For the PTT model, FLASH is several orders of magnitude more accurate while also being 1-2 orders of magnitude faster. For the TNM, the superiority of FLASH over NI is less pronounced. In terms of the time-domain residual, they both perform equally well, while FLASH outperforms NI by several orders of magnitude in terms of the frequency-domain residual. FLASH retains its speed advantage for the TNM. As the number of harmonics $H$ in the ansatz increases, it shows exponential convergence for the PTT model, which degrades for the TNM due to a first-order discontinuity in the CM.

The speed and accuracy of FLASH open new avenues for theoretical studies of CMs, such as thermodynamic stability and the presence of stable and unstable solution branches. Combined with the spectral method \cite{Shanbhag2021} for integral CMs, FLASH offers a compelling toolbox for interpreting LAOS data through the lens of a CM and for model calibration and selection.

\section{Appendix}

\subsection{Fourier Series}
\label{app:Fourier}

Any periodic function $f(t)$ with period $T$, $f(t+T) = f(t)$, can be approximated by a truncated Fourier series. It can be represented using either complex exponential or trigonometric basis functions.  In this work, we choose the latter because it facilitates numerical computation by relying only on ``real'' algebra. Thus,
\begin{equation}
f(t) \approx f^{H}(t) = \hat{f}(0) + \sum_{k=1}^{H} \hat{f}_c(k) \cos k\omega t + \hat{f}_s(k) \sin k\omega t,
\label{eqn:truncFS}
\end{equation} 
where $H$ determines the order of truncation. The $2H+1$ trigonometric basis functions  $\bm{B}_H(t) = [1, \cos \omega t, \cos 2 \omega t, \cdots, \cos H \omega t, \sin \omega t, \sin 2 \omega t \cdots, \sin H \omega t]$. The Fourier coefficients (FCs), for $k = 0, 1, \cdots, H$, are given by
\begin{align}
\hat{f}(0) & = \frac{1}{T} \int_{0} f(t) \, dt  \notag\\ 
\hat{f}_c(k) & = \frac{2}{T} \int_{0} \cos(k \omega t) \cdot f(t) \, dt   \notag\\ 
\hat{f}_s(k) & = \frac{2}{T} \int_{0} \sin(k \omega t) \cdot f(t) \, dt.
\label{eqn:FScoeff}
\end{align}
Following the order of the basis functions in $\bm{B}_H(t)$, we can arrange these FCs into a $2H+1$-dimensional vector $\hat{\bm{f}} = [\hat{f}(0),\hat{f}_c(1), \hat{f}_c(2), \cdots, \hat{f}_c(H), \hat{f}_s(1), \hat{f}_s(2), \cdots, \hat{f}_s(H)]$. Thus, the dot product $\hat{\bm{f}} \cdot \bm{B}_H(t) = f^{H}(t)$ expresses a linear combination of the basis functions determined by the FCs. This allows us to approximate and encode a continuous periodic time-domain function $f(t)$ as a discrete frequency domain representation $\hat{\bm{f}}$. This is attractive for the following reasons:

\begin{itemize}
\item \textbf{Spectral accuracy}: If the function $f(t)$ is analytic (infinitely differentiable), the FCs decay at an exponential rate, i.e $|\hat{f}_c(k)|$ and  $|\hat{f}_s(k)| \leq Fc^{-k}$, where $c > 1$ and $F$ is positive. Practically, this means that we can accurately approximate $f(t)$ with a relatively small number of harmonics $H$.

\item \textbf{Computational Efficiency}: $\hat{\bm{f}}$ can be calculated efficiently from $N$ discrete equispaced samples of $f(t)$ in the domain $[0, T)$ using fast Fourier transform (FFT) at a cost of $\mathcal{O}(N \log N)$. If $N$ is too small, improper resolution of high-frequency components of $f(t)$ results in distortion of low-frequency components, a problem known as aliasing. This can be prevented by choosing $N \geq 2H + 1$.

\item \textbf{Orthogonality of Basis Functions}: Let $B_{H, i}(t)$ and $B_{H, j}(t)$ be any two basis functions from the set $\bm{B}_H(t)$, where the subscripts $i$ and $j$ correspond to their positions in the set. Then, their inner product is given by,
\begin{equation}
\langle B_{H, i} \, B_{H, j} \rangle = \frac{2}{T} \int_{0} B_{H, i}(t) \, B_{H, j}(t)\, dt = \begin{cases}
2 & i = j = 1\\
\delta_{ij} & \text{otherwise}.
\end{cases}
\end{equation}
This relation allows us to conveniently pose the system of differential equations in weak form using a weighted residual method.

\item \textbf{Differential equations become algebraic equations}: The derivative of the periodic function $\dot{f}^H(t) = df^H(t)/dt$ is (see eqn. \ref{eqn:truncFS}),
\begin{align*}
\hat{\dot{f}}^H(t) & = \sum_{k=1}^{H} (k \omega) \hat{f}_c(k) (-\sin k\omega t) + (k \omega) \hat{f}_s(k) \cos k\omega t\\
 & = \sum_{k=1}^{H} \left[ (k\omega)\, \hat{f}_s(k)\, \cos k \omega t + (-k\omega)\, \hat{f}_c(k)\, \sin k \omega t \right].
\end{align*}
Thus, the Fourier representation $\hat{\dot{\bm{f}}}$ of $\dot{f}^H(t)$ can easily be obtained in terms of the FCs $\hat{\bm{f}}$ as, $\hat{\dot{\bm{f}}} = [0, \omega \hat{f}_s(1), \cdots,  H \omega \hat{f}_s(H), -\omega \hat{f}_c(1), \cdots,  -H \omega \hat{f}_c(H)]$. This allows us to transform the problem from solving nonlinear ODEs for $f(t)$ in the time domain to solving algebraic equations for $\hat{\bm{f}}$ in the frequency domain.

\end{itemize}

\subsection{HB equations for UCM model}
\label{app:UCM_HB}
The system of ODEs for the Upper Convected Maxwell model subjected to oscillatory shear strain deformation can be given by:
\begin{align}
\dot{\sigma}_{11} & + \frac{1}{\lambda}\sigma_{11} - 2 \dot{\gamma} \sigma_{12} = 0 \nonumber\\
\dot{\sigma}_{22} & + \frac{1}{\lambda} \sigma_{22} = 0 \nonumber\\
\dot{\sigma}_{33} & + \frac{1}{\lambda} \sigma_{33} = 0. \nonumber\\
\dot{\sigma}_{12} & + \frac{1}{\lambda} \sigma_{12} - \dot{\gamma} \sigma_{22} = G \dot{\gamma}
\label{eqn:ucm_ode}
\end{align}
Due to linearity, these equations can be solved analytically. Equations for $\sigma_{22}$ and $\sigma_{33}$ are decoupled from $\sigma_{11}$ and $\sigma_{12}$, and can be solved independently. At sufficiently long times, both these components decay to zero as $\sigma_{22}(t)/\sigma_{22}(0) = \sigma_{33}(t)/\sigma_{33}(0) = e^{-t/\lambda}$. Since $\sigma_{22}(t \rightarrow \infty) = 0$, the equation for the shear stress $\sigma_{12}(t)$ also decouples and has the classic solution given by eqn.\eqref{eqn:shearSAOS} similar to the linear Maxwell model. The PSS for the first normal stress difference, on the other hand, takes the form of eqn. \eqref{eq:normalSAOS}, where $F_0^{\prime\prime}=\Gp$, $F_2^{\prime}=-\Gp + \frac{1}{2} G^{\prime}(2 \omega)$ and $F_2^{\prime\prime}=\Gpp - \frac{1}{2} G^{\prime\prime}(2 \omega)$

The Fourier transform $\hat{\bm{f}}_{\nl} =\left[\hat{\bm{f}}_{\nl,1}, \hat{\bm{f}}_{\nl,2}, \hat{\bm{f}}_{\nl,3}, \hat{\bm{f}}_{\nl,4}\right]$ of the nonlinear term $\bm{f}_{\nl}$ given by eqn. \ref{eqn:fnl_ucm} can be determined analytically. The first component is given by
    \begin{align}
    \hat{\bm{f}}_{\nl,1}^{\text{UCM}} & = \left( \q_{1}(0)-\Wi\q_{c,4}(1) \right)\e_1 + \sum_{k=1}^{H} \q_{c,1}(2k)-\Wi\left(\q_{c,4}(2k-1)+\q_{c,4}(2k+1)\right) \text{ }\e_{k+1} 
    \\ & + \sum_{k=1}^{H} \q_{s,1}(2k)-\Wi\left(\q_{s,4}(2k-1)+\q_{s,4}(2k+1)\right) \text{ }\e_{k+H+1}. \label{eqn:fnl1_UCM}
\end{align}
For $i = $ 2 and 3,
    \begin{equation}
    \hat{\bm{f}}_{\nl,i}^{\text{UCM}} =  \q_{i}(0)\text{ }\e_1 + \sum_{k=1}^{H} \q_{c,i}(2k) \text{ }\e_{k+1} + \q_{s,i}(2k) \text{ }\e_{k+H+1}. \label{eqn:fnl23_UCM}
\end{equation}
Finally the term corresponding to the shear stress
    \begin{align}
    \hat{\bm{f}}_{\nl,4}^{\text{UCM}} & = \sum_{k=0}^{H} \q_{c,4}(2k+1) - \frac{\Wi}{2} \left( \q_{c,2}(2k)+\q_{c,2}(2k+2) \right) \text{ } \e_{k+1} 
    \\ & + \sum_{k=0}^{H} \q_{s,4}(2k+1) - \frac{\Wi}{2} \left( \q_{s,2}(2k)+\q_{s,2}(2k+2) \right) \text{ } \e_{k+H+2}. \label{eqn:fnl4_UCM}
\end{align}

We formulate the HB equation system by using eqns. \eqref{eqn:qndot_hat} -- \eqref{eqn:qsdot_hat} and \eqref{eqn:fnl1_UCM} -- \eqref{eqn:fnl4_UCM} to obtain equations of the form eqn. \eqref{eqn:residualHB}. 
\begin{gather}
    \hat{r}_{c,1}(2k)  = 2k\De q_{s,1}(2k) + q_{c,1}(2k) -\Wi \left(q_{c,4}(2k-1)+q_{c,4}(2k+1)\right)  \\  \label{eqn:r_c1}
    \hat{r}_{s,1}(2k) = -2k\De q_{c,1}(2k) + q_{s,1}(2k) -\Wi \left(q_{s,4}(2k-1)+q_{c,4}(2k+1)\right)  \\  \label{eqn:r_s1}
    \hat{r}_{c,2}(2k)  = 2k\De q_{s,2}(2k) + q_{c,2}(2k) \\
    \hat{r}_{s,2}(2k) = -2k\De q_{c,2}(2k) + q_{s,2}(2k) \\
    \hat{r}_{c,3}(2k)  = 2k\De q_{s,3}(2k) + q_{c,3}(2k) \\
    \hat{r}_{s,3}(2k) = -2k\De q_{c,3}(2k) + q_{s,3}(2k)
\end{gather}
for the ODEs describing the evolution of the normal stresses with $k\in(1,H)$, and
\begin{gather}
    \hat{r}_{c,4}(2k+1) =(2k+1)\De q_{s,4}(2k+1) + q_{c,4}(2k+1) -\frac{\Wi}{2} \left(q_{c,2}(2k)+q_{c,2}(2k+2)\right) - \delta_{k,0}  \label{eqn:r_c4}  \\
    \hat{r}_{s,4}(2k+1) = -(2k+1)\De q_{c,4}(2k+1) + q_{s,4}(2k+1) -\frac{\Wi}{2} \left(q_{s,2}(2k)+q_{s,2}(2k+2)\right) 
  \label{eqn:r_s4}
\end{gather}
for fourth ODE corresponding to $\sigma_{12}$ with $k\in(0,H)$. From inspection, we find that all these equations are identically zero except for the case of $k=0$ and $k=1$ in eqn. \eqref{eqn:r_c1}, $k=1$ in eqn. \eqref{eqn:r_s1}, and $k=0$ in eqns. \eqref{eqn:r_c4} and \eqref{eqn:r_s4}. This set of five equations can be solved analytically. Unsurprisingly, they yield the classic analytical solution.

\section*{Supplementary Material}
See supplementary material online for 
\begin{itemize}
    \item Validation of FLASH for the UCM model
    \item Additional results for the PTT and TNM models
    \item A framework to apply HB for other nonlinear constitutive models
    \item Thermodynamic stability- Corotational Maxwell model and TNM
    \item Nomenclature and abbreviations used
\end{itemize}

\section*{Acknowledgments}

SM acknowledges support from the Prime Minister Research Fellowship program and the Fulbright Nehru Doctoral Research program. YMJ further acknowledges financial support from the Department of Science and Technology, Govt. of India (Grant No. CRG/2022/004868). This work is based in part upon work supported by the National Science Foundation under grant no. NSF DMR-1727870 (SS).

\section*{Data Availability Statement}
The Python code for FLASH and instructions for using it are included as supplementary material. Other data that support the findings of this study are available from the corresponding author upon reasonable request.

\bibliography{hb,LAOS}

\end{document}

\begin{center}
\textbf{\huge{Supplementary Material}}\\
\vspace{0.5cm}
\textbf{\Large Harmonic Balance for Differential Constitutive Models under Oscillatory Shear}\\
\vspace{0.5cm}

\medskip

Shivangi Mittal$^{1}$, Yogesh M. Joshi$^{1}$, and Sachin Shanbhag$^{2,*}$\\ 

${}^{1}$Department of Chemical Engineering, Indian Institute of Technology
Kanpur, India\\
${}^{2}$Department of Scientific Computing, Florida State University, Tallahasee,
Florida, USA

\end{center}

\vspace{0.5cm}

\tableofcontents

\newpage

\section{Validation of FLASH for the UCM model}

The oscillatory shear (OS) response of the UCM model can be calculated analytically. This allows us to perform an initial validation test of FLASH. 

\begin{figure}[hb!]   
%\begin{center}
        \begin{tabular}{cc}
        \includegraphics[scale=0.28]{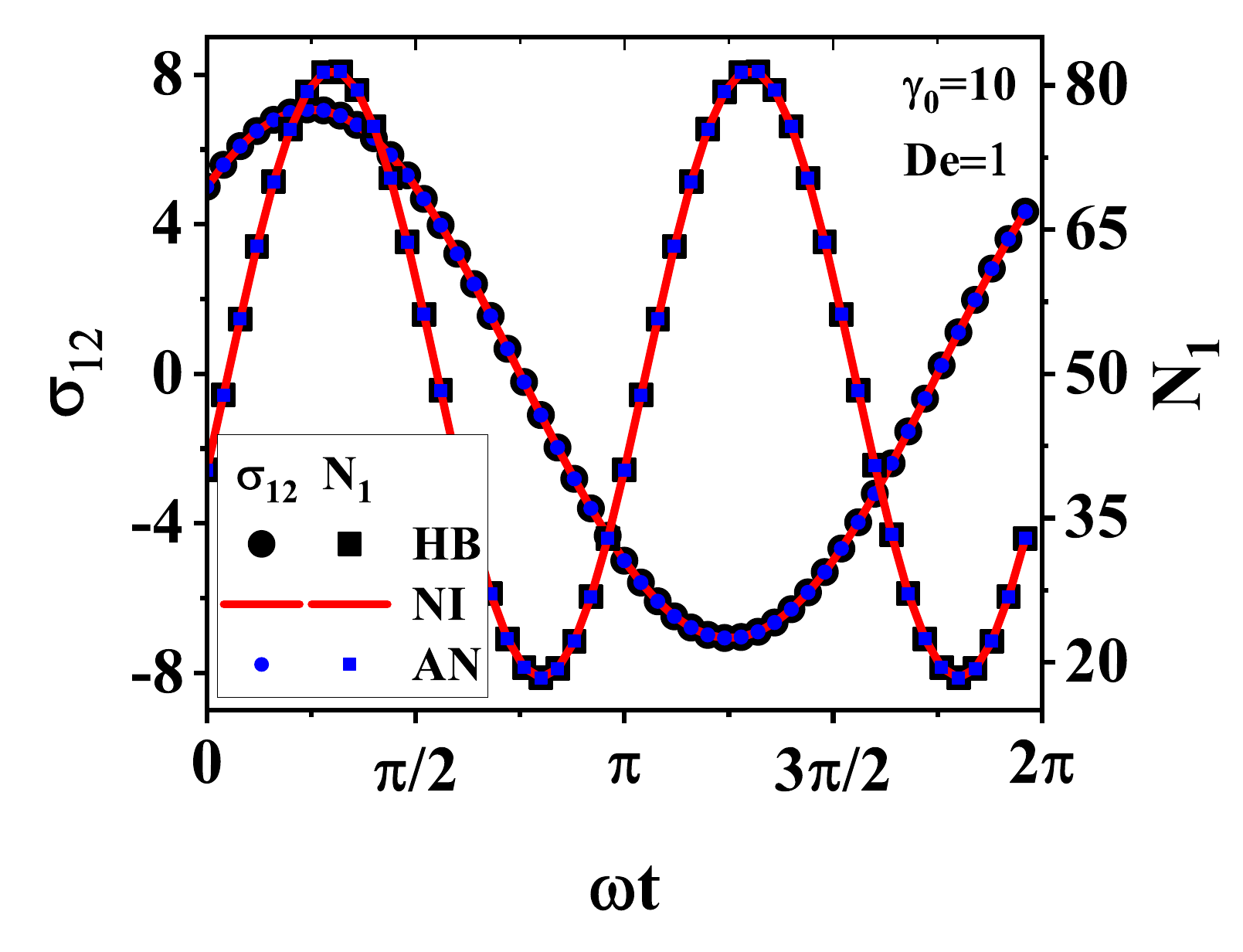} & \includegraphics[scale=0.28]{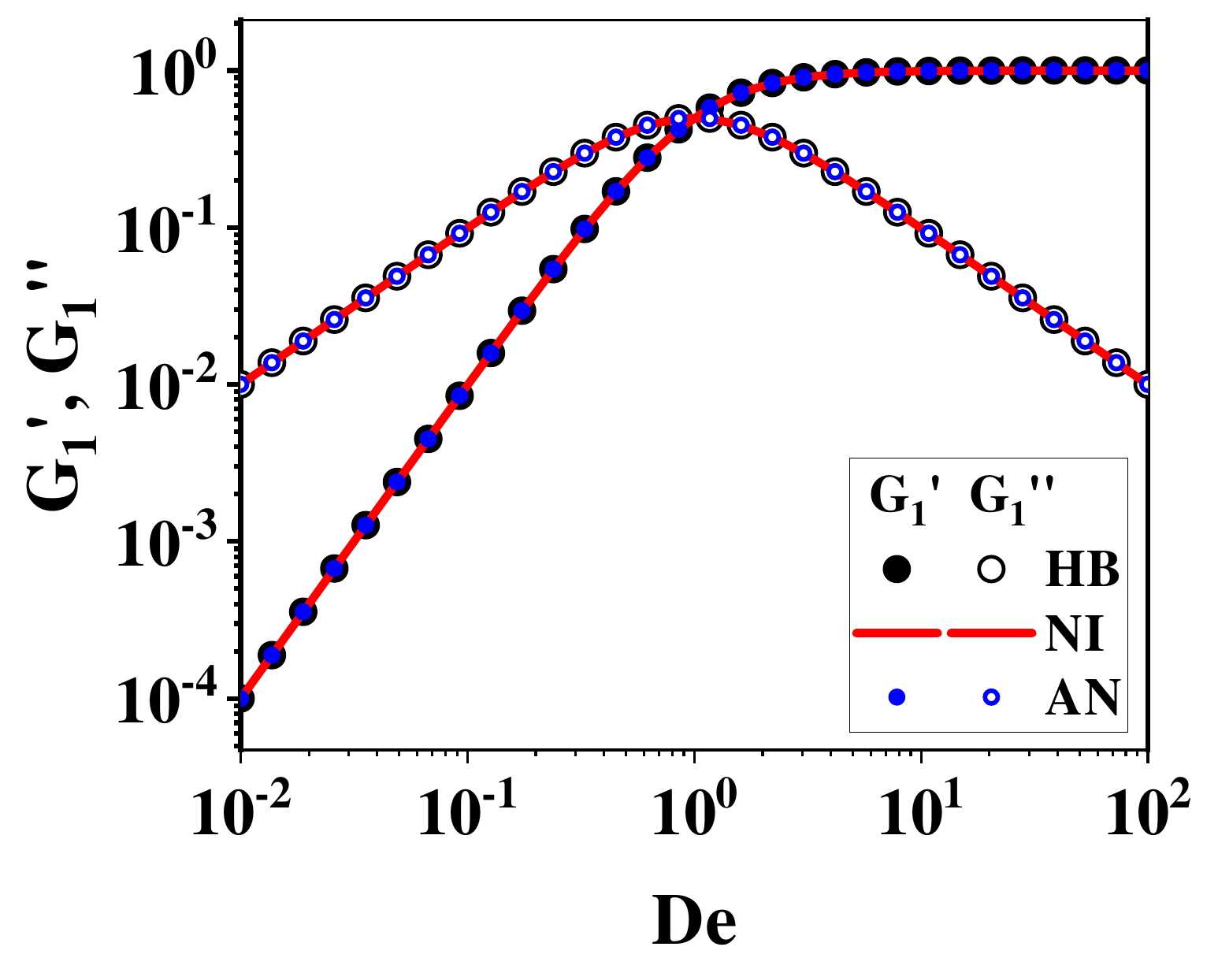}  \\ (a) & (b) \\
        \includegraphics[scale=0.28]{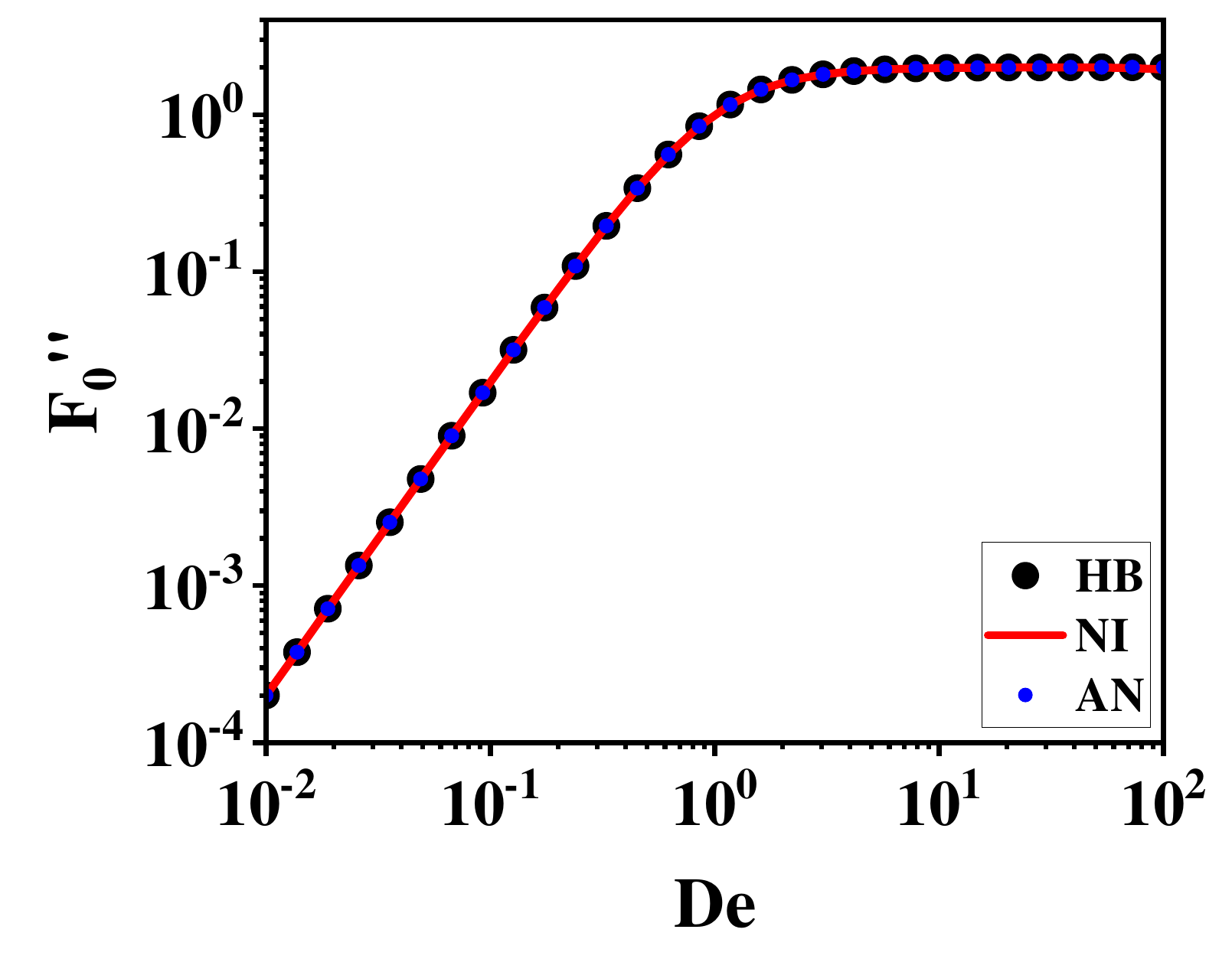} 
        & \includegraphics[scale=0.28]{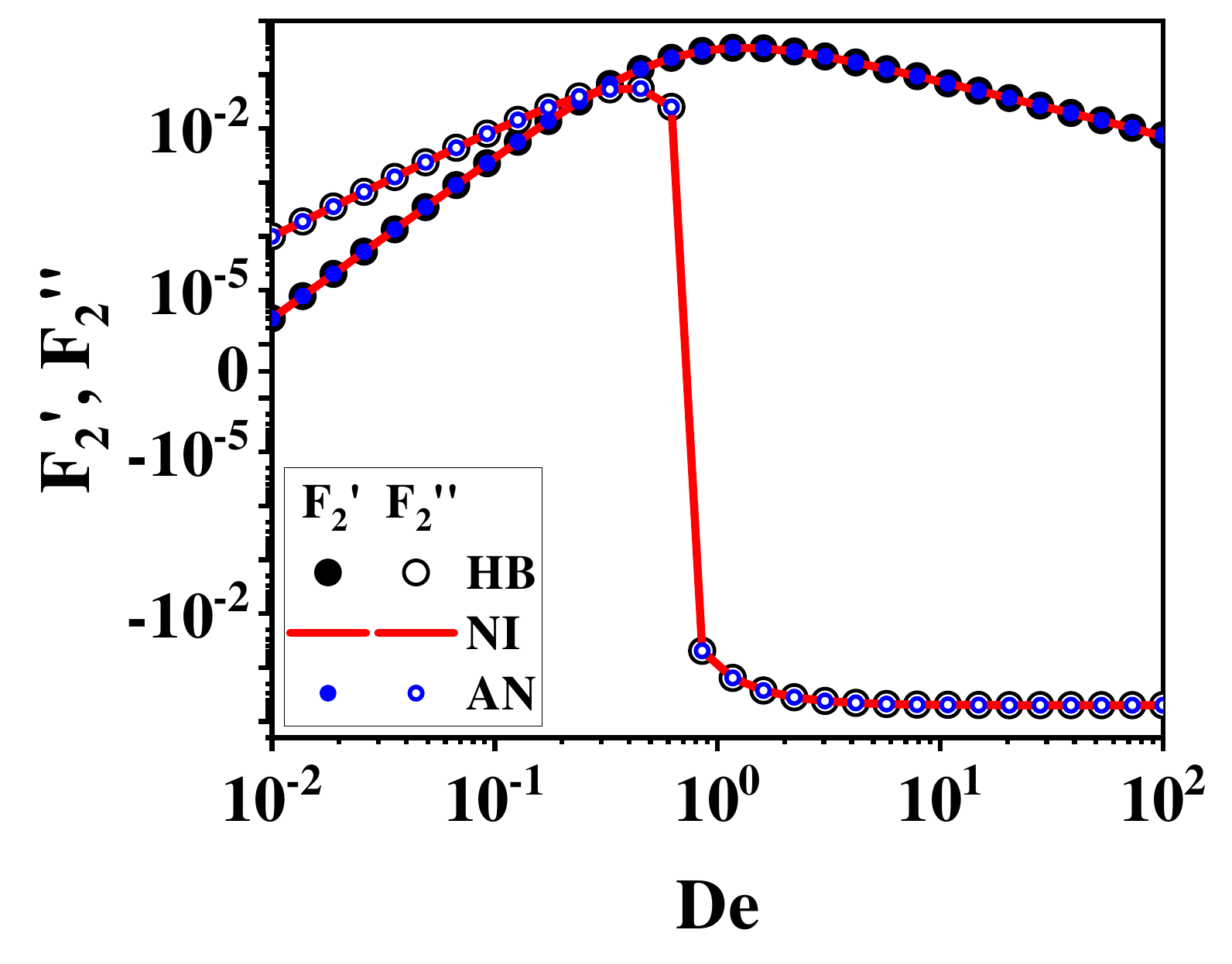}
        \\ (c) & (d)
    \end{tabular}
%    \end{center}
\caption{The (a) stress waveforms for the UCM model at $\gamma_0=10$ using FLASH and numerical integration (NI) overlap with the analytical solution (AN). Subfigures (b), (c), and (d) depict the Fourier coefficients corresponding to the shear stress and first normal stress difference.   \label{fig:UCM}}
\end{figure}

\medskip

For the UCM model $\sigma_{22} = \sigma_{33}=0$, $\sigma_{11} = N_1$, and,
\begin{equation}
\sigma_{12}(t)  = \gamma_0 \left(\Gp \sin \omega t + \Gpp \cos \omega t \right),
\label{eqn:ucm_s12}
\end{equation}
\begin{equation}
N_1(t) = \gamma_0^2 \left[\Gp + \left(-\Gp + \frac{1}{2} G^{\prime}(2 \omega) \right) \cos 2 \omega t + \left(\Gpp - \frac{1}{2} G^{\prime\prime}(2 \omega) \right) \sin 2 \omega t \right]
\label{eqn:ucm_N1}
\end{equation}
with $\Gp = G \omega^2 \lambda^2/(1 + \omega^2 \lambda^2)$ and $\Gpp = G \omega \lambda/(1 + \omega^2 \lambda^2)$ which is equivalent to the classic Maxwell solution.

\medskip

In figure \ref{fig:UCM}, we compare the analytical solution (AN) with the solutions obtained using FLASH and NI of the IVP. For FLASH, we used a single harmonic ($H = 1$) because the UCM model has a trivial LAOS response, as discussed in the Appendix. Thus, even at $\gamma_0 = 10$ the stress waveform, and moduli overlap as expected.

\clearpage

\section{Additional Results for the PTT model and the TNM}

\begin{figure}
    \centering
    \begin{tabular}{cc}
    {\large \bm{$\gamma_0=0.1$}} & {\large \bm{$\gamma_0=10$}} \\ 
    \includegraphics[scale=0.32]{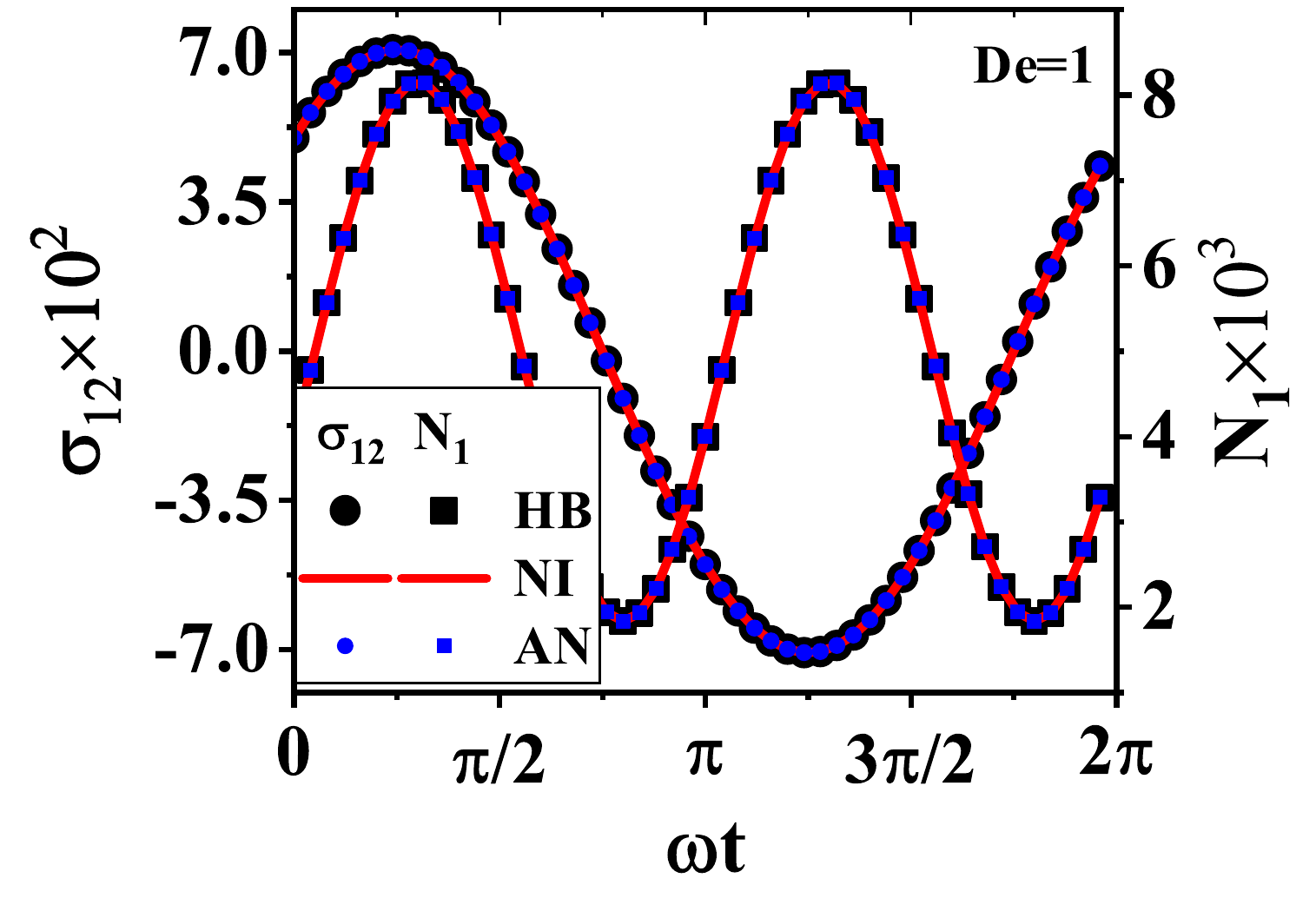} & \includegraphics[scale=0.28]{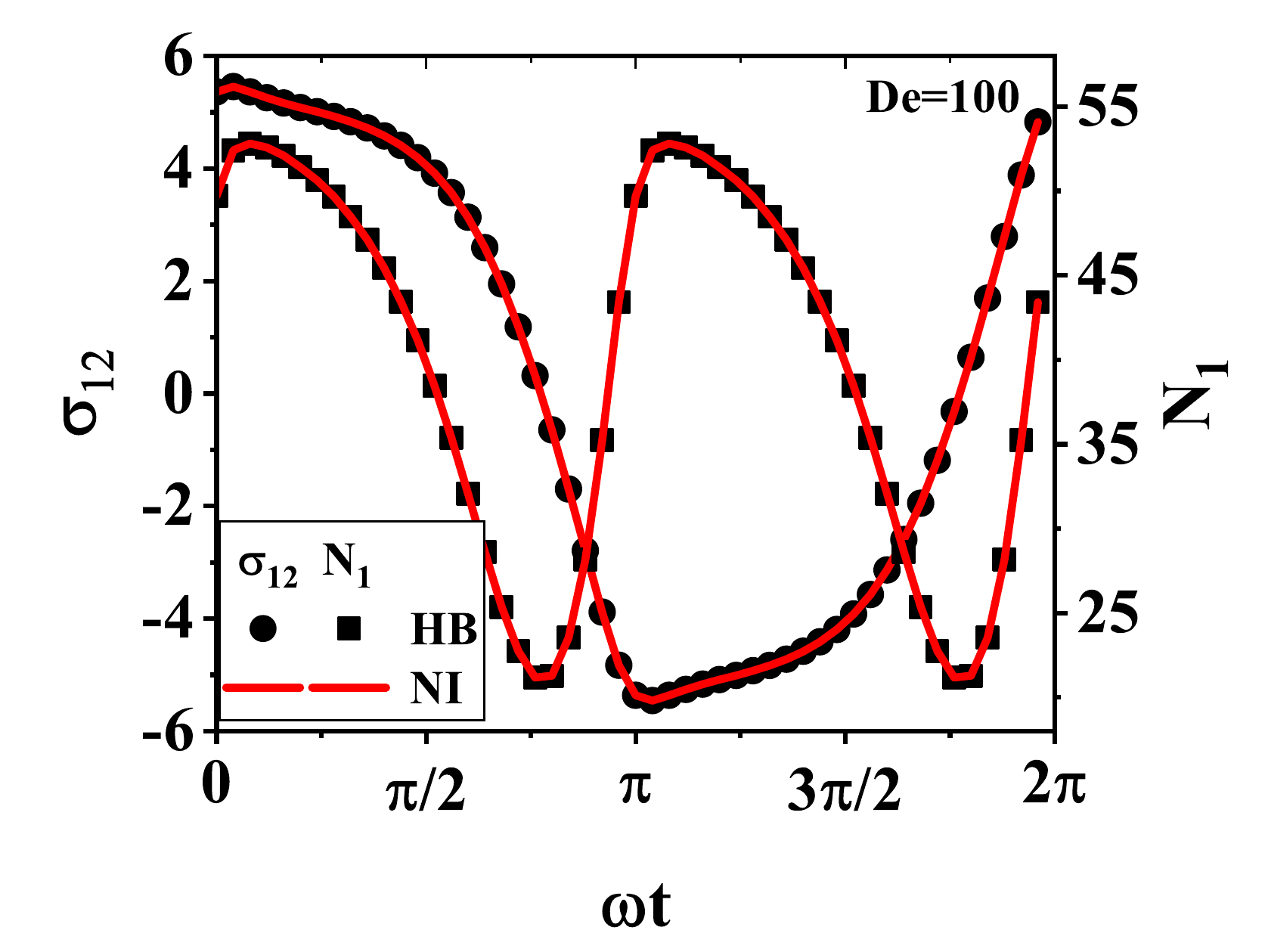} \\ (a) & (b) \\ \\ \includegraphics[scale=0.28]{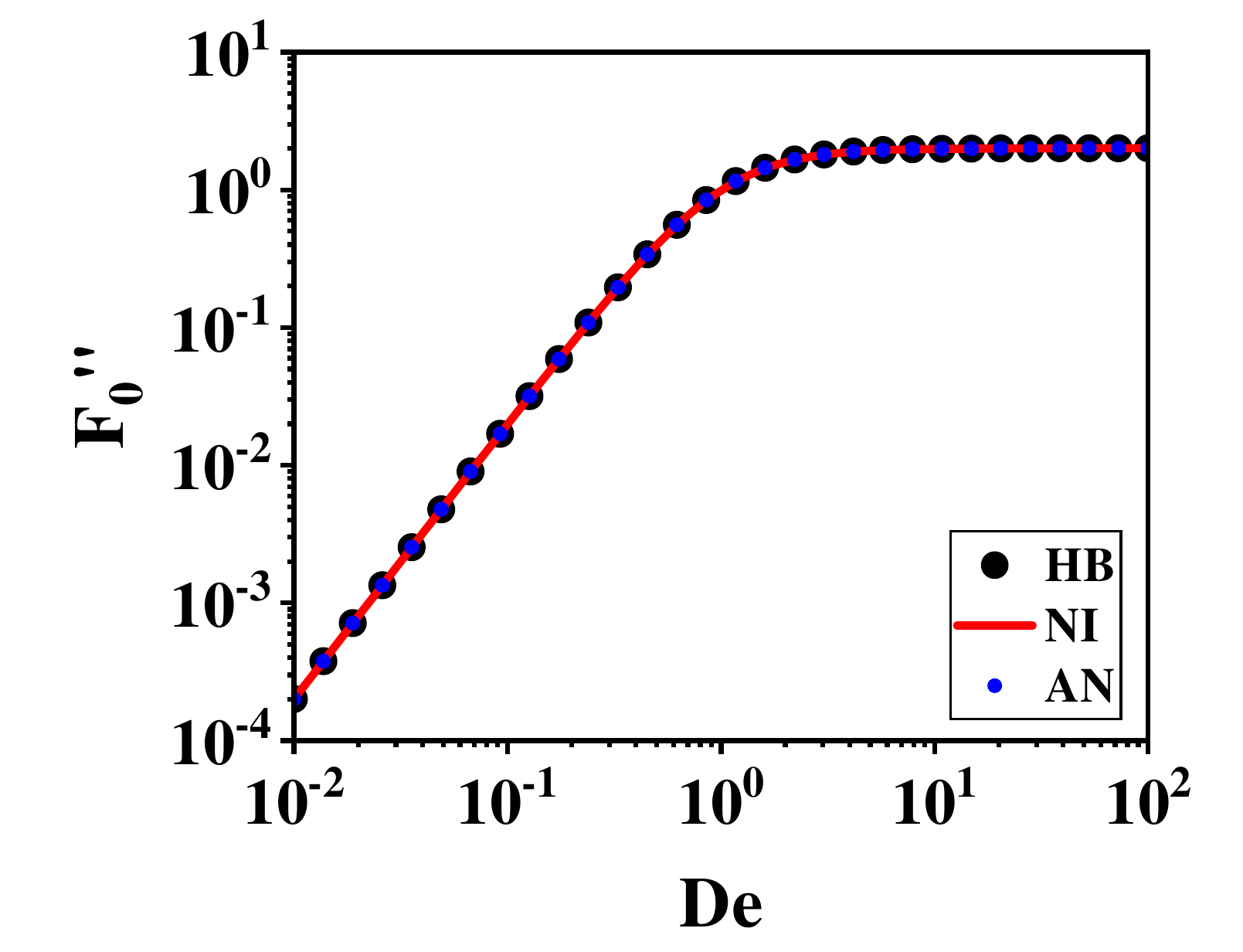} & \includegraphics[scale=0.28]{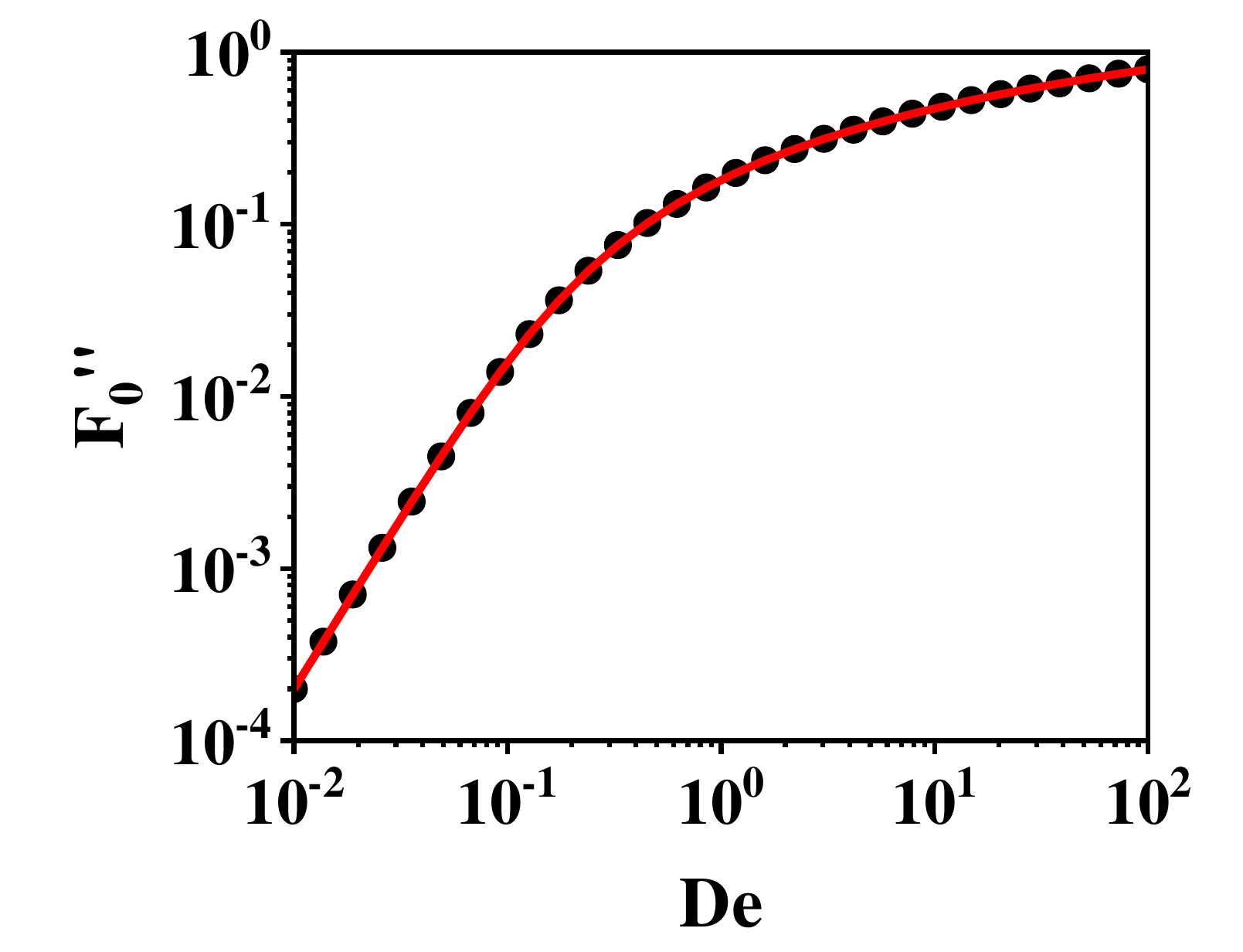}\\ (c) & (d)\\ \\ \includegraphics[scale=0.28]{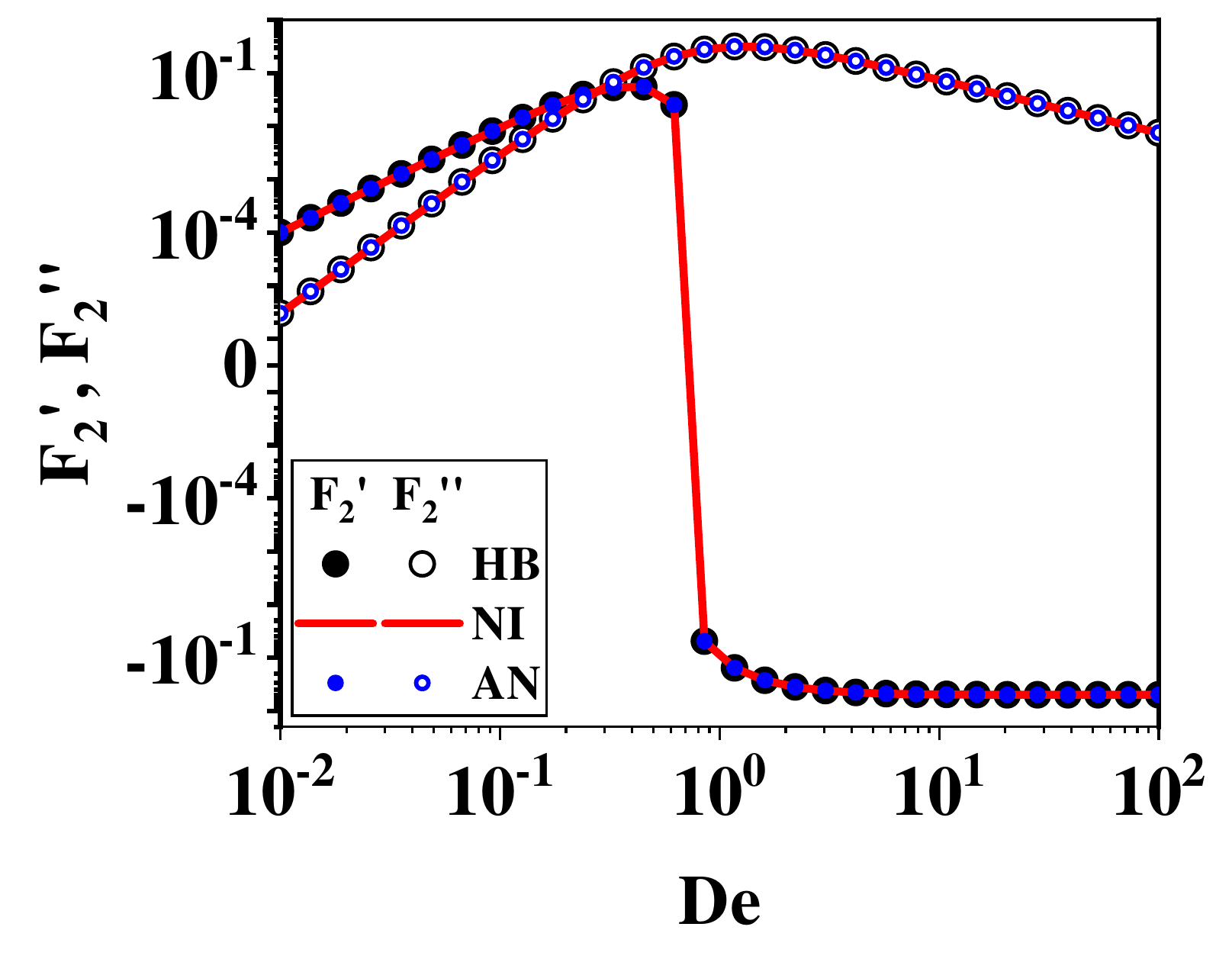} & \includegraphics[scale=0.28]{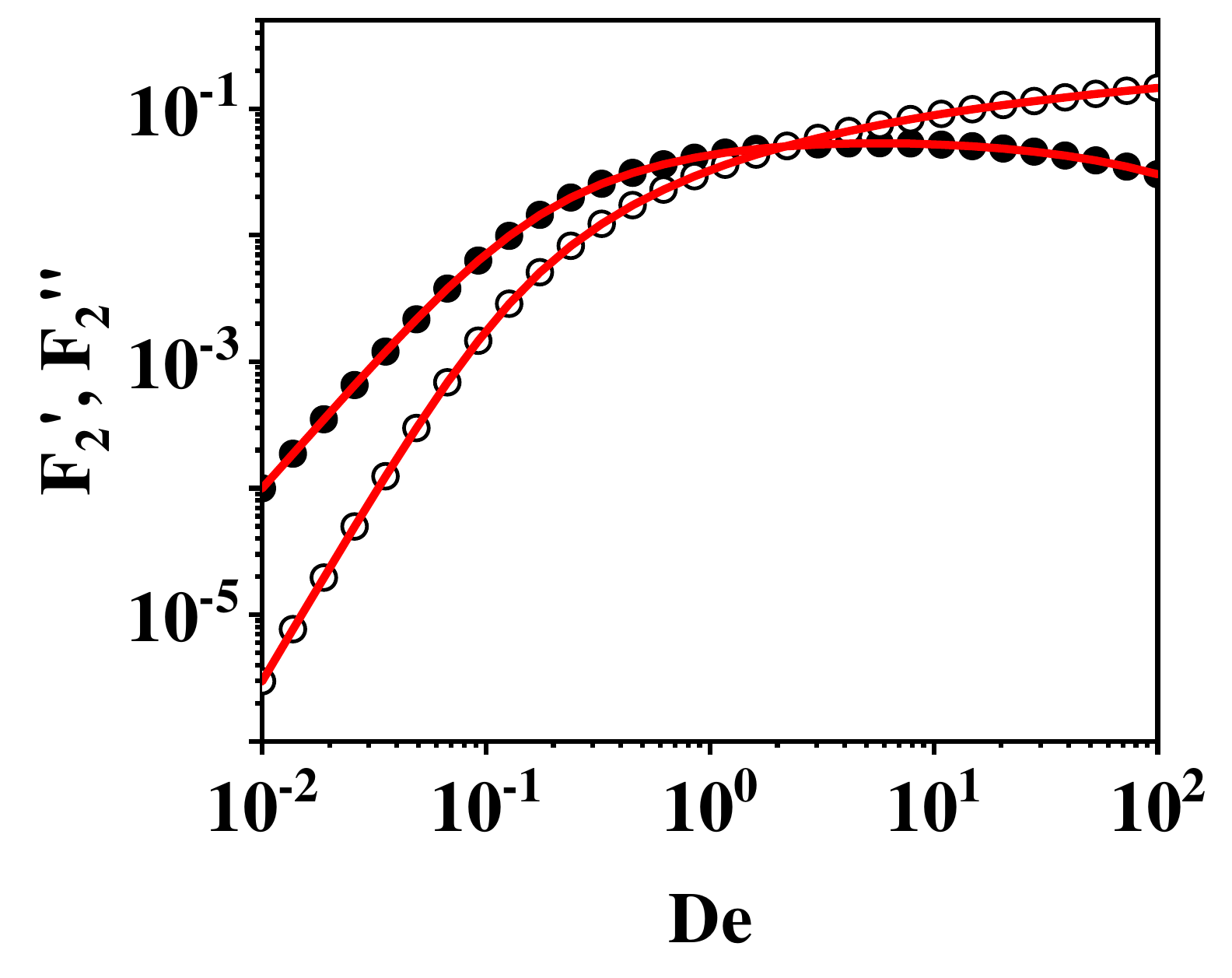} \\ 
    (e) & (f)
    \end{tabular}
    \caption{\textbf{PTT model}: A comparison for the (a),(b) stress waveforms, (c),(d) and (e),(f) and Fourier coefficients for normal stress difference at $\gamma_0=0.1,10$ obtained by HB, NI, and the MAOS analytical solution.}
    \label{fig:PTT_extra}
\end{figure}

In this section, we present some additional results for the LAOS behavior of the PTT model and the TNM.

\medskip

Figure \ref{fig:PTT_extra} shows the stress waveforms and leading Fourier coefficients of $N_1$ for the PTT model using $\varepsilon=0.1$ at two different strain amplitude $\gamma_0=0.1$ (left column) and $\gamma_0=10$ (right column), where the stress waveforms have been shown for specific frequencies $\De=1.0$ and $\De=100.0$, respectively.
%operating conditions: $(\gamma_0, \De) = (0.1, 1)$ (left column) and $(\gamma_0, \De) = (10, 100)$ (right column).

\medskip

Results obtained using FLASH are compared with the numerical solution of the IVP obtained by NI. The MAOS analytical solution is shown at the lower strain amplitude \cite{Song2020}. The curves for the first and third harmonics of the shear stress are provided in the main manuscript. Here, the first two harmonics corresponding to the first normal stress difference are presented for completeness. As expected, there is agreement between FLASH and NI.

\begin{figure}
    \centering
    \begin{tabular}{cc}
         \includegraphics[scale=0.28]{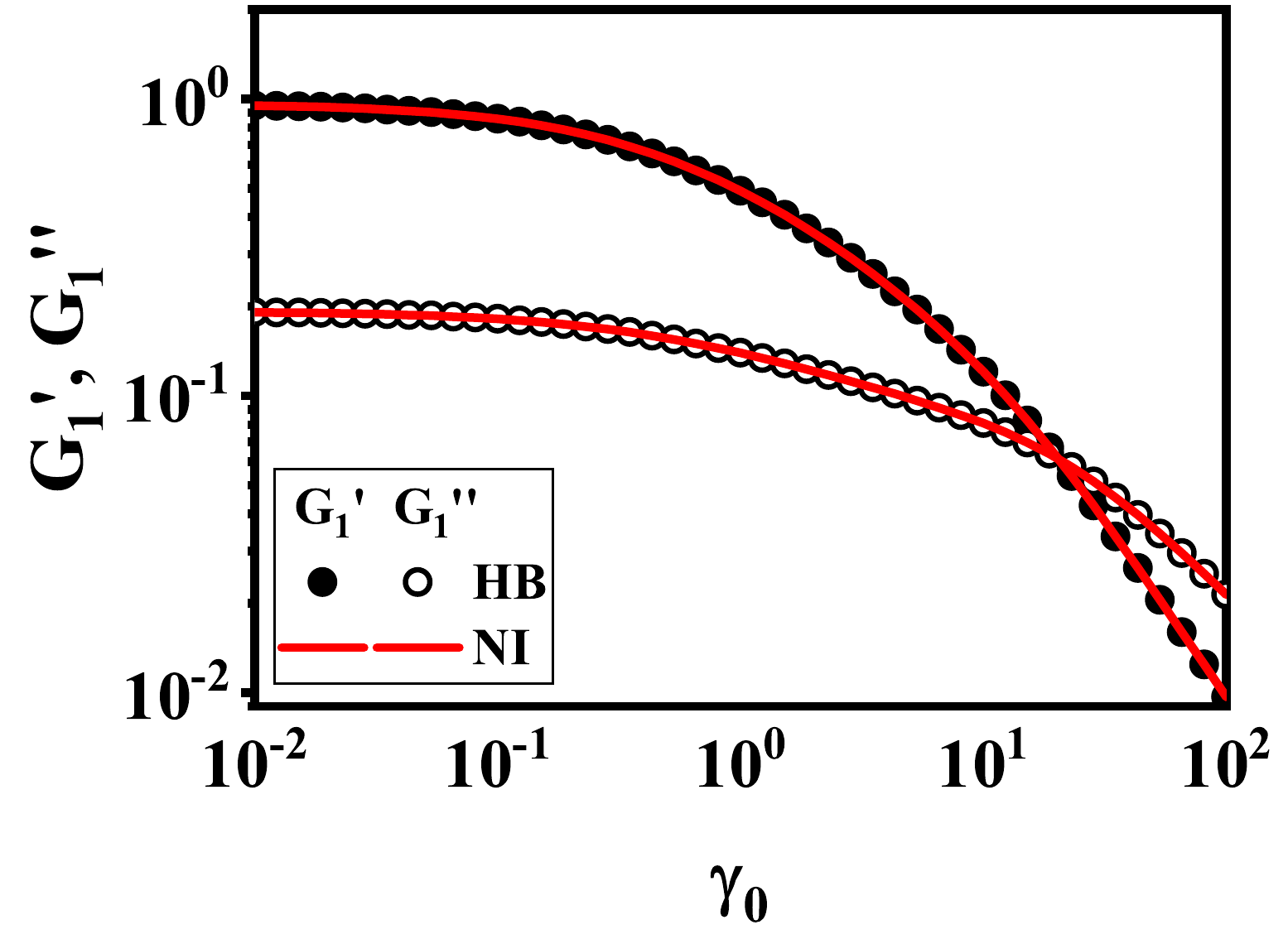}
         &
         \includegraphics[scale=0.28]{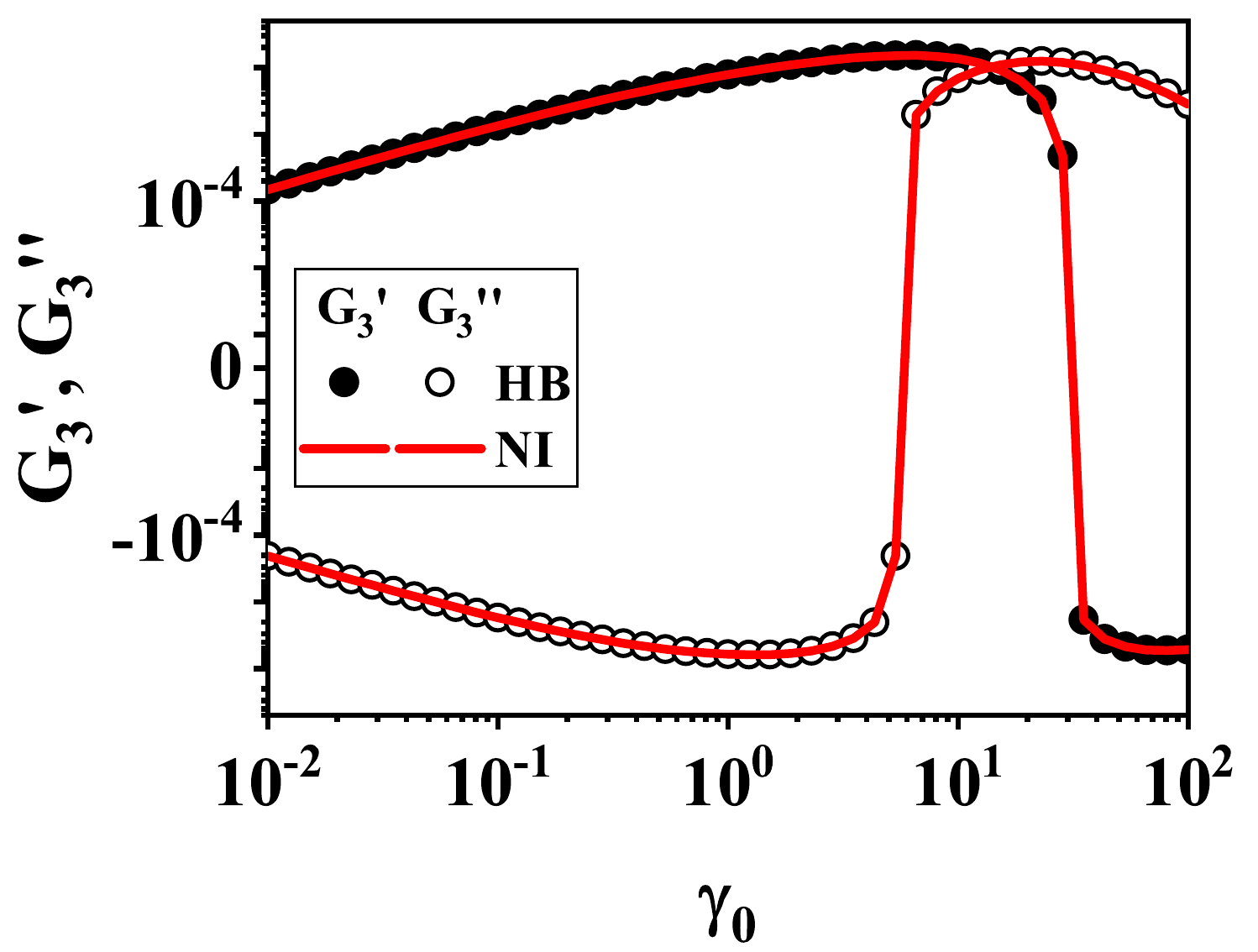}
         \\ (a) & (b) \\ \\
         \includegraphics[scale=0.28]{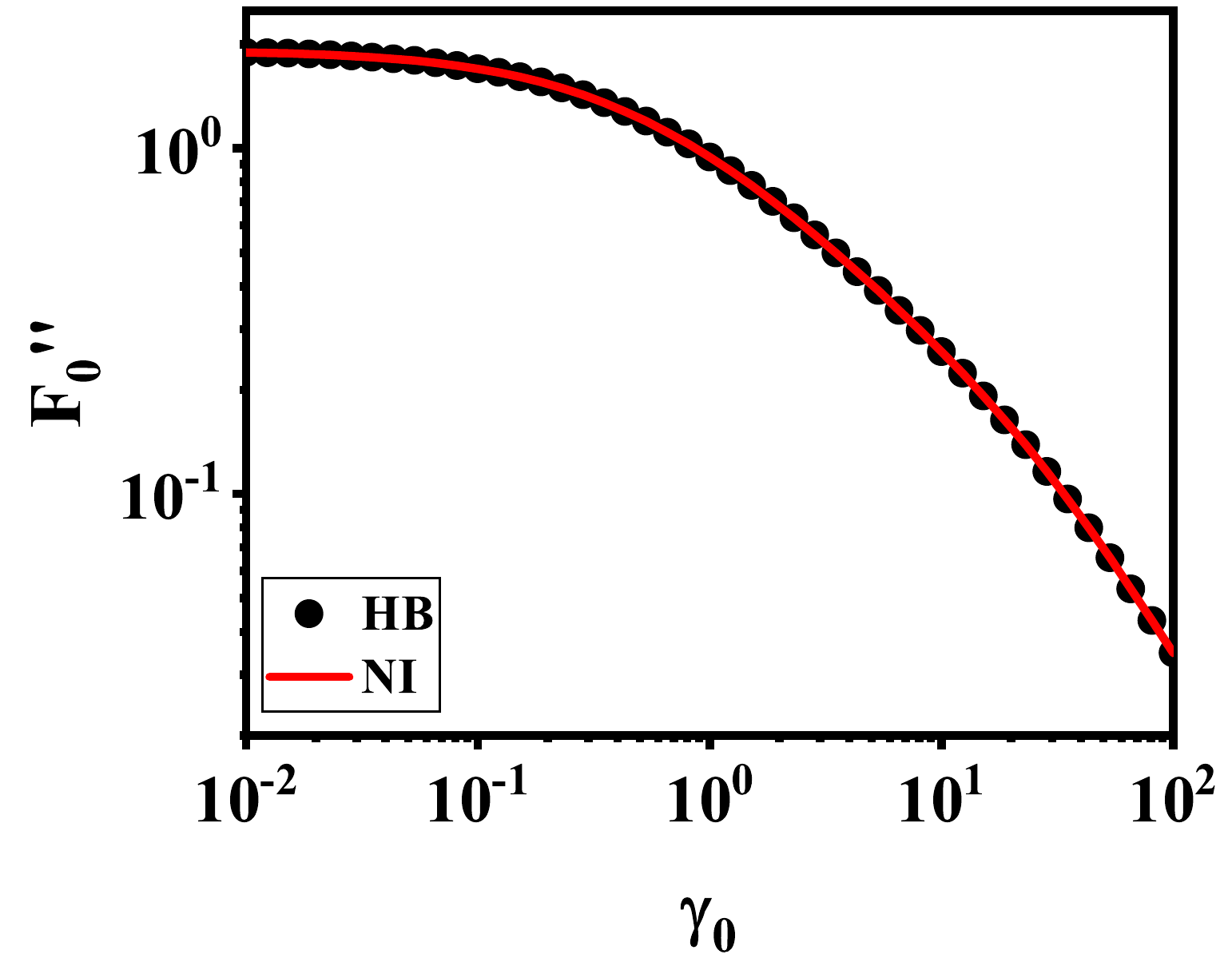}
         &
         \includegraphics[scale=0.28]{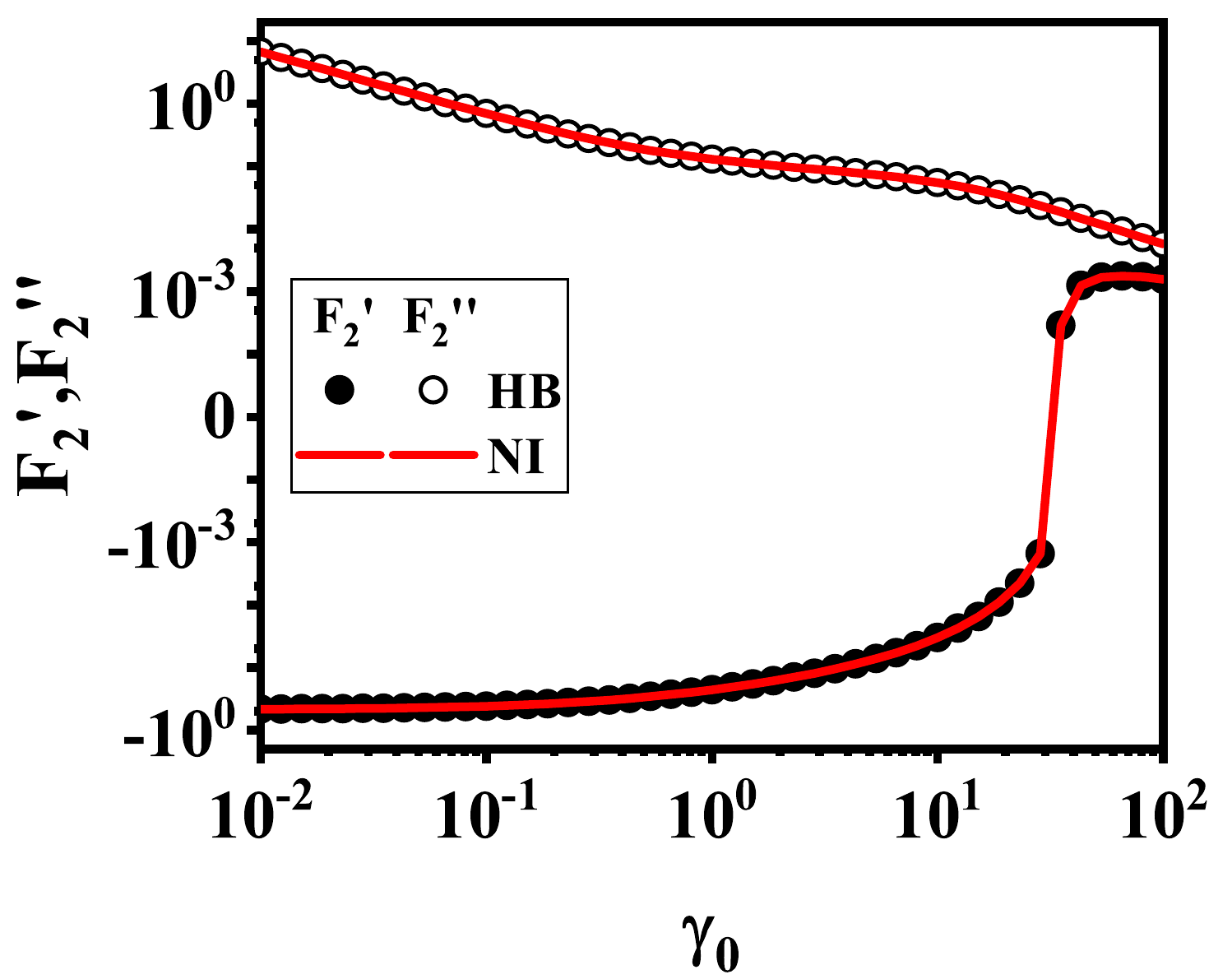} 
         \\ (c) & (d)
    \end{tabular}
    \caption{\textbf{Strain Softening}: The top (bottom) row shows the two leading moduli corresponding to the shear stress (first normal stress difference). Here, $a=-1.0$, $b = 1.0$, and $\De = 5$.}
    \label{fig:tnm_softening}
\end{figure}

\begin{figure}
    \centering
    \begin{tabular}{cc}
         \includegraphics[scale=0.28]{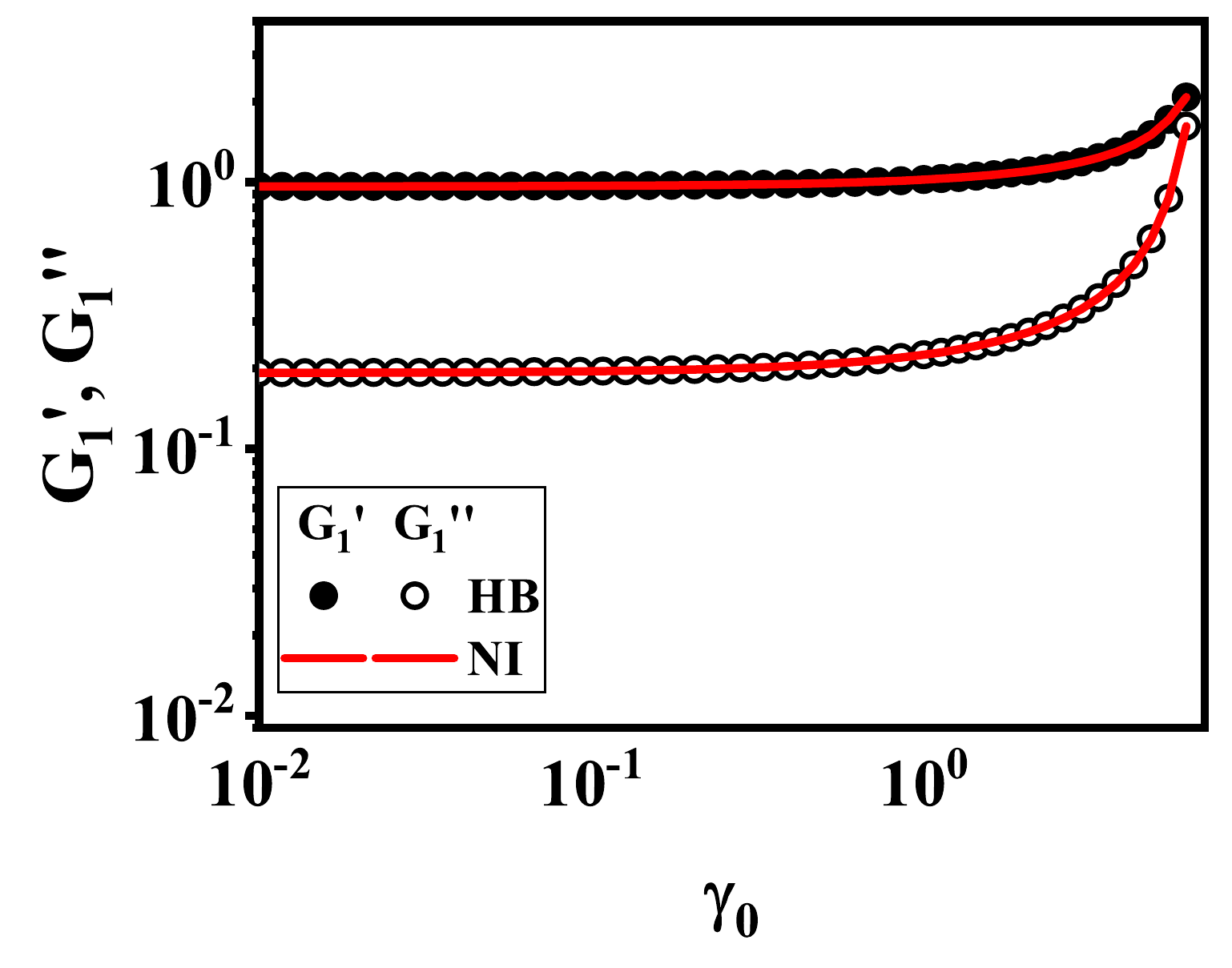}
         &
         \includegraphics[scale=0.28]{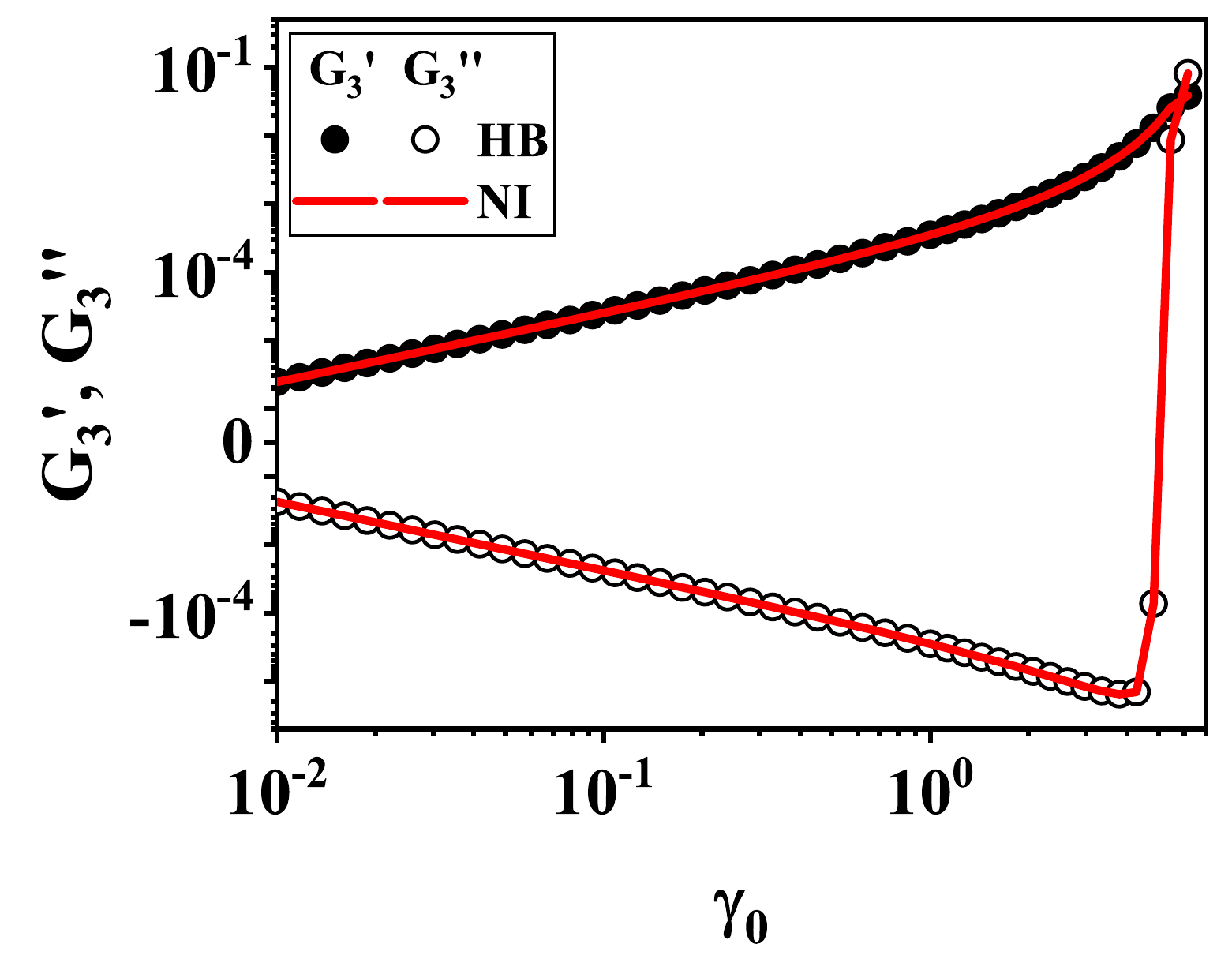}
         \\ (a) & (B) \\ \\
         \includegraphics[scale=0.28]{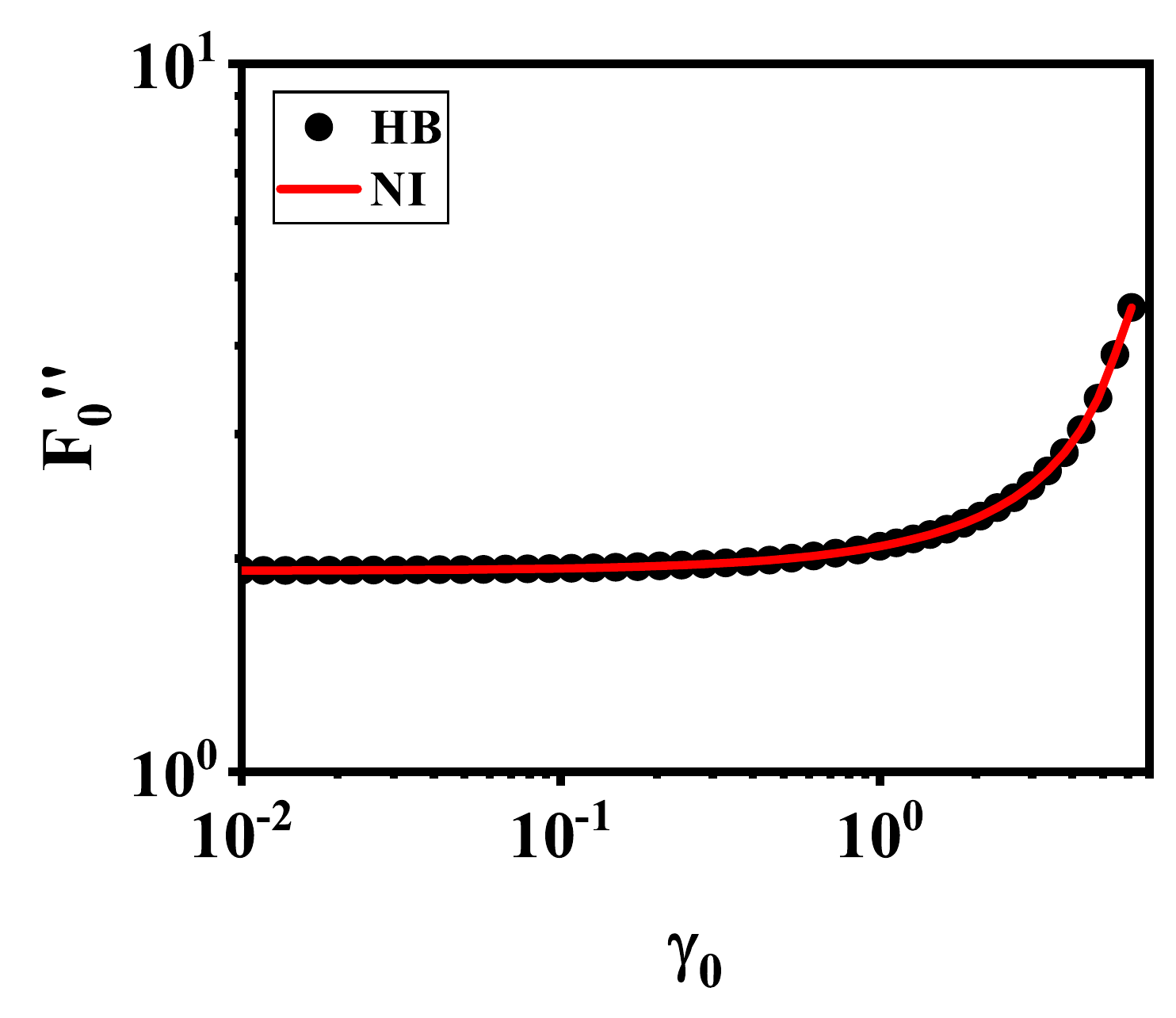}
         &
         \includegraphics[scale=0.28]{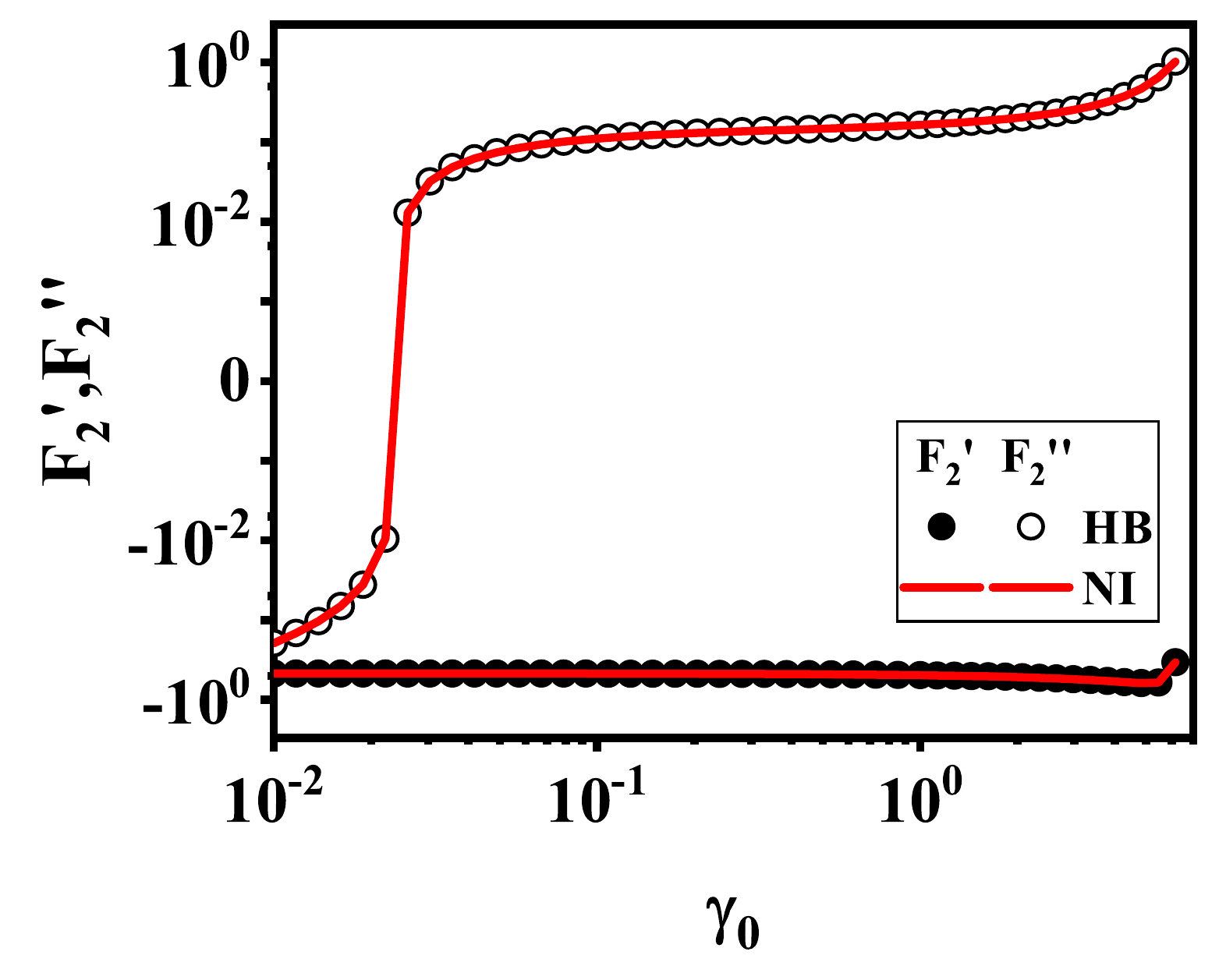} 
         \\ (c) & (d)
    \end{tabular}
    \caption{\textbf{Strain hardening}: The top (bottom) row shows the two leading moduli corresponding to the shear stress (first normal stress difference). Here, $a=0.2$, $b = 0.1$, and $\De = 5$.}
    \label{fig:tnm_hardening}
\end{figure}

\begin{figure}
    \centering
    \begin{tabular}{cc}
         \includegraphics[scale=0.28]{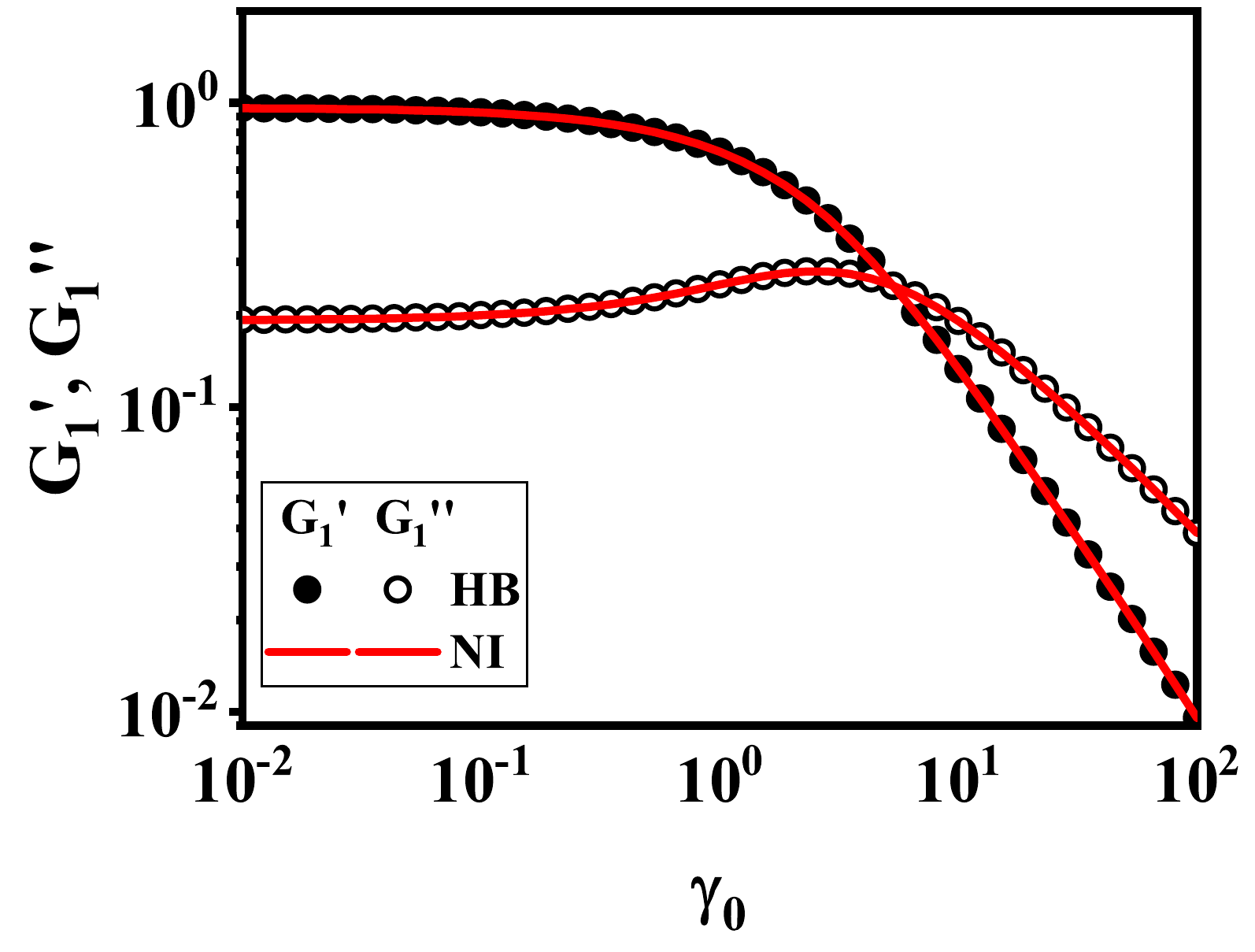}
         &
         \includegraphics[scale=0.28]{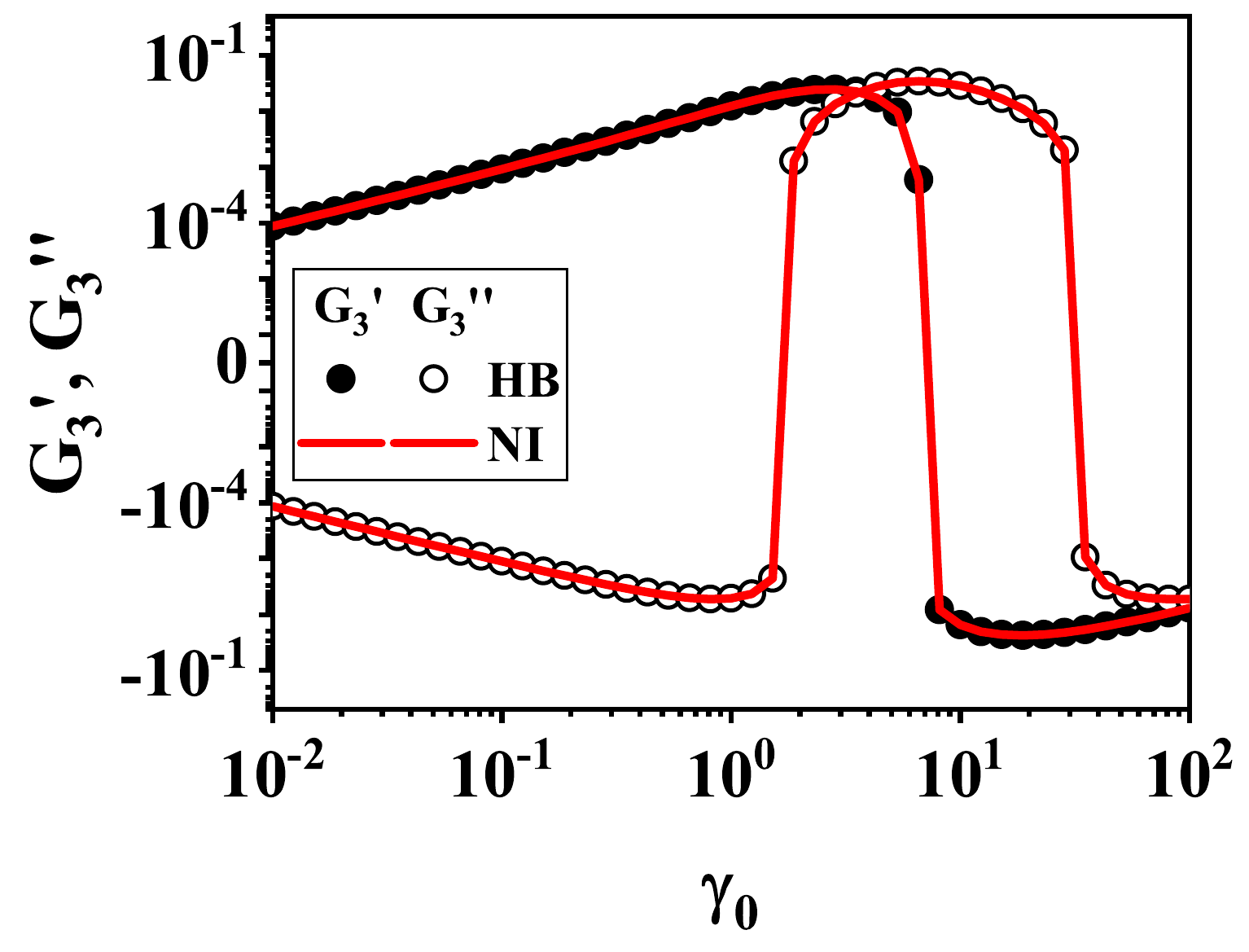}
         \\ (a) & (B) \\ \\
         \includegraphics[scale=0.28]{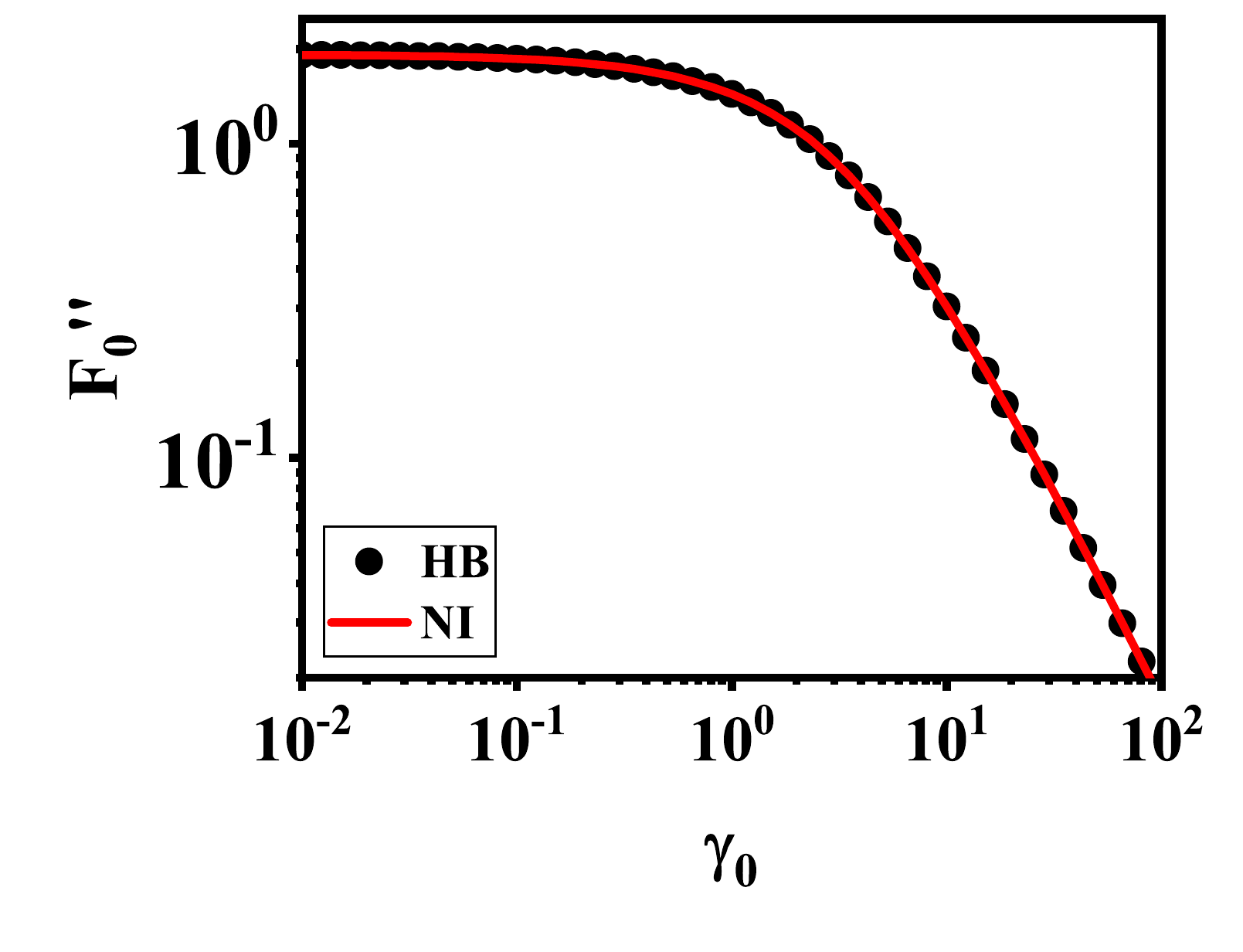}
         &
         \includegraphics[scale=0.28]{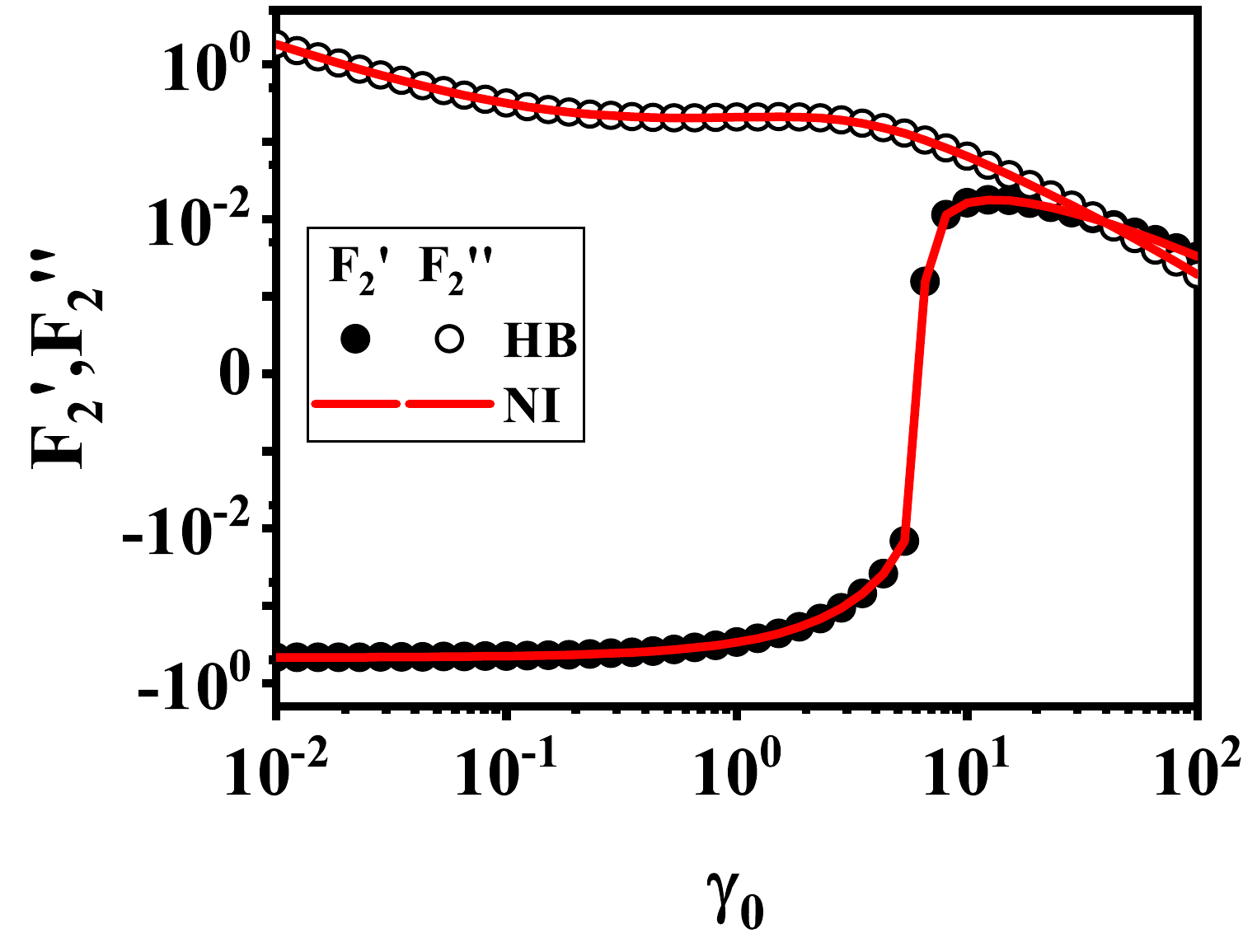} 
         \\ (c) & (d)
    \end{tabular}
    \caption{\textbf{Weak strain overshoot}: The top (bottom) row shows the two leading moduli corresponding to the shear stress (first normal stress difference). Here, $a=0.5$, $b = 1.0$, and $\De = 5$.}
    \label{fig:tnm_weak_overshoot}
\end{figure}

\begin{figure}
    \centering
    \begin{tabular}{cc}
         \includegraphics[scale=0.28]{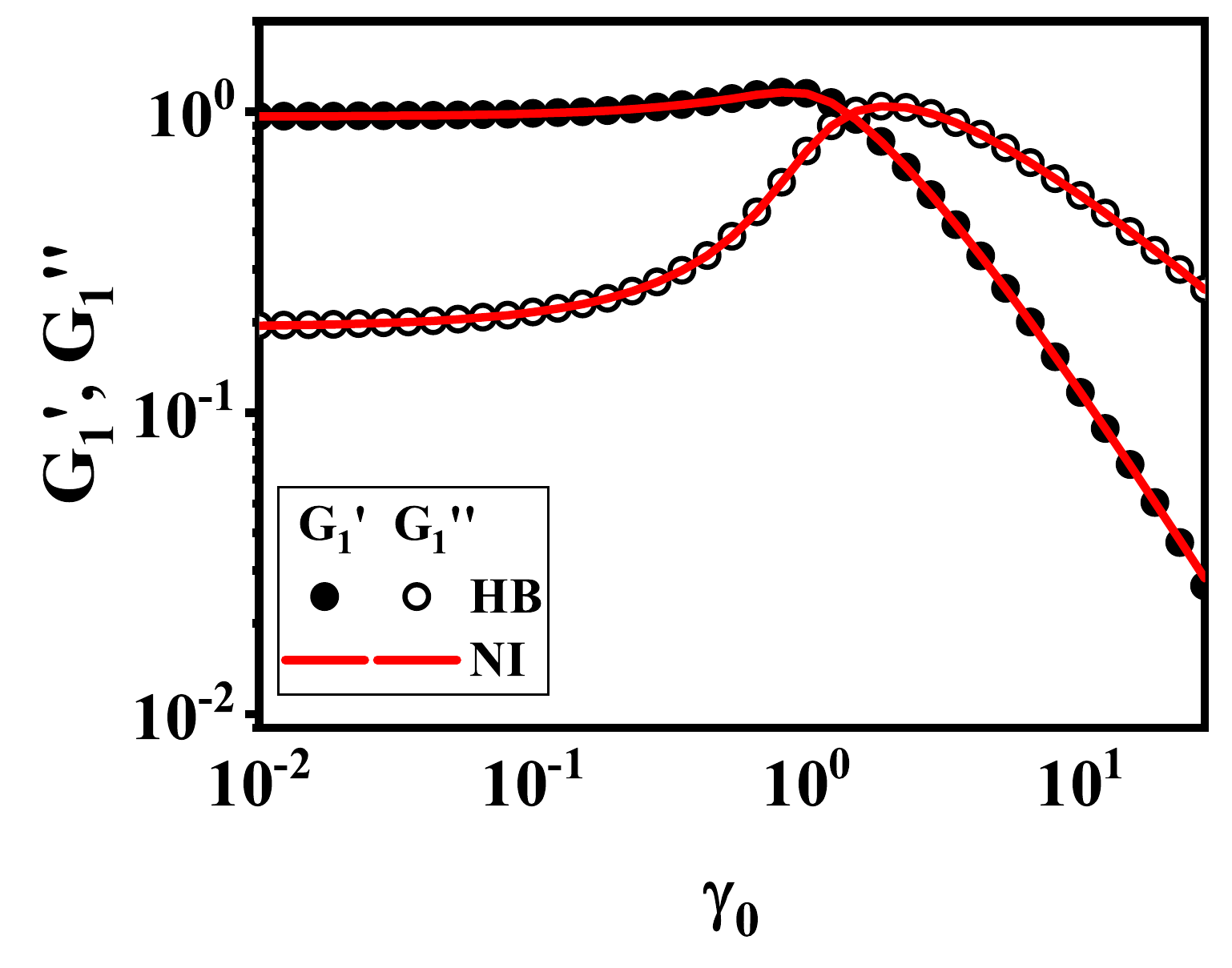}
         &
         \includegraphics[scale=0.28]{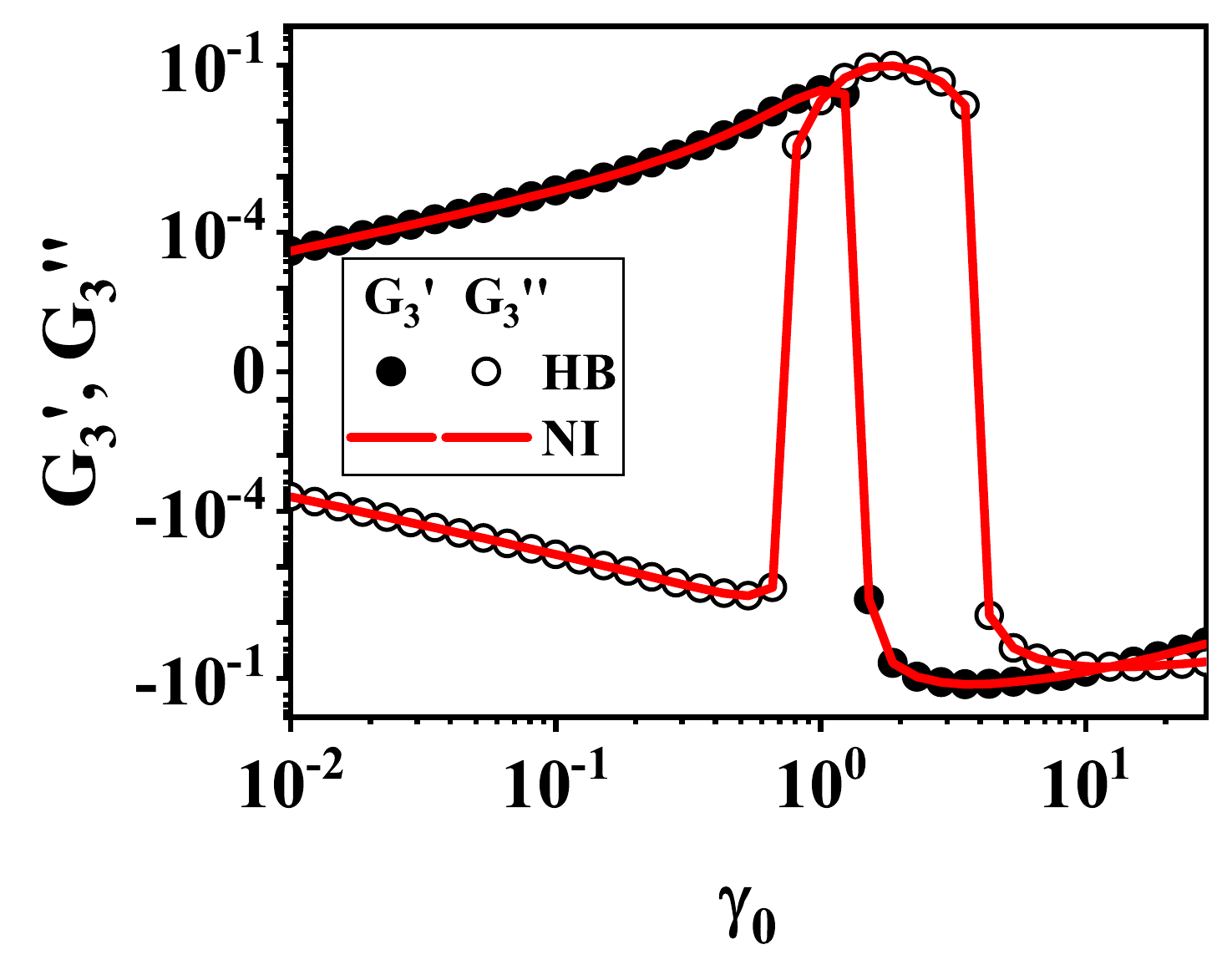}
         \\ (a) & (B) \\
         \includegraphics[scale=0.28]{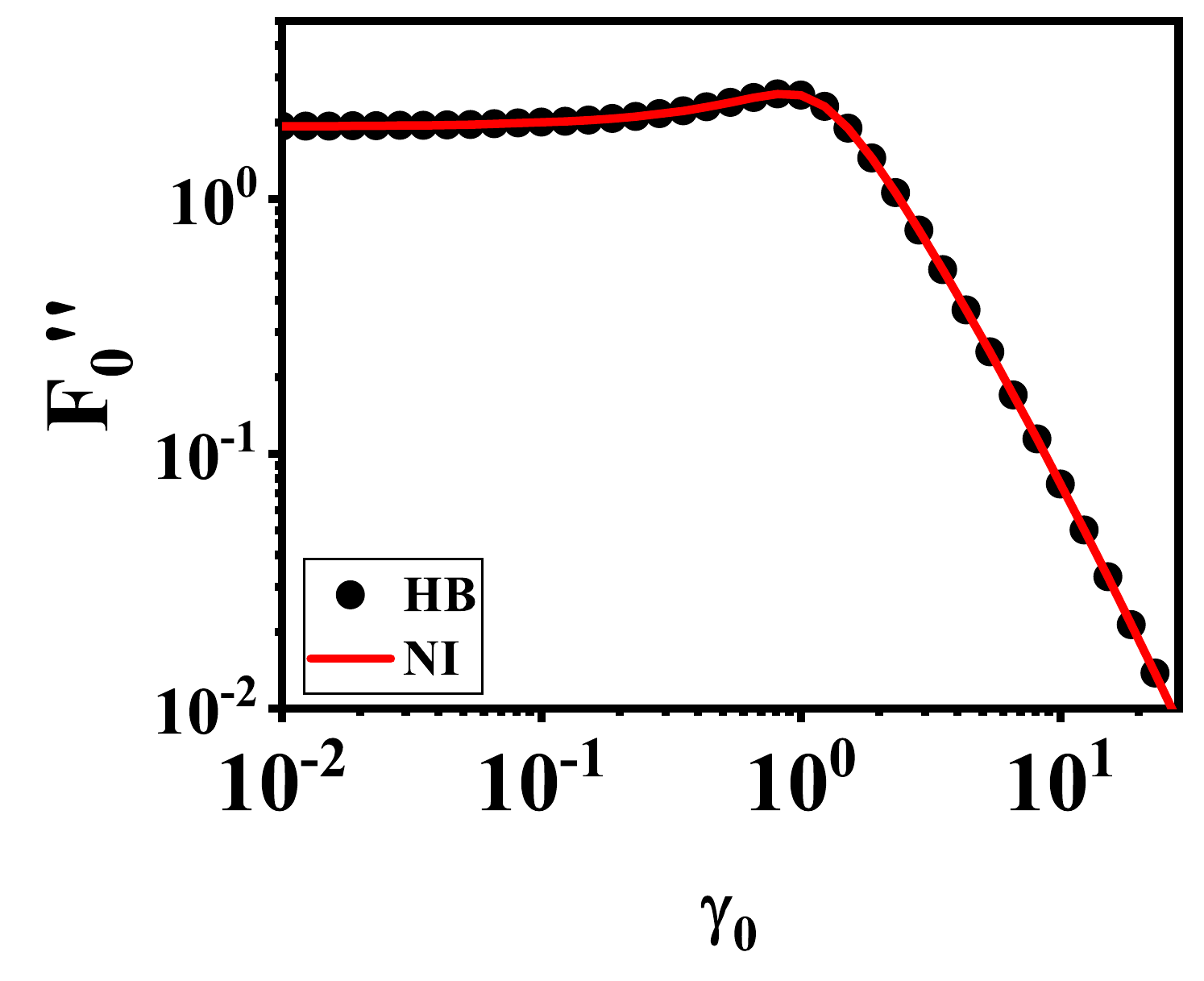}
         &
         \includegraphics[scale=0.28]{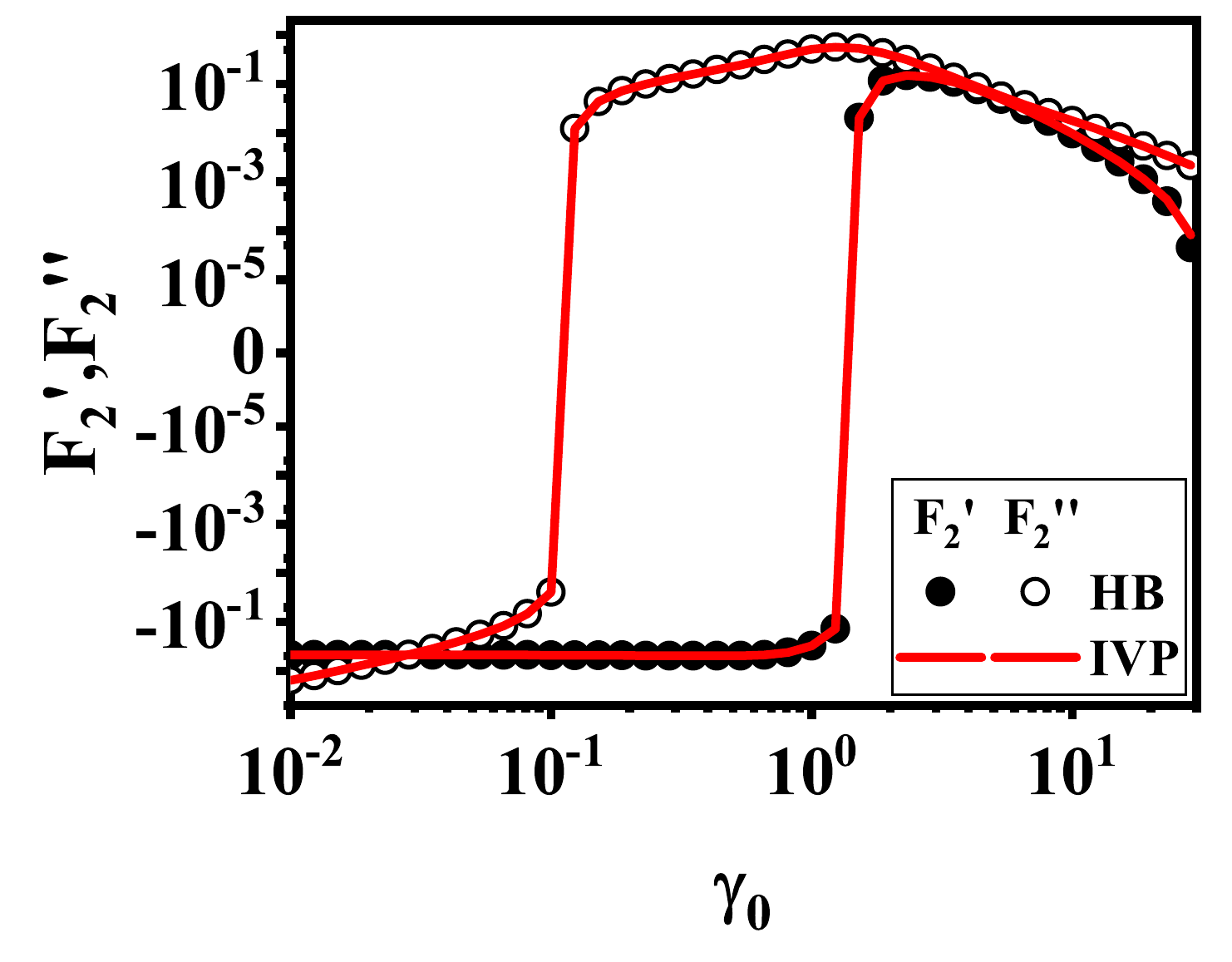} 
         \\ (c) & (d)
    \end{tabular}
    \caption{\textbf{Strong strain overshoot}: The top (bottom) row shows the two leading moduli corresponding to the shear stress (first normal stress difference). Here, $a=1.5$, $b = 1.0$, and $\De = 5$.}
    \label{fig:tnm_strong_overshoot}
\end{figure}

\medskip

Similarly, the following four figures present a comparison of the IVP solution and FLASH for the TNM for different parameter values corresponding to the four types of LAOS behavior.

\begin{table}[h!]
\begin{center}
\begin{tabular}{lrr}
\hline
type & $a$ & $b$\\ 
\hline
strain-softening & -1.0 & 1.0\\
strain hardening & -0.2 & 0.1\\
weak strain overshoot & 0.5 & 1.0\\
strong strain overshoot & 1.5 & 1.0\\
\hline
\end{tabular}
\end{center}
\caption{TNM parameter values corresponding to different types of LAOS behavior. \label{tab:laos_types}}
\end{table}

These parameter values are summarized in table \ref{tab:laos_types}.  Calculations are performed at $\De = 5$. Figs. \ref{fig:tnm_softening}, \ref{fig:tnm_hardening}, \ref{fig:tnm_weak_overshoot}, and \ref{fig:tnm_strong_overshoot} show the strain softening, strain hardening, weak strain overshoot, and strong strain overshoot behavior, respectively.

\medskip

These plots depict the two leading moduli corresponding to the shear stress and the first normal stress difference as a function of $\gamma_0$. They demonstrate the agreement between FLASH and NI.

\clearpage

\section{Nonlinear Constitutive Models}

As discussed in the ``Discussion'' section of the manuscript, we need to supply the $\bm{f}_{\nl}$ term corresponding to the CM in eqn. (16) to find the OS response of a nonlinear differential CM using FLASH. Here, we extend that discussion by presenting the $\bm{f}_{\nl}$ term for some popular CMs that are not special cases of the Oldroyd 8-constant model, or generalized Maxwell model discussed in the paper. Some CMs also require us to expand the definition of the unknown $\bm{q}(t)$; this is shown using a few illustrative examples.

\medskip

We describe the White-Metzner, Leonov, Larson, Pom-Pom, and the cDCR-CS models.\cite{Bird1987kt,Song2020} %The terms have their usual meaning, as described in the main manuscript.

\begin{enumerate}
\item \textbf{White-Metzner} 

The White-Metzner model is a modified version of the Upper Convected Maxwell model that not only incorporates the shear-rate dependence of viscosity similar to a generalized Newtonian fluid but also predicts a non-zero first normal stress. It is used to account for shear thinning behavior, which is a result of accelerated relaxation at high shear rates. \cite{Bird1987kt,larson2013constitutive}
    \begin{equation}
        \bm{\sigma} + \frac{\eta\left(\dot{\gamma}\right)}{G} \stackrel{\triangledown}{\bm{\sigma}}-  \eta\left(\dot{\gamma}\right)\bm{\gamma}_{(1)} = \bm{0}
    \end{equation}
   Here $\eta\left(\dot{\gamma}\right)$ is the shear rate dependent viscosity, while all other terms retain their usual meaning as given in the main text.
    The corresponding nonlinear term is
    \begin{equation}
        \bm{f}_\nl  = \left(\sigma_{11} \dfrac{G}{\eta \left( \dot{\gamma} \right)} - 2\dot{\gamma}\sigma_{12} \right)\textbf{ }\e_1 + 
        \sigma_{22} \dfrac{G}{\eta \left( \dot{\gamma} \right)}\text{ }\e_2 
        +\sigma_{33} \dfrac{G}{\eta \left( \dot{\gamma} \right)} \text{ }\e_3 + \left(\sigma_{12} \dfrac{G}{\eta \left( \dot{\gamma} \right)} - \dot{\gamma}\sigma_{22} \text{ }\right)\e_4.
    \end{equation}

\item \textbf{Leonov Model}

\medskip

The Leonov model is similar to the Giesekus model in its simplest form due to the presence of terms quadratic in stress. This model has its roots in thermodynamics rather than molecular modeling. Leonov's equation relates the shear stress to the elastic strain under free recovery, i.e. the strain recovered when all loads (both shear and normal) are retracted. \cite{larson2013constitutive}

Hereon, all following models are represented in terms of the conformation tensor\cite{Song2020} 
    \begin{equation}
        \bm{A}=\left[\begin{matrix}
            A_{11} & A_{12} & 0 \\
            A_{12} & A_{22} & 0 \\
            0 & 0 & A_{33}
        \end{matrix} \right],
        \label{eqn:conformation_tensor}
    \end{equation}
    where the extra stress tensor is related to $\bm{A}$ by $\bm{\sigma}=G(\bm{A}-\bm{I})$. 
    
    The Leonov model is given by
    \begin{equation}
        \stackrel{\triangledown}{\bm{A}} +\frac{1}{2\lambda}\phi(\text{tr }\bm{A})\left[\bm{A}^2-\bm{I} - k(\bm{A})\bm{A}\right] = \bm{0},         
    \end{equation}
     where $\phi(\text{tr }\bm{A})$ is responsible for the irreversible dissipation in the model. Several forms have been suggested for $\phi$ and accordingly, it may contain additional model parameters. \cite{leonov1999constitutive,Leonov1992,larson1983elongational,simhambhatla1995rheological,zatloukal2003differential} 
     
     Furthermore, 
     \begin{align}
         k(\bm{A}) &=\frac{1}{3}\left(\text{tr }\bm{A}+\text{tr }\bm{A}^{-1}\right), \nonumber
         \\
         &=\frac{1}{3} \left[A_{11}+A_{22}+A_{33}-\frac{1}{A_{33}} - \frac{A_{11}+A_{22}}{A_{11}A_{22}-A_{12}^2}\right].
         \label{eqn:leonov_k}
     \end{align}
     
     To use FLASH, the stress components in eqn. (16) are replaced with,
     \begin{equation}
         \bm{q} = \left[\begin{matrix}
             A_{11} & A_{22} & A_{33} & A_{12}
         \end{matrix}\right],
         \label{q_A}
     \end{equation}
     \begin{align}
         \bm{f}_\nl &= 
         \left[\frac{\phi}{2\lambda}\left(A_{11}^2 + A_{12}^2 -1 -kA_{11}\right) - 2\dot{\gamma}A_{12} \right] \e_1 +
         \frac{\phi}{2\lambda}\left[A_{22}^2 + A_{12}^2 -1 -kA_{22}\right]\e_2 
         \nonumber \\ 
         & + \frac{\phi}{2\lambda}\left[A_{33}^2 -1 -kA_{33}\right] \e_3 +
         \left[\frac{\phi}{2\lambda}A_{12}\left(A_{11} + A_{22}-k\right) - \dot{\gamma}A_{22} \right] \e_4,
     \end{align}
     \begin{equation}
         \bm{f}_\text{ex} = \left[\begin{matrix}
             0 & 0 & 0 & 0
         \end{matrix}\right],
         \label{fex_A}
     \end{equation}
     where $k(\bm{A})$ is given  by eqn. \eqref{eqn:leonov_k}. For this definition of $\bm{q}$, $\bm{f}_\text{ex}$ is identically zero. The effect of the applied oscillatory deformation is captured by the $\dot{\gamma}$ terms in $\bm{f}_\nl$.

\item \textbf{Larson model}

The Larson model was introduced to explain the deviation of the damping function from the DE-IAA model for commercial polymers. Unlike the De-IAA that allows rapid retraction of strand inside the deformed tube, the Larson model accounted for the suppressed retraction in the backbone due to long-chain branching, which leads to partial retraction when the reptation starts. \cite{larson1984constitutive} 

The model is expressed in terms of the conformation tensor, 
    \begin{equation}
        \stackrel{\triangledown}{\bm{A}} + \frac{2\xi}{3}\left(\dot{\bm{\gamma}}:\bm{A} \text{ }\bm{A}\right) + \frac{1}{\lambda} \left(\bm{A}-\bm{I}\right) = \bm{0},
    \end{equation}
where $\dot{\bm{\gamma}}= \dot{\gamma}\e_1 \e_2 + \dot{\gamma}\e_2 \e_2$ is the rate of deformation tensor, $\xi$ is a nonlinear parameter $(0<\xi<1)$ that measures the extent of retraction, and $\lambda$ is the relaxation time. The Larson model reduces to the UCM form for $\xi=0$ i.e., no retraction. On the other hand, $\xi=1$ indicates complete retraction.
    
    To apply HB, we use eqn. \eqref{q_A} for $\bm{q}$ and  $\bm{f}_{\text{ex}}=\bm{0}$. Then, $\bm{f}_\nl$ is given by, 
    \begin{align}
        \bm{f}_\nl &=\left[\frac{A_{11}-1}{\lambda} + \frac{4}{3}\xi\dot{\gamma}A_{12}A_{11} - 2\dot{\gamma}A_{12} \right]\e_1 
        +
        \left[\frac{A_{22}-1}{\lambda} + \frac{4}{3}\xi\dot{\gamma}A_{12}A_{22}  \right]\e_2 
        \nonumber \\ 
        & + \left[\frac{A_{33}-1}{\lambda} + \frac{4}{3}\xi\dot{\gamma}A_{12}A_{33}  \right]\e_3 
        +
        \left[\frac{A_{12}}{\lambda} + \frac{4}{3}\xi\dot{\gamma}A_{12}^2 - \dot{\gamma}A_{22} \right]\e_4 
    \end{align}

\item \textbf{Pom-Pom Model}

\medskip

The Pom-Pom model is used to capture the nonlinear rheology of branched polymers. It was originally proposed by McLeish and Larson \cite{mcleish1998molecular} for an idealized polymer represented by a ``Pom-Pom'' molecule consisting of an equal number of arms attached to both ends of a linear backbone. The model has undergone several modifications over the years. \cite{inkson1999predicting,blackwell2000molecular} The form considered here does not account for the backbone relaxation mechanism that may be significant for rapidly reversing flows. \cite{lee2001experimental}

The model is given in terms of the conformation tensor as
    \begin{equation}
        \stackrel{\triangledown}{\bm{A}}+\frac{1}{\lambda}(\bm{A}-\bm{I})=\bm{0} 
    \end{equation}
    which is identical to the UCM form. However, it is coupled to backbone stretch $L(t)$, which evolves with time as
    \begin{equation}
        \frac{\partial}{\partial t}L = L(\bm{\kappa}:\bm{S})-\frac{1}{\lambda_s}(L-1)\exp{\left(\frac{2}{p}(L-1)\right
        )},
    \end{equation}
    where $\bm{\kappa}$ is the velocity gradient tensor, $\lambda_s$ is the stretch relaxation time, $p$ is the number of dangling arms. The orientation tensor is given by
    \begin{equation}
        \bm{S}=\frac{\bm{A}(t)}{\text{tr }\bm{A}(t)}
    \end{equation}

    Thus, the terms in eqn. (16) are as follows:
    \begin{equation}
        \bm{q}= \left[\begin{matrix}
            A_{11} & A_{22} & A_{33} & A_{12} & L
        \end{matrix} \right]^{T}, 
        \label{q_pp}
    \end{equation}
    where we introduce a new variable to the list of unknowns. The external forcing vector is zero, and the nonlinear term is,
    \begin{align}
        \bm{f}_{\nl} & = 
            \left(\frac{1}{\lambda}\left(A_{11}-1\right) - 2\dot{\gamma}A_{12} \right) \text{ }\e_1 +
            \frac{1}{\lambda}\left(A_{22}-1\right) \text{ }\e_2 + 
            \frac{1}{\lambda}\left(A_{33}-1\right)\text{ }\e_3 
            \nonumber \\  
            & + \left(\frac{1}{\lambda}A_{12} - \dot{\gamma}\sigma_{22} \right) \text{ }\e_4 +
            \left( L\dot{\gamma}\frac{A_{12}}{A_{11}+A_{22}+A_{33}} - \frac{1}{\lambda_{s}}(L-1)\exp{\frac{2}{p}(L-1)} \right) \text{ }\e_5.
        \label{fnl_pp}
    \end{align}

After solving the HB equation set, the extra stress tensor for the Pom-Pom model is given by, \cite{Song2020} 
\begin{equation}
    \bm{\sigma} = 3G_0 L^2 \bm{S}.
\end{equation}

\item \textbf{cDCR-CS}: 

\medskip

The coupled double-convection-reptation with chain stretch (cDCR-CS) model was proposed by Marrucci and Ianniruberto \cite{marrucci2003flow} for entangled linear chains. The model accounts for reptation, contour length fluctuation, constraint release, and chain stretch. The authors proposed two variations of the model to consider stretching Gaussian and non-Gaussian chains.

For Gaussian chains, the model is given by, 
    \begin{align}
         \stackrel{\triangledown}{\bm{A}} & + \frac{1}{\lambda}\frac{1}{\left( 1+\beta_c \left(\text{tr }\bm{A} -1\right)\right)}\left(\bm{A}- \frac{1}{3}\bm{I} \text{tr }\bm{A} \right) + \frac{1}{3\lambda_R}\left(\text{tr }\bm{A} - 1\right)\bm{I} \\
         &+ \frac{\beta_c \left(\text{tr }\bm{A} -1\right)}{\lambda_R \left(1+\beta_c \left(\text{tr }\bm{A} -1\right)\right)}\left(\bm{A}-\frac{1}{3}\bm{I}\text{tr }\bm{A}\right) = \bm{0}, 
         \label{eqn:cm_cdcrcs}
    \end{align}
    where $\lambda_R$ is the Rouse relaxation time, and $\beta_c \geq 0$ sets the CCR strength. 

    To apply the HB formalism, we use $\bm{q}$ and $\bm{f}_\text{ex}$ given in eqns. \eqref{q_A} and \eqref{fex_A} respectively. The nonlinear terms, on the other hand, are given as follows, 
    \allowdisplaybreaks
    \begin{align}
        \bm{f}_\nl & = \left[ \begin{array}{r}
            \dfrac{2A_{11} + A_{22} +A_{33}}{3\left( 1+\beta_c \left(A_{11}+A_{22}+A_{33} -1\right)\right)}
         \left(\dfrac{1}{\lambda} + \dfrac{\beta_c \left(A_{11}+A_{22}+A_{33} -1\right)}{\lambda_R} \right) \\
         + \dfrac{1}{3 \lambda_R} \left(A_{11}+A_{22}+A_{33}-1 \right) - 2 \dot{\gamma}A_{12} 
        \end{array} \right] \e_1 
        \nonumber\\
        &+ \left[ \begin{array}{r}
            \dfrac{A_{11} + 2A_{22} +A_{33}}{3\left( 1+\beta_c \left(A_{11}+A_{22}+A_{33} -1\right)\right)}
         \left(\dfrac{1}{\lambda} + \dfrac{\beta_c \left(A_{11}+A_{22}+A_{33} -1\right)}{\lambda_R} \right) \\
         + \dfrac{1}{3 \lambda_R} \left(A_{11}+A_{22}+A_{33}-1 \right)  
        \end{array} \right] \e_2
        \nonumber\\
        &+ \left[ \begin{array}{r}
        \dfrac{A_{11} + A_{22} + 2A_{33}}{3\left( 1+\beta_c \left(A_{11}+A_{22}+A_{33} -1\right)\right)}
         \left(\dfrac{1}{\lambda} + \dfrac{\beta_c \left(A_{11}+A_{22}+A_{33} -1\right)}{\lambda_R} \right) \\
         + \dfrac{1}{3 \lambda_R} \left(A_{11}+A_{22}+A_{33}-1 \right)  
        \end{array} \right] \e_3
        \nonumber \\
        &+ \left[
        \dfrac{A_{12}}{\left( 1+\beta_c \left(A_{11}+A_{22}+A_{33} -1\right)\right)}
         \left(\dfrac{1}{\lambda} + \dfrac{\beta_c \left(A_{11}+A_{22}+A_{33} -1\right)}{\lambda_R} \right) - \dot{\gamma}A_{22} \right] \e_4 . 
         \label{eqn:fnl_cdcrcs}
    \end{align}
    
    Furthermore, the extra stress tensor for stretching Gaussian chains is given by, 
    \begin{equation}
        \bm{\sigma}= 3G \left(\bm{A} - \frac{1}{3} \bm{I}\right).
        \label{sigma_cdcrcs}
    \end{equation}
    
    For the case of non-Gaussian chains,  $\bm{A}$ is replaced by $f_C \bm{A}$ in all the equations mentioned above except the terms associated with the convected derivative. This implies that the eqn. \eqref{eqn:cm_cdcrcs} becomes,
    \begin{align}
         \stackrel{\triangledown}{\bm{A}} & + \frac{1}{\lambda}\frac{f_C}{\left( 1+\beta_c \left(f_C\text{tr }\bm{A} -1\right)\right)}\left(\bm{A}- \frac{1}{3}\bm{I} \text{tr }\bm{A} \right) + \frac{1}{3\lambda_R}\left(f_C\text{tr }\bm{A} - 1\right)\bm{I} \\
         &+ \frac{f_C\beta_c \left(f_C\text{tr }\bm{A} -1\right)}{\lambda_R \left(1+\beta_c \left(f_C\text{tr }\bm{A} -1\right)\right)}\left(\bm{A}-\frac{1}{3}\bm{I}\text{tr }\bm{A}\right) = \bm{0}, 
         \label{eqn:cm_cdcrcs_ng}
    \end{align}
     where,
     \begin{equation}
         f_C = \dfrac{b-1}{b-\text{tr }\bm{A}}.
         \label{eqn:cdcrcs_fc}
     \end{equation}
     Here $b$ is the square of maximum chain stretch ratio $(b>1)$ and $f_C$ gives a Warner-type expression for finite extensibility. 
     The corresponding  $\bm{q}$ and $\bm{f}_\text{ex}$ are still given by eqns. \eqref{q_A} and \eqref{fex_A}, However, the terms in $\bm{f}_\nl$ change and can be derived by replacing all $A_{ij}$ with $\frac{b-1}{b-\left(A_{11}+A_{22}+A_{33}\right)}A_{ij}$ in eqn. \eqref{eqn:fnl_cdcrcs}, except for the terms associated with the convected derivative i.e. $2\dot{\gamma}A_{12}$ and $\dot{\gamma}A_{22}$ in the first and fourth brackets respectively. 
\end{enumerate}

Thus, the method of harmonic balance is flexible and can be used to find the LAOS response of any nonlinear differential CM.

\clearpage

\section{Thermodynamic Stability of CMM and TNM}

\medskip
In this segment, we present the contour plots for assessing the thermodynamic stability of the Corotational Maxwell model and the Temporary Network model based on the Ziegler criteria.

\begin{figure}[h]
    \begin{subfigure}{0.49\textwidth}
        \includegraphics[width=\textwidth]{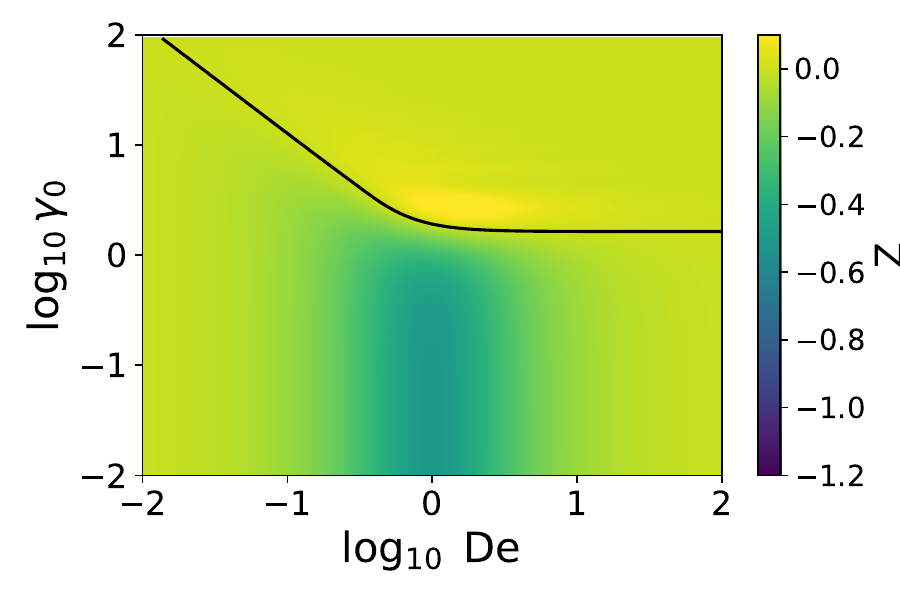}
        \caption{CMM, CPU $\approx$
        99 s}
        \label{fig:thermo_cmm}
    \end{subfigure}
    \begin{subfigure}{0.49\textwidth}
        \includegraphics[width=\textwidth]{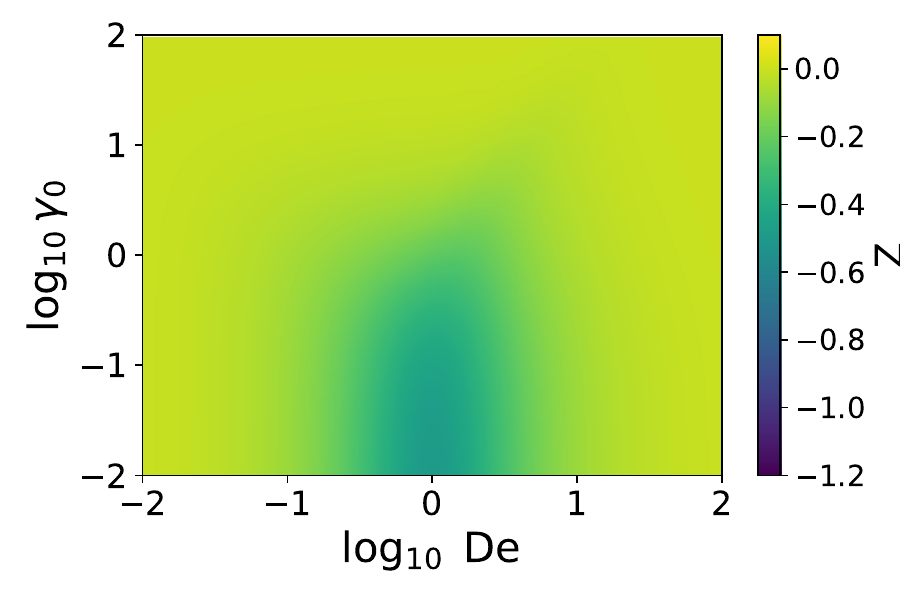}
        \caption{TNM, CPU $\approx$ 93 s}
        \label{fig:thermo_tnm}
    \end{subfigure}
    \caption{Contour plots of $Z$ for the CMM and TNM ($a$ =-1, $b$ = 1) equations. The thick black line corresponds to $Z=0$ that demarcates stability using the Ziegler criterion.}
    \label{fig:enter-label}
\end{figure}

Figures \ref{fig:thermo_cmm} and \ref{fig:thermo_tnm} represent the respective plots for the CMM and the TNM ($a$ = -1.0, $b$ = 1.0). The plots are created analogous to figs. 8a and 8b for the PTT and Giesekus models in the main text. The FLASH algorithm was implemented over 80 equispaced logarithmic points in the range $\gamma_0 \in [10^{-2},10^{2}]$ for 65 logarithmically spaced frequencies in $\De \in [10^{-2},10^{2}]$. Subsequently, $Z$ was calculated by interpolating the dataset at each frequency at 400 values of $\gamma_0$. The CPU time required to generate each plot is about 1.5 minutes. The CMM shows a large domain of instability and agrees well with the results of Saengow and Giacomin (their figure 4). \cite{Saengow2018} On the other hand, the TNM does not show any instability for the given $a$ and $b$ values across the probed $\gamma_0$ range. 

\clearpage

\section{Nomenclature and Abbreviations}
\begin{table}[h!]
    \centering
    \begin{tabular}{ll}
        \hline
         \textbf{Abbreviation} & \textbf{Description}
         \\
         \hline
         SAOS & Small Amplitude Oscillatory Shear \\
         MAOS & Medium Amplitude Oscillatory Shear \\
         LAOS & Large Amplitude Oscillatory Shear \\
         HB & Harmonic Balance \\
         AFT & Alternating Frequency Time \\
         IVP & Initial Value Problem \\
         NI & Numerical Integration\\
         AN & Analytical solution \\
         FLASH & Fast Large Amplitude Simulation using Harmonic balance \\
         UCM & Upper Convected Maxwell \\
         PTT & Phan-Thien Tanner \\
         TNM & Temporary Network Model \\
         CMM & Corotational Maxwell model \\
         CMs & Constitutive Models \\
         OS & Oscillatory Shear \\
         ODEs & Ordinary Differential Equations \\
         PSS & Periodic Steady State \\
         FT & Fourier Transform \\
         FFT & Fast Fourier Transform \\
         iFFT & Inverse Fast Fourier Transform \\ 
         \hline
    \end{tabular}
    \caption{List of abbreviations used in the text}
    \label{tab:abbr}
\end{table}

\begin{table}[]
    \centering
    \begin{tabular}{ll}
        \hline 
        \textbf{Symbol} & \textbf{Description} \\
        \hline
        $t$ & time \\
        $\omega$ & Frequency \\
        $T$ & Time period of one oscillation\\ 
        $\gamma_0$ & Strain amplitude \\
        $\dot{\gamma}$ & Applied strain rate \\
        $\sigma_{12}$ & shear stress \\
        $\sigma_{11},\sigma_{22},\sigma_{22}$ & Normal stresses \\
        $N_{1}$ ($N_2$)  & First (second) normal stress difference \\
        $G_n',G_n''$ & $n$th sine and cosine Fourier coefficients for shear stress \\
        $F_n',F_n''$ & $n$th sine and cosine Fourier coefficients for first normal stress difference \\
        $G',G''$ & Storage and loss modulus for the Maxwell model \\
        $\bm{\sigma}$ & Extra stress tensor \\
        $\bm{I}$ & Identity tensor\\
        $G$ & Modulus \\
        $\lambda$ & Relaxation time \\
        $g^{\text{PTT}}$ & Nonlinear term in the PTT model \\
        $\varepsilon$ & Nonlinear parameter of the PTT model \\
        $c(t),d(t)$ & Nonlinear terms in TNM \\
        $a,b$ & Nonlinear parameters in TNM \\
        $\De$ & Deborah number \\
        $\Wi$ & Weissenberg number \\
        $\bm{q}$ & Vector of unknown variables (stresses) \\
        $\bm{f}_{\nl}$ & Vector of nonlinear terms \\
        $\bm{f}_\text{ex}$ & Vector for external oscillatory forcing terms \\
        $\bm{q}^H$ & Truncated Fourier series form for $\bm{q}$  \\
        $H$ & No. of harmonics in the truncated series \\
        $\bm{B}^H$ & Vector of Fourier basis functions \\
        $B_n^H,B_s^H$ & Basis functions for normal and shear stress, respectively \\
        $\bm{r}$ & Residual in the time domain \\
        $\bm{r}^H$ & Truncated form of the residual\\
        $\hat{\bm{q}}^H$ & Fourier transform of $\bm{q}$ with $H$ harmonics \\
        $\hat{\bm{f}}_{\nl}$ & Fourier transform of $\bm{f}_{\nl}$ with $H$ harmonics\\
        $\hat{\bm{f}}_{\text{ex}}$ & Fourier transform of $\bm{f}_{\text{ex}}$ with $H$ harmonics\\
        $\hat{\bm{r}}^H$ & Fourier transform of $\bm{r}^H$ \\
        $\hat{\bm{r}}_c^H,\hat{\bm{r}}_s^H$ & Cosine and sine Fourier coefficients of $\bm{r}^H$ \\
        $\e_i$ & $i$th element of a vector \\
        $\tilde{\dot{\gamma}}$ & Dimensionless strain rate \\
        $\tilde{\sigma}_{ij}$ & Dimensionless stress \\
        $\tilde{t}$ & Dimensionless time \\
        $N$ & Number of data points used for AFT \\
        $\varepsilon_\omega$ & frequency-domain error metric\\
        $\varepsilon_t$ & time-domain error metric  \\
        \hline
    \end{tabular}
    \caption{List of symbols}
    \label{tab:nomenclature}
\end{table}

\clearpage

%\bibliographystyle{unsrt}
%\bibliography{hb,LAOS}

\printbibliography